\documentclass [11pt, reqno]{article}
\title{ \vskip -0.9in
An Extension of the Permutation Group Enumeration Technique\\
\large{(Collapse of the Polynomial Hierarchy: $\mathbf{NP = P}$)}
\vspace{ -0.2in}}
\author{\vspace{ -0.4in} Javaid Aslam\footnote{ \copyright Copyright Javaid Aslam 2009-2017, Santa Clara, CA 95054 All rights reserved.}
\ \ jaslamx@yahoo.com \date{\vspace{ -0.2in}\small{\today}}
\vspace{ -0in}
}

\usepackage{wrapfig}
\usepackage{float}
\usepackage{fancyhdr}
\usepackage{subfig}
\usepackage{calc}
\usepackage{amsthm}
\usepackage{amsfonts, stmaryrd}
\usepackage{amssymb, amsmath, latexsym}
\usepackage[utf8]{inputenc}
\usepackage[T1]{fontenc}
\usepackage{lmodern} 
\usepackage{setspace}
\usepackage[usenames]{color}
\usepackage{url}
\usepackage{lineno,hyperref}
\usepackage{algorithmic}
\usepackage[section]{algorithm}
\usepackage{eqparbox}

\usepackage[titletoc, title]{appendix}
\usepackage{graphicx}

\newtheorem{theorem}{Theorem}[section]
\newtheorem{fact}[theorem]{Fact}
\newtheorem{remark}[theorem]{Remark}
\newtheorem{corollary}[theorem]{Corollary}
\newtheorem{lemma}[theorem]{Lemma}
\newtheorem{claim}[theorem]{Claim}

\newtheorem{property}[theorem]{Property}
\theoremstyle{definition}
\newtheorem{definition}[theorem]{Definition}
\theoremstyle{remark}
\numberwithin{equation}{section}

\renewcommand{\abovecaptionskip}{7pt}
\renewcommand{\belowcaptionskip}{5pt}

\newcommand{\less} {\textless{}}
\newcommand{\more} {\textgreater{}}
\newcommand{\defn}{\overset{ \text{def}}{=}}
\newcommand{\vmpset}{%
V\hspace{-2pt}M\hspace{-1.50pt}P\hspace{-0.5pt}Set}%
\renewenvironment{proof}{{\bfseries Proof.}}{\par \qed \par}
\newcommand{\cvmp}{CV\hspace{-2pt}M\hspace{-1.0pt}P}
\newcommand{\vmp}{V\hspace{-2pt}M\hspace{-1.0pt}P}
\newcommand{\cms}{C\hspace{-1pt}M\hspace{-1.0pt}S\,}
\newcommand{\gms}{G\hspace{-1pt}M\hspace{-1.0pt}S\,}
\begin{document}
\raggedright
\maketitle
\vspace{ -.3in}
\begin{abstract}
\parindent 0.0in
\vskip -0.2in
\ \\
The distinguishing result of this paper  is a $\mathbf{P}$-time enumerable partition of all the $n!$ possible perfect matchings in a bipartite graph. This partition is a set of equivalence classes induced by the missing edges in  the  potential perfect matchings.\par \vskip 4pt
  We capture the behavior of these missing edges in a polynomially bounded representation of the exponentially many perfect matchings by a graph theoretic structure, called MinSet Sequence, where MinSet is a P-time enumerable structure derived from a graph theoretic counterpart of a generating set of the symmetric group. This leads to a polynomially bounded generating set of all the classes, enabling
the enumeration of  perfect matchings in polynomial time.
The sequential time complexity of this $\mathbf{\#P}$-complete problem is shown to be $O(n^{45}\log n)$.

\par \vskip 4pt
And thus we prove a result even more surprising than $\mathbf{NP = P}$, that is,
$\mathbf{\#P}=\mathbf{FP}$, where $\mathbf{FP}$ is the
class of functions, $f: \{0, 1\}^* \rightarrow \mathbb{N} $, computable in polynomial time on a deterministic model of computation.
\par
\vspace{5pt}
\end{abstract}

\section{Introduction}
\vskip -5pt
Enumeration problems \cite{GJ79} deal with counting the number of solutions in a given instance of
 a search problem, for example, counting the total number of perfect matchings in a bipartite graph. Their complexity poses unique challenges and surprises. Most of them are
$\mathbf{NP}$-hard, and therefore, even if $\mathbf{NP=P}$, it does not imply a polynomial time solution of an $\mathbf{NP}$-hard enumeration problem.
\par
$\mathbf{NP}$-hard enumeration problems fall into a distinct
 class of polynomial time equivalent problems called the $\mathbf{\#P}$-complete problems \cite {Val79a}. As noted by Jerrum \cite{Jerrum94counting},
 problems in $\mathbf{\#P}$ are ubiquitous- those in $\mathbf{FP}$ are more of an exception.
What has been found quite surprising is that the enumeration problem for perfect matching in a bipartite graph is $\mathbf{\#P}$-complete ~\cite{Val79} even though the associated
search problem has long been known to be in $\mathbf{P}$ ~\cite{kuhn55, egervary, Edmonds65}.
\par
Enumeration of a permutation group has long been known to be in $\mathbf{FP}$ (\cite{Butler91, LNCS82}).
The basic technique for enumerating a permutation group $G$ (any subgroup of the symmetric group $S_n$) is based on creating a hierarchy of the \emph{Coset Decompositions} over a sequence of the subgroups of $G$,
where the smallest subgroup is the trivial group $I$.
\par
\begin{wrapfigure}{r}{190pt}
\center
\vskip -32pt
 \includegraphics[scale=0.35]{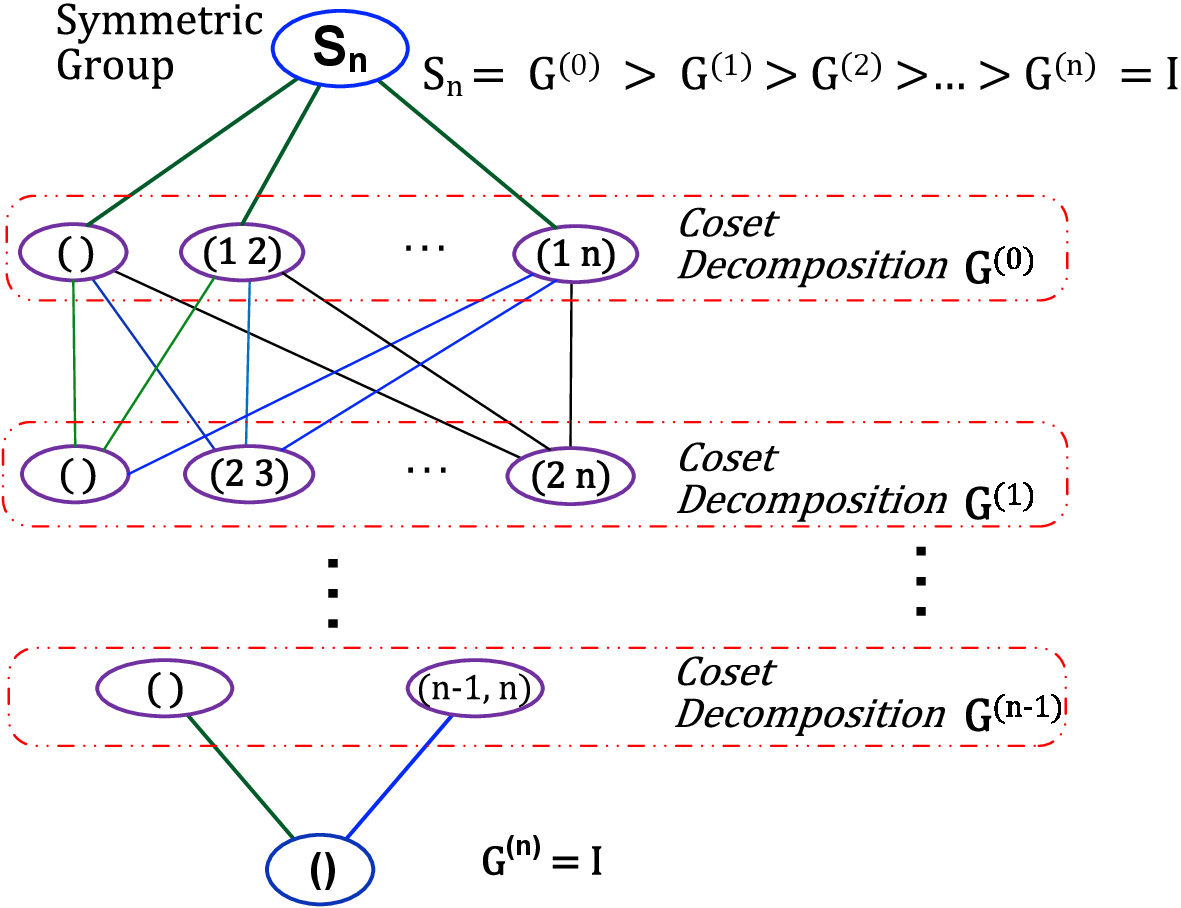}
  \caption{{A Hierarchy of the Coset \\Decompositions of $\mathbf{S_n}$}}\label{FG:CosetHier}
\vskip -00pt
\end{wrapfigure}

{A Coset Decomposition of $G$ is essentially a} {set of equivalence classes defining a \emph{Partition} of $G$} {for a subgroup $H$ of $G$, induced by} a set of group elements called
 \emph{Coset Representatives}(CR). Here each element $\psi$ in CR represents a unique subset of
 $G$, called Coset of $H$ in $G$, obtained by multiplying each element in $H$ by $\psi$, in certain (right or left) order.
For the symmetric group, $S_n$, the partition hierarchy for a fixed subgroup sequence is shown as an $n$-partite directed acyclic graph in
Figure \ref{FG:CosetHier}, where the nodes in each partition are the elements in CR representing the subsets of a group $G^{(i)}$ in the subgroup
sequence $G^{(0)} > G^{(1)} \cdots > G^{(n)} $.
 The edges represent a disjoint subset relationship.\\
\par
%

\par
\begin{wrapfigure}{r}{165pt}
\vskip -00pt
 \includegraphics[scale=0.29]{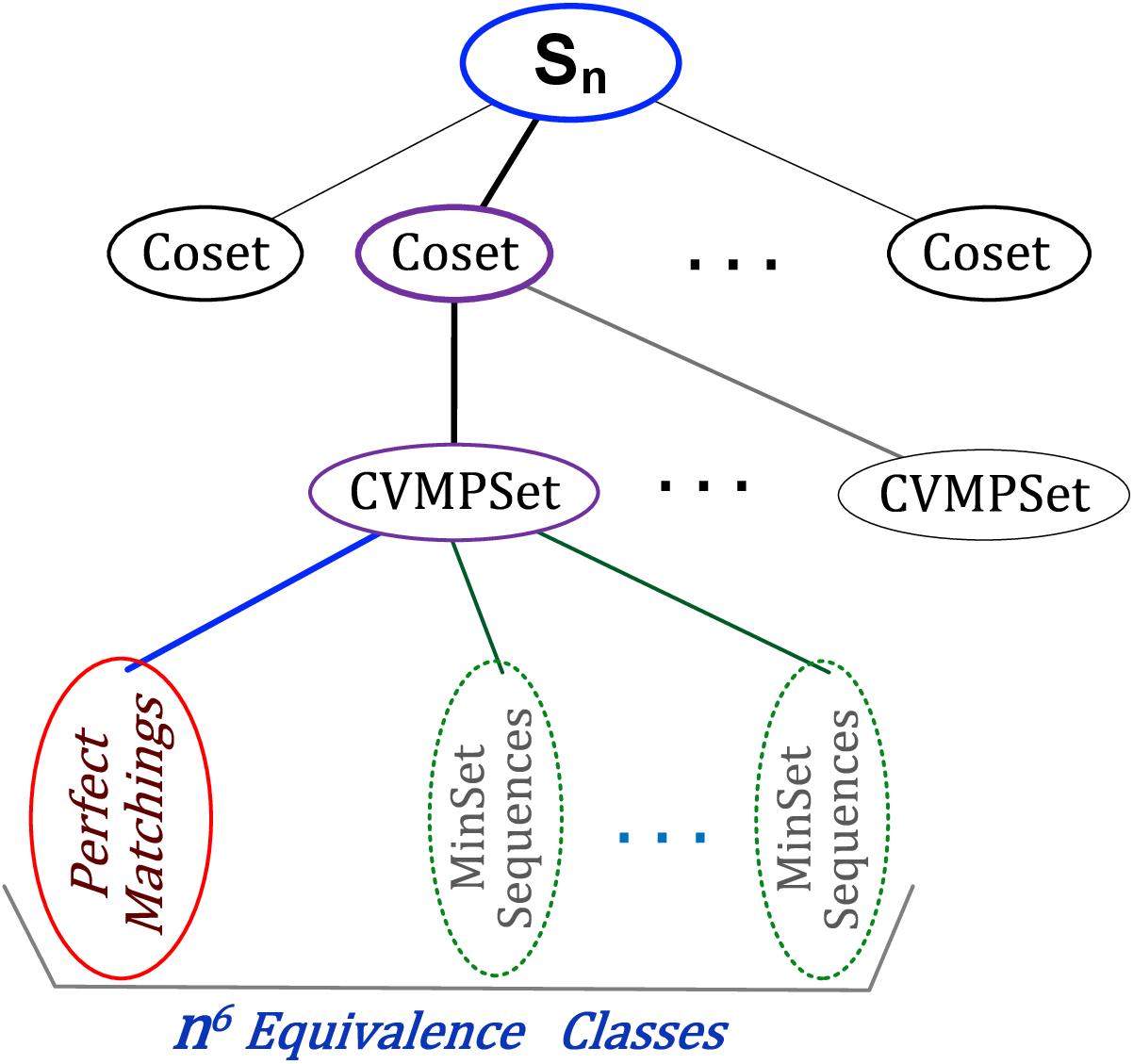}
\vskip -00pt
 \caption{{The Extended Partition Hierarchy}}\label{FIG:equiHierSimple}
\vskip -011pt
 \end{wrapfigure}

The enumeration technique for perfect matchings\\
 extends the above coset decomposition scheme\\ by further partitioning each coset
into a family of
polynomially many equivalence classes.
This extended
partition hierarchy (Figure \ref{FIG:equiHierSimple}) then captures the perfect matchings as an equivalence class in this
 partition, where each such class allows the P-time enumeration uniformly for all
$n \ge 3$.
\raggedright
 \vskip -000pt
\vspace {-0pt}
\par
The associated equivalence relation over a coset is induced by a graph theoretic
 attribute called \emph{edge requirements} which confirms a potential perfect matching subset in each equivalence class.
\par
The hierarchy of the various classes for a bipartite graph holds the following containment relationship:\\
\vspace {8pt}
\hskip 0.1in $S_n \supset Coset \supset [ other ~ equiv ~classes ] \supseteq \text{ Perfect Matchings}
$

The extended partition hierarchy contains ``other equivalence classes'' \emph{CVMPSets} and \emph{MinSet Sequences}, described below.
\par
We map a specific generating set of the
symmetric group $S_n$ to a graph theoretic ``generating set'', such that each coset representative of a (group, subgroup) pair is mapped to a set of graph theoretic coset representatives. This mapping is then used to construct a \emph{generating graph} for generating all the
perfect matchings as directed paths in the generating graph which is a directed acyclic $n$-partite graph of size $O(n^3)$.
\par

Each perfect matching in a bipartite graph with $2n$ nodes is expressed as a
unique directed path of length
 $n\!-\!1$, called Complete \emph{Valid Multiplication Path} (\cvmp) in the generating graph. The condition for a \cvmp~ of length $n\!-\!1$ to
represent a unique perfect matching in the given bipartite graph is captured by an
attribute of the CVMP, called
\emph{Edge Requirement} (ER).
\par
The graph theoretic coset representatives induce disjoint subsets of the Cosets, called
CVMPSets, an equivalence class containing the CVMPs.
\\
Each C\vmpset~ is further partitioned into polynomially bounded classes called \emph{MinSet Sequences} induced by the ER of each CVMP, where a MinSet is the set of all Valid Multiplication Paths (VMPs) of common ER.
 \par
A judicious choice of the common ER of these VMPs allows a MinSet and any sequence of the MinSets to be P-time enumerable,  and which makes the perfect matchings also P-time enumerable as follows.
\par
These MinSet sequences can be viewed as an instance of a perfect matching subproblem, where a sequence containing only one MinSet would represent a set of perfect matchings when each CVMP is of length $n\!-\!1$ with common $ER=\emptyset$.
\par
\begin{wrapfigure}{r}{220pt}
 \includegraphics[scale=0.50]{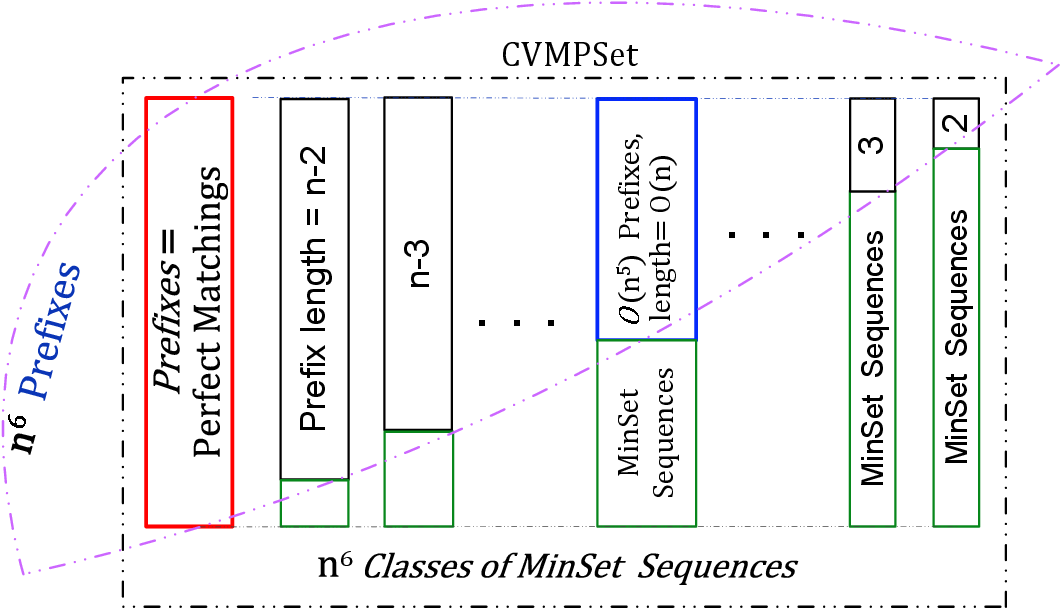}
\vskip -00pt
 \caption{Partition of a CVMPSet into\\ MinSet Sequences}\label{FIG:cosetPartitions}
 \end{wrapfigure}

There are exponentially many MinSet sequences in a CVMPSet, and all of them can be covered by only $O(n^6)$ unique prefix MinSets. And thus these unique prefixes induce polynomially many
equivalence classes each containing exponentially many MinSet sequences (Figure \ref{FIG:cosetPartitions}).\\
Further, each class size decreases exponentially with $n$, and thus this hierarchy enables enumeration of all the equivalence classes in polynomial time.
 \par
\par\par
Main results of this paper are summarized as follows:
\begin{enumerate}
\item The Mapping : Section 3 develops the concepts leading to the graph theoretic
counterparts of the permutation group generating set.
It defines the mapping (Lemma \ref{L:mapGenset}) of a specific generating set of $S_n$ to a set of graph theoretic structures.

\item {The Extended Partition Hierarchy:}\\
Section 4 describes how to partition all the potential perfect matchings into a set of equivalence classes induced by the missing edges in all the  potential perfect matching in the given bipartite graph. \par
 Lemma \ref{L:CountofCMS} states  how exponentially many MinSet sequences in any CVMPSet are partitioned into polynomially many equivalence classes of polynomially many prefixes to the MinSet sequences.

\item Polynomial Time Enumeration: Algorithm \ref{ALG:countMatchings} describes a high level logic to count all the perfect matchings in time $O(n^{45}\log n)$.
\end{enumerate}

Section \ref{S:conclusion} provides the conclusion, collapse of the Polynomial Time Hierarchy.

\newpage
\section{Preliminaries: Group Enumeration}\label{S:basicConcepts}
\vspace{-7pt}
The following concepts can be found in many standard books (including \cite{Butler91}) on permutation group theory. The notations and definitions used here are taken mostly from ~\cite{LNCS82}.
\par
\subsection{The Permutation Group}
 \vspace{-7pt}
Let $G$ be a finite set of elements, and let ``$\cdot$'' be an associative binary operation, called \emph{multiplication}. Then $G$ is a group if satisfies the following axioms:
\begin{enumerate}
\item $\forall x, y \in G,~ x \cdot y \in G$.
    \item there exists an element, $e \in G$, called the identity, such that $\forall x \in G,~ x\cdot e = e \cdot x = x$.
    \item $\forall x \in G$, there is an element $x^{-1} \in G$, called the inverse of $x$, such that $x\cdot x^{-1} = x^{-1} \cdot x = e$.
\end{enumerate}
Let $H$ be a subgroup of $G$, denoted as $H < G$. Then $ \forall g \in
G$ the set $H\cdot g = \lbrace h\cdot g | h \in H \rbrace $ is
called a right \emph{coset} of $H$ in $G$.
 Since any two cosets of a subgroup are either disjoint or equal, any group $G$ can be
 partitioned into its right (left) cosets. That is, using the right cosets of $H$ we can
 partition $G$ as:
\begin{equation}\label{e:cosetRep0}
 G ~=~ \biguplus_{i=1}^{r} H\cdot g_i
 \end{equation}
 The elements in the set $\lbrace g_1, g_2, \cdots, g_r \rbrace$ are called the \emph{right coset representatives} (AKA a \emph{complete right traversal}) for $H$ in $G$.
\par
 In this paper we will deal with only one specific type of finite groups called \emph{permutation groups}.\\
  A \emph{permutation} $\pi$ of a finite set, $\Omega= \{ 1, 2, \cdots, n \}$, is a 1-1 mapping from $\Omega$ onto itself, where for any $i \in \Omega$, the image of $i$ under $\pi$ is denoted as $ i ^\pi$. The product of two permutations, say $\pi$ and $\psi$, of $\Omega$ will be defined by $i ^{\pi\psi} = (i^\pi)^\psi$.\\
   A permutation group contains permutations of a finite set $\Omega$ where the binary operation, the multiplication, is the the product of two permutations. The group formed on all the permutations of $\Omega$ is a distinguished permutation group called the \emph{Symmetric Group} of $\Omega$, denoted as $S_n$.
\par
We will use the cycle notation for permutations. That is, if a permutation $\pi = (i_1, i_2, ~\cdots~ i_r),$ where $~ i_x \in \Omega$, and $ r \le n$, then $i_x ^ \pi = i_{x+1}$, for $1 \le x <r$ and $i_r ^ \pi = i_{1}$. Of course, a permutation could be a product of two or more disjoint cycles.

\subsection{The Enumeration Technique}

A permutation group enumeration problem is essentially finding generators for all the permutations in the group, and thus leading to finding \emph{order} of the group. It can also be viewed as an enumeration problem corresponding to any search problem ~\cite {GJ79} over a finite universe.
\par
A \emph{generating set} of a permutation group $G$ is defined to be the set of permutations, $K \subset G$, such that all the elements in $G$ can be written as a (polynomially bounded) product of the elements in $K$.
\par
Let $G^{(i)}$ be a subgroup of $G$ obtained from $G < S_n$ by fixing all the points in $\lbrace 1, 2, \cdots, i \rbrace$. That is, $\forall\,\pi \in G^{(i)}$, and $\forall j \in \lbrace 1, 2, \cdots, i \rbrace$, $~j^\pi = j$. Then it is easy to see that $G^{(i)} < G^{(i-1)}$, where $1 \le i \le n$ and $G^{(0)} = G$.
The sequence of subgroups
\begin{equation}\label{e:tower}
I = G^{(n)} < G^{(n-1)} < ~ \cdots ~ <G^{(1)} < G^{(0)} = G
\end{equation}
 is referred to as a \emph{stabilizer chain} of $G$.
\par

The stabilizer chain in \eqref {e:tower} gives rise to a generating set given by the following Theorem ~\cite{LNCS82}.
\begin{theorem}
~\cite{LNCS82} Let $U_i$ be a set of right coset representatives for $G^{(i-1)} $ in $G^{(i)}, ~ 1 \le i \le n$. Then a generating set $K$ of the group $G < S_n$ is given by
\begin{equation}
 K = ~\bigcup_{i=1}^n U_i,
\end{equation}
\end{theorem}
Group enumeration by a generating set creates a canonic representation of the group elements, i.e., a mapping $f$ defined as
\begin{equation}\label{e:gen-group1}
f: \underset{{i=n}}{\overset{1}{\mathsf{X}}} U_i \rightarrow G = \{ \psi_n \psi_{n-1}\psi_{n-2} \cdots \psi_i\psi_{i-1} \cdots \psi_2 \psi_1 \mid \psi_i \in U_i \}.
\end{equation}
The order $|G|$ can then be easily computed in time $O(n^2)$ by
\begin{equation}\label{e:group-order}
\mid\! G\! \mid = \prod_{i=1}^n \mid\! U_i \!\mid
\end{equation}

These generating sets are not unique, and the one we are interested in is derived from those coset representatives that are transpositions (except for the identity). That is, for $G = S_n$ and for the subgroup tower in Eqn. \eqref{e:tower}, the set of coset representatives $U_i$ are \cite{LNCS82}
\begin{equation}\label{e:cosetReps}
U_i = \lbrace I, (i, i+1), (i, i+2), ~\cdots, ~ (i, n)\rbrace, ~ 1
\le i < n.
\end{equation}
Then the generating set $K$ of $S_n$ is given by
\begin{equation}\label{e:cosetRep1}
K=\bigcup U_i = \lbrace I, (1, 2), (1, 3), ~\cdots, ~(1, n), (2, 3), (2, 4), ~\cdots,
~ (2, n), ~\cdots, ~ (n-1, n)\rbrace
\end{equation}
The partition hierarchy of the coset decompositions for $S_n$ is shown above in Figure \ref{FG:CosetHier}.
\newpage
\subsubsection*{Example}
 All the coset representatives $U_i$ for the stabilizer chain \eqref{e:tower} of the symmetric group $S_4$ are shown in Table \ref{TBL:Gen_S_4}. Each permutation in $S_4$ can be expressed as a unique ordered product, $\psi_4\psi_3\psi_2\psi_1$, of the four permutations $\psi_1 \in U_1,~ \psi_2\in U_2,~ \psi_3\in U_3$ and $\psi_4\in U_4$.\par
\vspace{0.1cm}
\begin{table}[h]
\centering
\small{
\begin{tabular}{|c|c|c|c|} \hline
\em $ U_1 $ & \em $ U_2$ &$ U_3$ &\em $ U_4$ \\
 \hline \hline
$\{ I, (1,2), (1,3), (1,4)\}$ & $\{ I, (2, 3),(2,4)\}$ & $\{ I, (3,4)\}$ & $\{ I\}$\\ \hline

\end{tabular}
}
\vspace{0.2cm}
\caption{{The Generators of $S_4$}}\label{TBL:Gen_S_4}
\end{table}
\vspace{-0.0cm}
For example, the permutation (1,3,2,4) in $S_4$ has a unique canonic representation, $\psi_4\psi_3\psi_2\psi_1 = I*(3,4)*(2,4)*(1,3)$; the element $(1,2)$ is represented as $I*I*I*(1,2)$.\\
 Also, note that under this enumeration scheme the order of $S_4$ is the product, $|U_1|* |U_2|* |U_3|* |U_4|$.

\subsubsection*{Perfect Matchings in a Bipartite Graph}
\vspace{-5pt}
Let $BG = (V\cup W, E)$ be a  bipartite graph on
$2n$ nodes, where, $|V| = |W|$, $E = V \times W$ is the edge set, and both the node sets $V$ and $ W$ are labeled from $\Omega = \lbrace 1, 2, \cdots, n \rbrace$ in the same order. Under such an ordering of the nodes, the node pair $(v_i, \,w_i) \in V \times W $ will often be referred to as simply the \emph{node pair at position $i$} in $BG$.
\par
A \emph{perfect matching} in  $BG$ is a subset $M \subseteq E$ of the edges in $BG$ such that each node in $BG$ is incident with exactly one edge in $M$. A perfect matching $M$ in $BG$ represents a permutation $\pi$ in $S_n$, and hence a 1-1 onto correspondence in a natural way. That is, for each edge $(v, w) \in M \Longleftrightarrow v^\pi = w$.
\par
We will use the above group generating set concepts in developing a combinatorial structure, i.e., a graph theoretic analog of the generating set $K$, for generating all the perfect matchings in a bipartite graph.

\newpage
\vspace{-7pt}
\section{The Mapping: Graph Theoretic Generators}
\vspace{-11pt}
In this Section we develop the basic framework for defining the partition hierarchy of the Symmetric group applicable to the perfect matchings in a bipartite graph.
We will develop the concepts leading to the graph theoretic counterparts of the permutation group generating set. Specifically, we choose the
generating set of the Symmetric group, $S_n$, to be the set of transpositions given by
Eqn. \eqref {e:cosetRep1}, with the \emph{Right Coset Decomposition} (Eqn. \eqref{e:cosetRep0}) for group enumeration.
\par
First we develop the concepts for defining the permutation multiplication in a bipartite graph.

\vspace{-8pt}
\subsection{Permutation Multiplication in a Bipartite Graph}
\vspace{-7pt}
Let $BG'$ be a bipartite subgraph of $K_{n,n}$ on $2n$ nodes, and let $E(\pi)$ denote a perfect matching in $BG'$ realizing a permutation $\pi \in S_n$.
\begin{theorem}\label{TH:multBasic1}
Let $\pi \in S_n$ be realized as a perfect matching $E(\pi)$ in $BG'$. Then for any transposition, $\psi \in S_n$, the product $\pi\psi$ is also realized by $BG'$ iff $BG'$ contains a cycle of length 4 such that
the two alternate edges in the cycle are covered by $E(\pi)$ and the other two by $E(\pi\psi)$.
\end{theorem}
(The proof is in Appendix \ref{ap:multBasic1})
\par

\textbf{Examples.}
\par
\begin{figure}[h]
\vspace{-5pt}
\center
\includegraphics[scale=0.70]{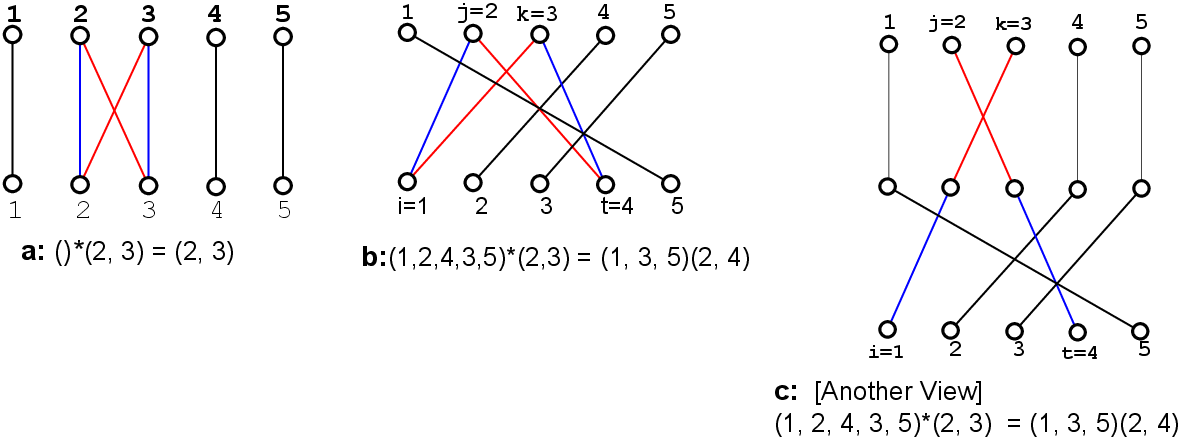}
\vspace{0pt}
  \caption{\textbf{Permutation Multiplication in a Bipartite Graph}}\label{FG:mult1}
\vspace{-15pt}
\end{figure}

Figure \ref{FG:mult1}(a) shows an instance of $BG'$ having two perfect matchings, and realizing
the product $\pi\psi$ of the permutations $\pi = I$ (identity permutation) and $\psi = (2,3)$.
Figure \ref{FG:mult1}(b) shows the two perfect matchings, realizing $\pi$ and $\pi\psi$, and the associated graph cycle $(1,2,4,3) $, representing the multiplier (2, 3), responsible for the multiplication. \\
Figure \ref{FG:mult1}(c) shows the same multiplication $\pi\psi = (1, 2, 4, 3, 5)*(2, 3)$ as a cascade of the two associated perfect matchings in the two bipartite graphs.
\par

\newpage
 \vspace{-9pt}
\newpage
\subsection{Graph Theoretic Coset Partition}
\vspace{-5pt}

We can now define a set of graph theoretic counterparts of the Coset Representatives, called \emph{partition representatives}. These are obtained by mapping the symmetric group $S_n$ generators in \eqref{e:cosetRep1} to a set of edge pairs in $K_{n,n}$ with the help of Theorem \ref{TH:multBasic1}.
\par
Let
$\mathbb{M}(BG')$ denote the set of permutations realized as perfect matchings in a bipartite graph $BG'$.
Let $BG_i$ denote the sub
(bipartite) graph of $BG=K_{n,n}$, induced by the subgroup $G^{(i)}$, such that $\mathbb{M}(BG_i) = G^{(i)}$.
 That is, $\forall t \in \{t \mid 1 \le t \le i\}$, there is exactly one edge $v_t w_t$ incident on the nodes $v_t$ and $w_t$. Moreover, $BG_i $ contains a complete bipartite
graph, $K_{n-i, n-i}$, on the nodes at positions $i+1, i+2, ~..., ~ n$.

 \par
The following is a direct Corollary of Theorem \ref{TH:multBasic1}, and is the basis for mapping the Coset representatives in $U_i$ to a set of edge pairs in bipartite graph $K_{n,n}$.\\
Let $K_{n,n} = (V \cup W, V \times W)$, where $V = W = \{1, 2,\, \cdots\,, n\}$.\\
Let $(v_i w_k, v_t w_i) \in (V \times W, V \times W)$ be an ordered pair of edges in $K_{n,n}$, where $n \ge t >i$.
\renewcommand{\abovecaptionskip}{5pt}
\renewcommand{\belowcaptionskip}{5pt}
\begin{figure}[h]
\vspace{5pt}
\center
  \includegraphics[scale=0.63]{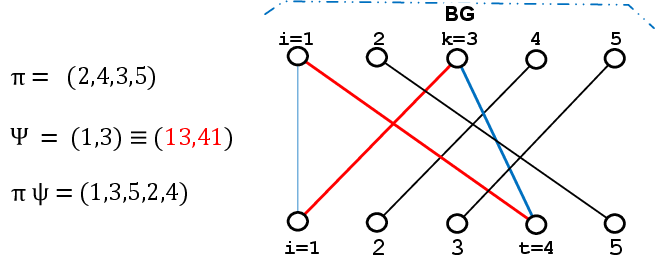}
   \caption{\textbf{Multiplication by a Coset Representative $\mathbf{\psi=(1,3)}$}}\label{FG:multCoset}
\vspace{-9pt}
\end{figure}
\renewcommand{\abovecaptionskip}{5pt}
\renewcommand{\belowcaptionskip}{5pt}


\begin{corollary}\label{CR2:cosetEdge}
Let $BG =K_{n,n}$ be a complete bipartite graph, and $U_i $ be a set of right coset representatives of $G^{(i)}$ in $G^{(i-1)}$ given by Eqn. \eqref{e:cosetReps}. Then,
for each $\psi =(i, k) \in U_i $, and for each $\pi\in G^{(i)}$, $\pi\psi$ is realized by $BG_{i-1}$ using a unique edge pair, $a_i(\pi, \psi)=(v_i w_k, v_t w_i)$ in $BG_{i-1}$, which forms a cycle of length 4 with $E(\pi)$ and represents
$\psi$ for a given $\pi$, such that $i^\psi = k = t^\pi$.
\par
When the coset representative $\psi$ is an identity, we have a special case of the above
behavior where the edge pair $a_i(\pi, \psi)$ reduces to one edge $v_i w_i$
for each $\pi \in G^{(i)}$.
\end{corollary}

\begin{definition}\label{D: mGenset}
A \emph{partition representative}, $g(i)$, $1 \le i \le n$, for the subgraph $BG_i$ in $K_{n,n}$ is defined as:
\begin{equation}\label{e:mGenset}
 g(i) \,\defn\, \big\{(i k, t i) \, \mid \, k,t \in \{i+1,\, \cdots\,, n\} \big\}
 \bigcup \big \{(ii, ii) \big\},
\end{equation}
where $(v_i,w_k,v_t,w_i)$ is a cycle of length 4 in $K_{n, n}$.
\end{definition}
The following Property states how each $g(i)$ is partitioned into $n-i+1$ equivalence classes induced by each $\psi \in U_i$.
\begin{property}\label{P:partitionSubset}
For each $(\pi, \psi) \in G^{(i)} \times U_i$, if $\psi = (i,k)$ is fixed, then the subset of $g(i)$ that can realize $\pi\psi$ is:
\begin{equation}\label{e:subsetMult}
   \{(ik, ti)\mid\, i^{\psi} = k =t^\pi\},\text{ where } k,t \in \{i+1,\, \cdots\,, n\}
\end{equation}
\end{property}
\newpage
The following Lemma states the exact mapping, showing how $ g(i)$ is effectively the graph theoretic counterpart of $U_i$, where the element $(ii,ii)\in g(i)$ is the ID element corresponding to $I \in U_i$. We will denote each $(ii,ii) \in g(i)$ by $I$.
\par
\begin{lemma} \label{L:mapGenset}
There exists a 1-1 mapping
\vspace{ -7pt}
$$
 h : G^{(i)} \times U_i \longrightarrow g(i) \times M(BG_i),\vspace{ -0pt}
 $$
\par
\vspace{ -8pt}
 \hskip 0.0in s.t., $\forall (\pi,\psi) \in G^{(i)} \times U_i$,
   $\pi\psi $ is realized by a unique pair $ (x_i, pm_i) \in g(i) \times M(BG_i)$ using a unique cycle $(v_i,w_k,v_t,w_i)$ of length 4 in $BG_{i-1}$, defined by $x_i = (ik,ti)$,  such that the edge pair $x_i$ is covered by $\pi\psi$, and the other two alternate edges in the cycle are covered by $\pi$.\\
   When $\psi =I$, the identity in $S_n$, the cycle collapses to one edge $x_i =(ii,ii)$ covered by $\pi$ and $\pi\psi$ both.
\end{lemma}
 \begin{proof}
 Follows from Corollary \ref{CR2:cosetEdge} and noting that the ID element in $U_i$ is not mapped by a cycle in $K_{n-i+1,\, n-i+1}$ but corresponds to the edge $(v_i,w_i) \in V\times W$.
\vskip -5pt
\end{proof}

Now we can define a generating set for enumerating all the perfect matchings in $K_{n,n}$, analogous to a generating set for the Symmetric group $S_n$. \\
Let $\psi(a_i)$, $a_i \in g(i)$, denote a transposition $\psi \in U_i $ realized by the edge pair $a_i$.
\begin{definition}\label{D:GSet}
A generating set, denoted as $E_M(n)$, for generating all the $n!$ perfect matchings in a complete bipartite graph $ K_{n,n}$ is defined as
\begin{equation}\label{e:GSet}
E_M(n)~ \defn ~ \bigcup_{i=1}^{n} g(i)
\end{equation}
\end{definition}
\vspace{ -11pt}
\subsubsection{The Multiplicative Behavior of the Generators}
\vspace{ -4pt}
Group enumeration by a generating set creates a canonic representation of the group elements, i.e., a mapping $f$ defined as
\begin{equation}\label{e:gen-group1}
f: \underset{{i=n}}{\overset{1}{\mathsf{X}}} U_i \rightarrow \{ \psi_n \psi_{n-1}\psi_{n-2}\, \cdots\, \psi_i\psi_{i-1} \cdots \psi_2 \psi_1 \mid \psi_i \in U_i \}= G.
\end{equation}
This suggests the need for a set of similar rules that would govern the multiplication of $g(i)$ elements. The binary relations in the next sub sections are aimed to do precisely that.

\subsubsection*{Binary Relations over the Generating Set $\mathbf{E_M}$}
\vspace{ -7pt}
We now formulate two binary relations, $R$ and $S$, over the generating set $E_M$. These two relations will be used to capture a far more complex relationship than what exists in a permutation group generating set in order to enumerate the  group elements.

\par
First we define some more terms.
\par
Let $\pi(a_i), ~a_i \in g(i)$, represent a class of permutations defined as follows.
\begin{equation}
\pi(a_i) \defn \pi | \pi \in G^{(i-1)} \text{ and } E(\pi) \text{ covers } a_i.
\end{equation}
For brevity we will often qualify a permutation $\pi \in G$ as ``$\pi$ \emph{covers a set of edges e}'' whenever the corresponding perfect matching, $E(\pi)$ in $K_{n,n}$ covers $e$.
\par
Let $\psi(a_i)$ denote the coset representative of $G^{(i)}$ in $G^{(i-1)}$ realized by the edge pair $a_i$ for some $\pi \in G^{(i)}$ such that $\pi(a_i) = \pi\psi(a_i) \in M(BG_{(i-1)})$.
\par
Corresponding to the identity coset representative $I \in U_i$ we will call the edge pair $(v_i w_i, v_i w_i)$ at node pair $i$ as \emph{identity} edge pair, denoted by $id_i$.

%
\subsubsection*{The Transitive Relation $R$ over $E_M(n)$}
\vskip -9pt
The following definition of the relation $R$ specifies the condition under
which two coset representatives, $\psi(a_i)$ and $\psi(b_j)$, corresponding to the two edge pairs $a_i \in g(i)$ and $ b_j \in g(j), ~i < j$, may realize the product, $\pi(b_j)\psi(a_i)$ by the bipartite graph $BG=K_{n,n}$. 
\begin{definition}\label{D:TRelation}
 \vskip -0pt

 Two edge pairs $a_i \in g(i)$ and $ b_j \in g(j), ~ 1\le i < j \le n$,
 are said to be related by the relation $R$, denoted as $a_i R b_j$, if
$\forall \pi(b_j) \in G^{(j-1)}$, $\exists \psi(a_i) \in G^{(i-1)}$, such that $\pi(b_j)\psi(a_i) \in G^{(i-1)}$ is realized by $K_{n,n}$, conforming to Corollary \ref{CR2:cosetEdge}.
\end{definition}
\vskip 0pt
 \vskip -5pt
\textbf{Example}.
 \vskip -9pt
 \begin{figure}[h]
\center
  \includegraphics[scale=0.65]{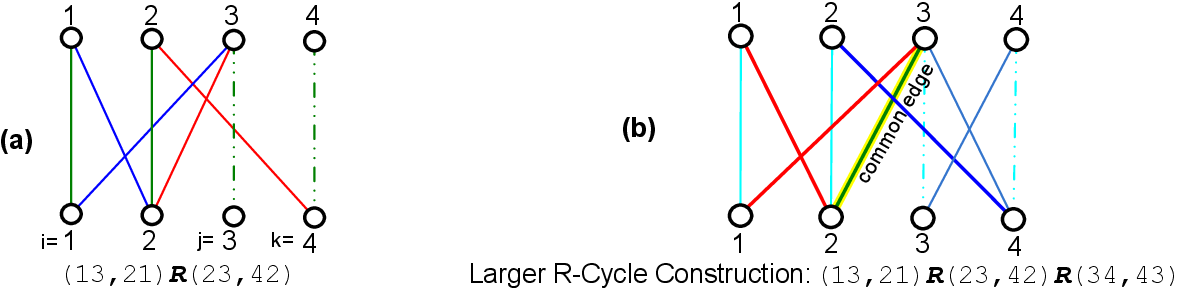}
\vskip 5pt
  \caption{\textbf{The Edge Pairs Forming the Cycles and the Relation $R$}}\label{FG:R-cycle-TR}
\vskip -5pt
\end{figure}

\begin{lemma}\label{PR:tr-R}
The relation $R$ is transitive.
\end{lemma}
\begin{proof}
Let $a_i R b_{j}$ and $b_j R c_k$.\\
First we consider the case when each of these edge pairs, $(a_i, b_j)$ and $(b_j, c_k)$, are related by the virtue of two graph cycles, $C_{ab}$ and $C_{bc}$ respectively, satisfying the condition of Corollary \ref{CR2:cosetEdge}. Then, for $j-i=1$, it is easy to see that we can merge $C_{ab}$ and $C_{bc}$ by deleting the edge, say $e \in b_j$, common to $C_{ab}$ and $C_{bc}$, and replacing it with the 2 edges in $a_i$ [Figure \ref{FG:R-cycle-TR}(b)].\\
Therefore, $a_i R c_k$.\\
Clearly this process of merging the two cycles can be iterated for each $(j,k)$, $i <j<k\le n$.
The result then follows from the induction on $j-i$.
\par
In the event that $a_i =I$, the transitivity is trivial as $I$ would multiply each permutation.\\
In the event that $\psi$ in the above definition is composed of 2 or more disjoint permutation cycles, the corresponding graph cycles are also disjoint and hence their composition is always valid.
Therefore, $a_i R c_k$.
\end{proof}

\parindent 0in
\subsubsection*{The Disjoint Relationship}
\vspace {-0.08in}
\begin{definition}\label{D:disjoint1}
Any two edge pairs $a$ and $b$ in
$E_M$ are said to be \emph{disjoint} if (i) the corresponding edges in the bipartite
graph $BG$ are vertex-disjoint, and (ii) $aRb$ is false.
\par
When the disjoint edge pairs $a$ and $b$ belong to two
adjacent node partitions, i.e., $a \in g(i)$ and $b \in g(i+1)$, $1 \le i < n,$ $a$ and $b$ are said to be related as $aSb$.
\end{definition}
\subsection{The Generating Graph}
\vspace{-8pt}
We now develop graph theoretic concepts to represent each permutation in $S_n$ by a directed path in a directed acyclic graph, called \emph{generating graph}, denoted as $\Gamma(n)$. This \emph{generating graph} models the multiplicative behavior of the elements in the set $E_M$ (Eqn. \eqref {e:GSet}).
\begin{definition}
The generating graph $\Gamma(n)$ for a
complete bipartite graph $K_{n,n}$ on $2n$ nodes is defined as
\vspace {-0.050in}
\[
\Gamma(n) \defn (V, ~E_R \cup E_S),
\]
\vspace {-0.25in}
\begin{align*}
\text{where } V &= E_M = \cup g(i)~~ \text{ (Eqn. \eqref {e:GSet})},\\
E_R &= \{ a_i a_j\,|\, a_i R a_j, ~ a_i \in g(i), a_j \in g(j), ~|a_i R a_j | = 1, ~1 \le i < j \le n \}, \text{ and}\\
E_S &= \{ b_i b_{i+1} \,| \, b_i S b_{i+1}, ~b_i \in g(i) \text{ and }
b_{i+1} \in g(i+1),~1 \le i < n \}.
\end{align*}
\end{definition}
\begin{figure}[h]
\center
  \includegraphics[scale=0.550]{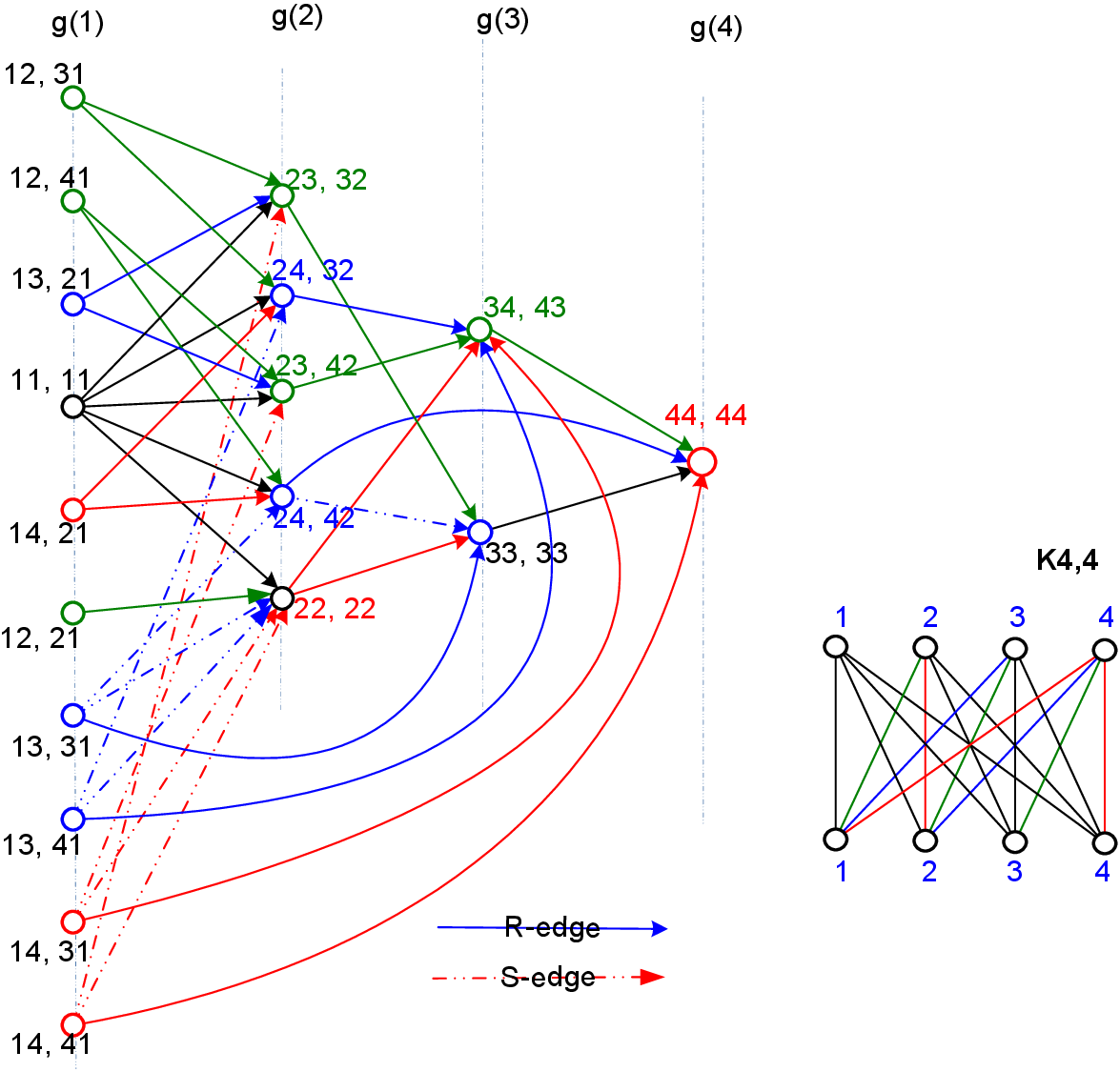}
  \caption{\textbf{The Generating Graph $\Gamma(4)$ for $K_{4,4}$}}\label{FG:GenK44}
\vspace {0pt}
 \end{figure}
\vspace {-0.10in}

Thus the generating graph is an $n$-partite directed acyclic
graph where the nodes in the partition $i$ are from $g(i),~ 1 \le i \le n$ (Eqn. \eqref{e:mGenset}), representing the right coset representative $U_i$ of $G^{(i)}$ in $G^{(i-1)}$, and therefore, are labeled naturally by the corresponding edge pairs, $g(i)$. \\
The edges in $\Gamma(n)$ represent either the transitive relation $R$ (by a solid directed line) between the two nodes, or the \emph{disjoint} relationship between the two nodes (by a dotted directed line) in the adjacent partitions. Each edge is a directed edge from a lower partition node to the higher partition node. Figure \ref {FG:GenK44} shows a generating graph $\Gamma(4)$ for the complete bipartite graph $K_{4,4}$.
\par
The edges in $E_R$ will be referred to as $R$-edges. Similarly, the edges in $E_S$ will be referred to as $S$-edges. An $R$-edge between two nodes that are not in the adjacent partitions will be called a \emph{jump} edge, whereas those between the adjacent nodes will sometimes be referred to as \emph{direct edges}. Moreover, for clarity we will always represent a jump edge by a solid curve.
\begin{definition}
An $R$-path, $a_i, a_{i+1}, \, \cdots\, , b_j \in \Gamma(n)$ is a sequence of adjacent $R$-edges between the two nodes $a_i, b_j \in \Gamma(n), j > i$ such that $a_i R b_j$.
\end{definition}
Note: The treatment of a ``path'' formed by a sequence of adjacent $R$ and $S$-edges (generally called as $RS$-path) is more complex and will be discussed in the next sub Section.

\begin{property} \label{PR:gen-graph-attr}
The following attributes of the generating graph $\Gamma(n)$ follow from the Properties stated in Appendix \ref{a:genGraphProps}.
\begin{subequations}

\begin{equation}
\text{Total number of nodes in the partition $i$ is } |g(i)| = (n-i)^2 + 1, ~ 1 \le i \le n
\vspace {-5pt}
\end{equation}
\begin{equation}
\text{Total number of nodes in } \Gamma(n)= O(n^3)
\vspace {-5pt}
\end{equation}
\vspace {-0pt}
\begin{equation}
 \text{Maximum $R$-outdegree of any node (except the ID) at partition } i = n-i
\vspace {-4pt}
\end{equation}
\begin{equation}
\text{Maximum $S$-outdegree of any node at partition } i = (n-i-2)^2 + 1, ~ 1 \le i <n-1
\vspace {-5pt}
\end{equation}
\begin{equation}
\text{Total number of R-edges in } \Gamma(n) = O(n^4)
\vspace {-5pt}
\end{equation}
\begin{equation}
\text{Total number of S-edges in } \Gamma(n) = O(n^5)
\vspace {-00pt}
\end{equation}
\end{subequations}
\end{property}

\begin{definition}
An $R$-path, $a_i, a_{i+1}, \, \cdots\, , b_j \in \Gamma(n)$ is a sequence of adjacent $R$-edges between the two nodes $a_i, b_j \in \Gamma(n), j > i$ such that $a_i R b_j$.\\
\emph{Length} of an $R$-path $aRb$ is the number of adjacent $R$-edges between the node pair $(a,b) \in \Gamma(n)$.
\end{definition}
\vspace{-9pt}
Two $R$-edges, $x_i x_t$ and $x_j x_t$, $i\ne j$, meeting at a common node $x_t \in \Gamma(n)$ are said to be \emph{disjoint} if the associated cycles in $K_{n,n}$ are vertex disjoint.
\par
Now we can define a distinguished path called \emph{valid multiplication path} (VMP) in a generating graph $\Gamma(n)$, in order to represent a perfect matching in a bipartite graph.
\vspace{-8pt}
\subsubsection{\textbf{Valid Multiplication Path (VMP)}}
\vspace{-6pt}
\begin{definition}\label{D:CVMP}
 Let $p= x_i x_{i+1} ~\cdots~ x_{j-1}x_j$ be any path formed by adjacent $R$- and $S$-edges in $\Gamma(n)$ such that exactly one node $x_r$ is covered in each node partition $r$, where $x_r \in g(r)$, $1 \le i \le r \le j$.
Then $p$ is a \emph{ valid multiplication path} if $\forall (x_r, x_s)$ on $p$, $s > r$, we have either $x_r R x_s$ or the edge pairs $x_r $ and $x_s$, are vertex-disjoint in $K_{n,n}$ and  $x_r R x_s$ is false.\par
Further, $p$ is a \emph{complete valid multiplication path} (CVMP) if every $R$-edge, $x_t R x_r$(direct or jump edge) in $p$, $i \le t < r \le j$, is covered by $p$.
\vspace{-6pt}

\end{definition}

Note that a \emph{VMP}, $p' = x_i x_{i+1} ~\cdots~ x_{j-1}x_j$, is any sub-path of a $\cvmp$,
$p = x_r x_{r+1} ~\cdots~ x_{n-1}x_n$ which contains $p'$.
\vspace{-6pt}
\subsubsection*{The Multiplying DAG: General Specification for Multiplying two Nodes}
\vspace{-6pt}
We can now define an inductive structure called \emph{Multiplying Directed Acyclic Graph} (abbr. \emph{mdag}) that can be used to specify a VMP of length $r+1$ using a VMP of length $r$.

\begin{definition}
A \emph{Multiplying Directed Acyclic Graph} (MDAG), denoted as $mdag(x_i, x_{i+1}, x_t)$, is a general specification for ``multiplying'' two nodes $x_i$ and $x_{i+1}$ in adjacent node partitions where $x_i S x_{i+1}$, and  $x_i R x_t$ defines an  $R$-edge such that all three nodes, $x_i$, $x_{i+1}$ and  $x_t$ are covered by a common \vmp. Clearly, in the  extreme case $mdag(x_i, x_{i+1}, x_t)$ reduces to an $R$-edge defined by $x_i R x_{i+1}$, with $x_{i+1}= x_t$.
\end{definition}
\begin{figure}[h]
\center
\includegraphics[scale=0.32]{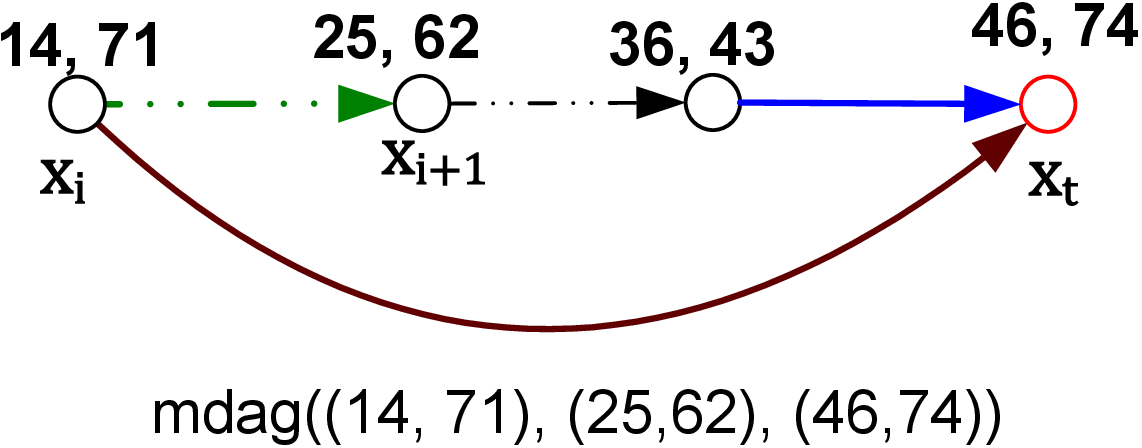}
 \caption{A Simple mdag}\label{FG:mdagBasic}
\vspace{-6pt}
\end{figure}

\vspace{-5pt}
\subsubsection{\textbf{Perfect Matching Represented by a \cvmp}}
The following Lemma follows directly from Lemma \ref{a:L:CVMP-perm}
\begin{lemma}\label{L:matchingSpecs}
Every $\cvmp$, $p= x_1 x_{2} ~\cdots~ x_{n-1}x_n$ in $\Gamma(n)$, of length $n-1$, where $x_r \in g(r)$, $ r \in [1..n]$ represents a unique
permutation $\pi \in S_n$ realized as a perfect matching $E(\pi)$ in $K_{n,n}$ given by
\begin{equation} \label{e:CVMP-perm1}
\pi = \psi(x_n) \psi(x_{n-1}) ~\cdots ~ \psi(x_{2}) \psi(x_1),
\end{equation}
where $\psi(x_r) \in U_r$ is a transposition defined by the edge pair $x_r$, and $U_r$
 is a set of right coset representatives of the subgroup $G^{(r)}$ in $G^{(r-1)}$ such
 that $U_n \times U _{n-1} \cdots U_2 \times U_1$ generates $S_n$.
\end{lemma}
 \vspace{-10pt}
\section{The Extended Partition Hierarchy}
 \vspace{-10pt}
We extend the permutation group enumeration technique to the set of perfect matchings by essentially identifying two additional levels of the partition hierarchy of $S_n$, consisting \\ of C\vmpset\!s and MinSet sequences:
\vspace {-11pt}
$$ S_n \supset Coset \supset C\vmpset \supseteq \text{ MinSet Sequences.} \vspace {-8pt} $$
In this Section we develop these two key structures and the associated enumeration algorithms for the perfect matchings.

\subsection{The C\vmpset}
\vspace{-10pt}
A C\vmpset \ is essentially a collection of CVMPs to represent the partition of a Coset of a subgroup in $S_n$.
First we define a more general structure, the VMP Set, a set of all the VMPs.\par
Let $mdag\langle a_i \rangle = mdag(a_i, x_{i+1}, x_{r}), r > i+1$, denote a family of mdags.
We also note that each mdag, $mdag(a_i, x_{i+1}, x_{r})$, can reduce to an $R$-edge when $a_i R x_{i+1}$.\\
For brevity we will use the notation $m_i$ to represent an mdag, $mdag\langle a_i \rangle $ at some node $a_i$ in the node partition $i$ of $\Gamma(n)$ for a bipartite graph on $2n$ nodes.\\

\begin{definition}
A $\vmpset(m_i, m_j)$ is a set of all the VMPs between an mdag pair $(m_i , m_j)$,  in the node partition pair $(i,j)$,$1 \le i < j \le n$, in the generating graph $\Gamma(n)$.
\end{definition}
\vspace{-15pt}
 \subsubsection*{\textbf{Representation of a \vmpset}}
\vspace{-10pt}
A polynomial size representation of a \vmpset, $\vmpset(m_i, m_j)$, is a subgraph of the generating graph $\Gamma(n)$. This subgraph contains all the VMPs between the mdags $(mdag\langle a_i \rangle, mdag\langle b_j\rangle)$ at the node pair $(a_i, b_j)$.\\

A data structure, for representing a $\vmpset$ is a collection, called ``Struct", of various primitive components defined using Algol kind of notation as follows.
\par
$EdgesAtNode ~(\text{\textbf{Node}, \{incident edges\}})$;\\
$NodePartition \text{ \textbf{Array}[ ] \emph{of }} EdgesAtNode$;\\
\vspace{-8pt}
\begin{equation}
\vspace{-8pt}
\begin{split}
\mathbf{\vmpset(mdag\langle a_i \rangle, ~mdag\langle b_j\rangle) =} \textbf{ Struct } \{\\
&MdagPair \text{ (mdag\less $a_i$\more, ~mdag\less $b_j$\more);}\\
&PartitionList ~\text{\textbf{Array}[$i\,\cdot\cdot\,(j\!+\!1)$]\, \emph{{of}}}\,\text{NodePartition;}\\
&Count ~\textbf{integer}; \text{// the count of all the contained VMPs} \\
&\}
\vspace{-8pt}
\end{split}
\end{equation}
\ A CVMPSet can analogously be represented by the same structure.
\par
The following Figure [\ref{FG:CVMPSet}] shows a subset of a C{\vmpset} between two fixed mdags at nodes (16,31) and (57,85).
\par
\begin{figure}[h]
\vspace{-5pt}
\center
  \includegraphics[scale=0.48]{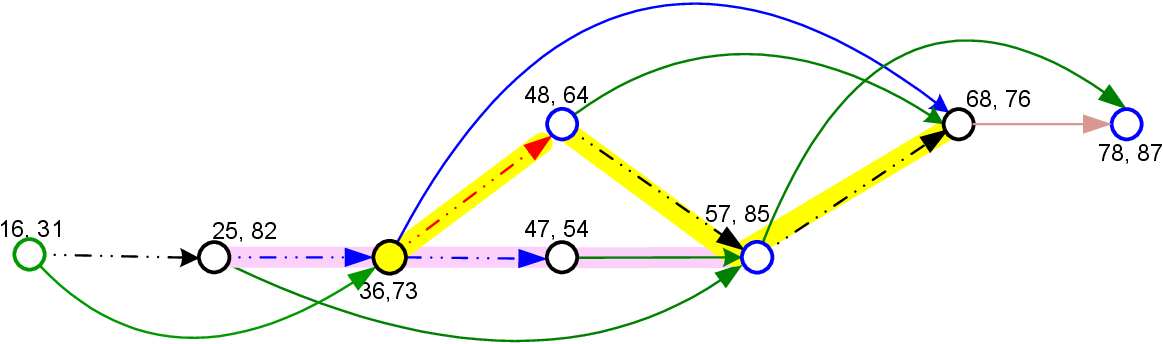}
   \caption{\textbf{A Subset of $C\vmpset(mdag\langle (16, 31) \rangle, mdag\langle(57, 85)\rangle)$}}\label{FG:CVMPSet}
\end{figure}
\vspace{-10pt}
\newpage
\subsubsection*{Product of two VMPs}
\vspace{-7pt}
Let $q\in C\vmpset(x_{i+1}, x_{n-1})$ and $\pi(q)$ be the corresponding permutation in
$G^{(i)}$, where $x_i \in g(i)$.\\ Then by $x_i\centerdot q$ we imply the product
$\pi(q)\psi(x_i) \in G^{(i-1)}$, where $x_i\centerdot q \in C\vmpset(x_{i}, x_{n-1})$.\par
Further, by induction, we can use the term product $p\centerdot q$ of two adjacent VMPs, $(p,q)$, to denote the multiplication of the associated permutations, noting that when $p = mdag(x_i, x_{i+1}, x_{i+2})$ it is same as $x_i\centerdot q$.\par
The product of two \vmpset ~then can be defined as,\\
\vspace{-11pt}
\begin{equation}\label{e:prodVMP}
\hspace{-10pt}
\vmpset(m_i, m_t) \centerdot \vmpset(m_t, m_j) \defn \big\{ p\centerdot q \,\big |\, (p, q) \in \vmpset(m_i, m_t) \times \vmpset(m_t, m_j) \big\}
\end{equation}
 Note that again, all $q \in \vmpset(m_t, m_j)$ cannot be multiplied by all $ p \in \vmpset(m_i, m_t)$.
\vspace{-10pt}
\subsubsection*{\textbf{C\vmpset ~Properties}}
\vspace{-10pt}
Let $x_i $ be any node in the $i$th node partition $g(i)$ of the generating graph $\Gamma(n)$, where $1\le i \le n-1$.
Let $\psi(x_i)$ and
$ \psi(y_i)$ be the two coset representatives in $U_i$ realized by the nodes $x_i$ and $y_i$ respectively, where $U_i$ is the set of coset representatives for the subgroup $G^{(i)} < G^{(i-1)} < S_n$. \\
\begin{property}\label{PR:disjointCVMP}
 All CVMP sets, $C\vmpset(mdag\less x_i\more,~ mdag\less x_{n-1}\more)$, are mutually disjoint for each mdag pair $(mdag\less x_i\more, ~mdag\less x_{n-1}\more)$ and each $i,~1 \le i \le n-2$.
\end{property}
\begin{proof}
Note that each CVMP set, $C\vmpset(mdag\less x_i\more, mdag\less x_{n-1}\more)$, is uniquely defined by its two mdags.
Moreover, every $x_i \in g(i)$ as well as every $R$-edge and every mdag at $x_i$ are unique. And therefore, if every CVMP, $p \in C\vmpset(mdag\less x_i\more, mdag\less x_{n-1}\more)$, is unique, so are all $x_i\cdot p \in C\vmpset(x_{i}, x_{n-1})$. That is, for each $(x_i, y_i) \in g(i)$ and for each $p_r \in C\vmpset(mdag\less x_{i+1}\more, mdag\less x_{n-1}\more)$,
\vspace{-11pt}
\begin{align*}
\vspace{-9pt}
x_i &\ne y_i \Longrightarrow x_i \cdot p_r \ne y_i \cdot p_r, \text{ and}\\
p_1 &\ne p_2 \Longrightarrow x_i \cdot p_1 \ne x_i \cdot p_2. \text{ \vspace{-5pt}}
\end{align*}
And hence, by induction, the disjointness of $C\vmpset(mdag\less x_{i}\more, mdag\less x_{n-1}\more)$ follows from the disjointness of the contained CVMPs. 
\vspace{-9pt}
\end{proof}
\vspace{-20pt}
\newpage
\begin{property}\label{P:polynomialCVMP}
There are at the most $O(n^5)$ CVMP sets, $C\vmpset(mdag\less x_{i}\more, mdag\less x_{n-1}\more)$, between any node pair $(x_i, x_{n-1})$ in a generating graph $\Gamma(n)$.
\end{property}
\begin{proof}
We simply note that there are $O(n^2)$ nodes $x_i\in g(i)$ at any  node partition $i$ in $\Gamma(n), and $there are $O(n^3)$ mdags at each node. Moreover, there are only two nodes in the node partition $n-1$.
\end{proof}

\subsubsection{Partition of a Coset into C\vmpset s}
\vspace{-7pt}
\begin{lemma}\label{L:Coset-CVMP}
Each Coset $G^{(i)} \psi_k$ of a subgroup $G^{(i)} < G^{(i-1)}$ can be partitioned
 into a family of equivalence classes, viz., the {C\vmpset}s, as follows:
\begin{equation}\label{e:cosetRep-cvmpset}
G^{(i)} \psi_k = \biguplus_{\substack {x_{n-1}, \\ \psi(a_i) = \psi_k}} \hspace{-10pt}C\vmpset\!(mdag\langle a_i\rangle, mdag\langle x_{n-1}\rangle),
\vspace{-0pt}
\end{equation}
where $\psi_k= (i,k) \in U_i$, $a_i \in g(i), ~x_{n-1} \in g(n-1)$, $1 \le i < n-1$.
\end{lemma}
\begin{proof}
The proof follows essentially from Property \ref{PR:disjointCVMP} and an observation that the multiplication $mdag\langle a_{i}\rangle \cdot cvmp$ where $cvmp \in C\vmpset(mdag\langle a_{i+1)}\rangle, mdag\langle x_{n-1}\rangle)$, defines an equivalence relation over the coset $G^{(i)} \psi_k$.
\par
Clearly, each distinct cvmp in a $C\vmpset(mdag\less a_{i+1}\more, mdag\less x_{n-1}\more)$ represents a unique permutation $\pi \in G^{(i)}$.
By the definition of a {C\vmpset} we also know that it contains precisely those CVMPs from $C\vmpset(mdag\langle a_{i+1}\rangle, mdag\langle x_{n-1}\rangle)$
which can be multiplied by a common mdag at the node $a_i$.\\
 Since the set $\{a_i | \psi(a_i) = \psi_k\}$ represents the coset representative $\psi_k \in U_i$,
  each CVMP in $C\vmpset(mdag\langle a_i\rangle, mdag\langle x_{n-1}\rangle)$ represents the product $\pi\psi(a_i)$ in the coset $G^{(i)}\psi(a_i)$.\\
   Therefore, a union of all these
  (disjoint) subsets satisfying $\psi(a_i) = \psi_k$ gives $G^{(i)}\psi(a_i)$. \\
\end{proof}
\textbf{Note}. \emph{For each node $a_i \in g(i)$ there are $O(n^3)$ distinct $mdag\less a_i\more$ over which the above union operation in \eqref{e:cosetRep-cvmpset} is performed.}

\vspace{-0pt}
\subsubsection{Coverage of the Symmetric Group $\mathbf{S_n}$}
\vspace{-0pt}
\begin{property}\label{P:countCVMPSet}
All the CVMPSets of length $n\!-\!1$ in $\Gamma(n)$ jointly contain $n!$ unique CVMPs representing precisely the $n!$ permutations in $S_n$. That is,
\begin{equation}\label{e:countCVMPSet}
 \biguplus_{\substack {a_1 \in g(1),\\ x_{n-1} \in g(n-1)}} \hspace{-15pt}C\vmpset(mdag\langle a_1\rangle, mdag\langle x_{n-1}\rangle) = S_n
\end{equation}
\end{property}
 \begin{proof}
The proof follows from \eqref{e:cosetRep-cvmpset} of Lemma \ref{L:Coset-CVMP}.
\\
Note that $g(1)$ is the set of all the edge pairs that collectively represent the \emph{generators} (elements in the set of right coset representatives), $U_1$, and further all the associated CVMP sets are mutually disjoint (Property \ref{PR:disjointCVMP}).\\
Therefore,
\vspace{-7pt}
\begin{align*}
\begin{split}
 S_n = \biguplus_{\psi_k \in U_1} G^{(1)}\psi_k
& = \biguplus_{ \substack {a_1 \in g(1),\\ x_{n-1} \in g(n-1)}} \hspace{-15pt} C\vmpset(mdag\langle a_1\rangle, mdag\langle x_{n-1}\rangle).
\end{split}
\vspace{-0pt}
\end{align*}
\end{proof}
 \vspace{9pt}
\subsubsection{Counting Perfect Matchings in $\mathbf{K_{n,n}}$}
\vspace{-11pt}
To demonstrate the basic counting technique, we first present a counting algorithm that enumerates seemingly a trivial set, viz., the set of all $n!$ perfect matchings in $K_{n,n}$ in polynomial time. We will then augment the same algorithm to allow the counting in any bipartite graph, while still maintaining the polynomial time bound.  The following algorithm describes the high level steps which realize the group enumeration technique as captured by Property \ref{L:Coset-CVMP} and \ref{P:countCVMPSet}.\\
\vspace{-00pt}
\begin{algorithm}[H]
 \caption{ CountAllPerfectMatchings ($K_{n,n}$)}%
 \begin{algorithmic}\label{ALG:basic}
 \STATE \textbf{Input:} generating graph $\Gamma(n)$;
\STATE \textbf{Output:} count of Prefect Matchings in $K_{n,n}$;
 \end{algorithmic}
\hrule
 \vspace{-5pt}
 \begin{description}
  \item[Step (a): \emph{Inductively compute all } C\vmpset\hspace{-1pt}s]
   \end{description}
\vspace{-10pt}
 \begin{algorithmic}[1] 
\FOR{ node partitions $ i=n-3$ to $ 1 $ }
    \FORALL{\emph{nodes} $ a_i \in g(i) $}
        \FORALL { $ (a_{i+1}, x_{n-1}) \in g(i+1) \times g(n-1)$ }
           \STATE \textbf{compute all} $C\vmpset(mdag\langle a_{i}\rangle, mdag\langle x_{n-1}\rangle )$ \\ $= \{mdag\langle a_{i}\rangle \cdot C\vmpset(mdag\langle a_{i+1}\rangle, mdag\langle x_{n-1} \rangle)\}$;
        \ENDFOR
\ENDFOR
\ENDFOR
\end{algorithmic}
 \vspace{-10pt}
\begin{description}
  \item[Step (b): \emph{Sum the counts in each CVMPSet}]
\begin{align*}
 \hspace{-2.5in}the ~Count\,=\,\sum_{(a_1, \,x_{n-1})} \hspace{-10pt} C\vmpset(mdag\langle a_{1}\rangle, mdag\langle x_{n-1}\rangle )\cdot Count\\
\end{align*}
\end{description}
 \vskip -27pt
 \end{algorithm}
\vspace{-00pt}
As we shall see that the most complex step in the above algorithm \ref{ALG:basic} is the step (4), incrementing a $C\vmpset(mdag\langle a_{i+1}\rangle, mdag\langle x_{n-1} \rangle)$
by an $mdag\langle a_{i}\rangle$.

 \vspace{-07pt}
\subsubsection*{The Disjoint Subsets of CVMP Sets}
 \vspace{-07pt}
Unlike the product $\pi\psi $ in a coset $G^{(1)} \psi$ of the group $G^{(1)} $, most $x_i \in g(i)$ \emph{cannot} multiply all the CVMPs
in any $C\vmpset(x_{i+1}, x_{n-1})$, but only a subset thereof induced by $x_i$.
 Therefore, we need to find the disjoint subsets of $C\vmpset(m_{i+1}, m_{n-1})$ in which all CVMPs can be multiplied by a common $x_i \in g(i)$. The structure,
$prod\vmpset ( m_i, m_t, m_{n-1})$, defined below is one accomplishes this capability.
\par
Let $x_r \in g(r)$ be node in the node partition $i$ of $\Gamma(n)$ for $1 \le r \le n$;
let $m_{r} = mdag(\less x_r \more)$ denote an mdag at the node $x_r$, and
let $C\vmpset(m_i, \,m_{n-1})$ be any subset of the coset $G^{(i)} \psi_k$ in \eqref{e:cosetRep-cvmpset}.
\vspace{-0pt}
 We define $prod\vmpset ( m_i, m_t, m_{n-1})$ as a
 family of disjoint subsets of $C\vmpset ( m_i, m_{n-1})$ as follows.
\begin{definition}
  Let $m_t= mdag(x_t, x_{t+1}, y_l)$, $1 < i \le t \le n-1$. Then
 \vspace{-7pt}
\begin{multline}\label{e:partitionCVMPSet}
prod\vmpset ( m_i, \,m_t, \,m_{n-1}) \defn
 \{ p \in C\vmpset (m_i, \,m_{n-1})\mid \text{$p$ covers $m_t$
such that }\\
\text{the node $x_{t+1}$ in $m_t$ is incident by $R$-edges, $R_{x_{t+1}}$,}
\text{ with common $S\hspace{-2pt}E(R_{x_{j+1}})$\}}
\end{multline}
\end{definition}
\begin{figure}[h]
\vspace{-5pt}
\center
\vspace{-0pt}
  \includegraphics[scale=0.73]{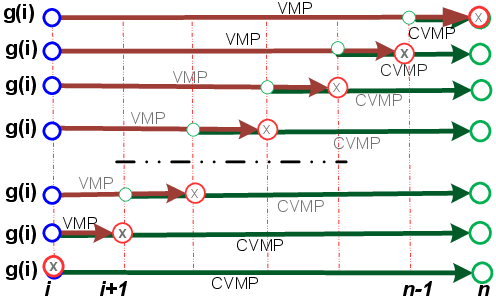}
\vspace{5pt}
 \caption{$\mathbf{prod\vmpset ( m_i, \,m_t, \,m_{n-1})}$\textbf{: Disjoint Subsets of $\mathbf{C\vmpset ( m_i, \,m_{n-1})}$ }}\label{FG:VMP-CVMP-Parts}
\end{figure}
\vspace{-6pt}
\emph{\textbf{Remark}}. In general a node $x_{t+1}$ in $m_t$ covered by $p$ can have 0 to 2 $R$-edges incident upon it. Thus for each $m_t$, there are 4 possible disjoint subsets induced by the 2 incident $R$-edges.\\

The following Property follows directly from the above definition.
 \begin{property}\label{P:propVMPxCVMP}
For each fixed $ t,~ 1 \le t \le n-1$, the associated disjoint subsets,
$prod\vmpset ( m_i, \,m_t, \,m_{n-1})$, are related to $C\vmpset ( m_i, \,m_{n-1})$ as follows:
\vspace{-2pt}
 \begin{equation}\label{e:VMPxCVMPb}
   C\vmpset(m_i, \,m_{n-1})~= ~\biguplus_{m_t} \hspace{-0.1pt}prod\vmpset ( m_i, \,m_t, \,m_{n-1})
\end{equation}
\end{property}

 \begin{property}\label{Pr:CVMP.VMP}
 \vspace{-3pt}

 \begin{equation}\label{e:CVMP.VMP}
   prod\vmpset ( m_i, \,m_t, \,m_{j}) =
      \vmpset(m_i, \,m_{t})\centerdot \vmpset (m_t, \,m_{j})
\end{equation}
 \vspace{-20pt}
\end{property}
\begin{proof}
Folllows from the definition of the product of two VMPs in \eqref{e:prodVMP}.
\end{proof}
\newpage
\subsection{Perfect Matching Generators}
\vspace{-9pt}
In this section we develop generators for  a subproblem of perfect matchings, the MinSet sequences. This subproblem is an equivalence class of the $n!$ perfect matchings, represented as a sequence of a P-time enumerable structure called MinSets which is induced by the missing edges in each potential perfect matching.  
\par
We first define MinSets which are collections of VMPs induced by the ``missing" edges, and which in turn are modeled as an attribute called \emph{Edge Requirements}(ER) of a CVMP that would represent an existing perfect matching under certain conditions on the ER. These MinSets become generators for the MinSet Sequences, and hence for the perfect matchings.

\vskip -7pt
\subsubsection{Edge Requirements: the CVMP Qualifier}\label{SS:ER}
Edge Requirement is an algebraic formulation of the perfect matching behavior that every node in the graph is incident with exactly one edge, i.e., the matched edge.
To define the edge requirement of a CVMP (VMP), $p$, we will define two terms, viz. the \emph{edge requirement}, $ER(x_i)$, of a node, $x_i$ in $p$, and \emph{surplus edge}, $SE(x_t x_i)$, of an $R$-edge $x_t x_i$.
  \par
  Let $p = x_1 x_2 \cdots x_{n-1} x_n$ be CVMP in $\Gamma(n)$ for a bipartite graph $BG$.
The (initial) Edge Requirement, $ER(x_i)$ of a node $x_i \in g(i)$ in $p$ is
\begin{equation}
ER(x_i) \defn \{ e~|~ e \in x_i ~\text{ and } e \notin BG \}
\end{equation}
$x_i$ represents an initial assignment of the matched edges incident on the node pair $(v_i, w_i)$.\\
We first consider the perfect matchings in $K_{n,n}$.
There are exactly two ways each node pair $(v_j,w_j) \in K_{n,n}$ can be incident with 2 matched edges in a perfect matching in $K_{n,n}$:\\
\vskip -5pt
\begin{enumerate}
\vskip -10pt
  \item The node pair $(v_j,w_j)$ is incident with an edge pair represented by $x_j \in g(j)$, and \\
\vspace{ -5pt}
\item one or both the edges in $x_j$ incident on $(v_j,w_j)$ are replaced by 2 alternate edges of a cycle formed between $x_i \in g(i)$ and $x_j$ such that $x_i R x_j$, where $x_i$ is in $p$.
\vskip -10pt
\end{enumerate}
\vskip -10pt

\par
\begin{figure}[h]
\center
  \includegraphics[scale=0.72]{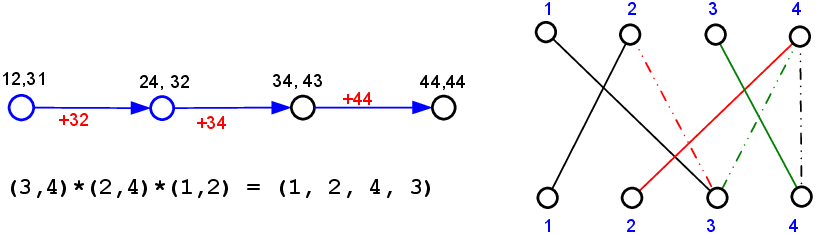}
  \caption{\textbf{A Perfect Matching:} \emph{satisfying the Edge Requirements}}\label{FG:exampleER1Fig5}
\end{figure}

For each instance of an $R$-edge $x_i R x_j$ in $p$, the edge in the edge-pair $x_j$ being replaced by the $R$-cycle $x_i R x_j$ will be called a \emph{Surplus Edge}, $SE(x_i x_j)$, and is defined as follows:\vspace{ -5pt}\vskip -10pt
\begin{equation}
\vspace{ -5pt}
SE(x_i x_j) \defn \text{ the edge $e \in x_j$ covered by the associated $R$-cycle defined by $x_i R x_j$. }
\vspace{ -0pt}
\end{equation}
When the given graph is not a complete bipartite graph, the edge requirement of a node $x$ in $\Gamma(n)$ can be met by the surplus edge(s) as determined by the $R$-edges incident on $x$. For example, in Figure \ref{FG:exampleER1Fig5}, for the CVMP
$p = (12, 31)\cdot(24, 32)\cdot(34, 43)\cdot(44, 44)$, the initial $ER$ of various nodes on $p$, i.e.,$\{ 44, 34, 32\}$, is satisfied by the SE of the incident $R$-edges.
\vskip -05pt
\par
The Edge Requirement $ER(p)$ of a VMP, $p$ in $\Gamma(n)$, is the collection of each of the nodes' Edge Requirement that is not satisfied by the SE of the $R$-edge incident on that node. That is,
\begin{equation}\label{e:ER-path}
ER(p) \defn \bigcup_{x_i \in p} ER(x_i) - \big( \{ SE(x_j x_k)~| ~x_j, x_k \in p \} \bigcap \big( \bigcup_{x_i \in p} ER(x_i)~\big)~\big)
\vspace{ -10pt}
\end{equation}
\vskip -5pt
(\emph{The intersection in the second term is needed simply to avoid any ``negative edge requirements'' resulting from any redundant edges incident on any of the nodes.})
\par
As a VMP $p$ in  $\Gamma(n)$ grows, its edge requirements $ER(p)$ can change.
 We will show later (Lemma \ref{L:ER-cvmp1}) that the edge requirements of a VMP, $p$ in $\Gamma(n)$ is null iff $p$ leads to a perfect matching in $BG'$.
 Note that $ER(p)$ is specific to a given bipartite graph $BG'$ since the ER of a node in $p$ depends on $BG'$, whereas SE of an $R$-edge is fixed for all $BG'$.
\par

The following corollary follows directly from Lemma \ref{a:L:CVMP-ER}
\begin{corollary}\label{CR:CVMP-ER}
Every $\cvmp\!\!,$  $ p= x_1 x_{2} ~\cdots~ x_{n-1}x_n$ in $\Gamma(n)$, represents a unique perfect matching $E(\pi)$ in $K_{n,n}$ given by
\begin{equation}\label{e:CVMP-perm2}
 E(\pi) = \bigcup_{ x_i \in p} {x_i} - \{SE(x_j x_k)\mid x_jR x_k,\, (x_j, x_k) \in p\}.
\end{equation}
\end{corollary}
\subsubsection*{The Condition for a CVMP to be a Perfect Matching} \begin{lemma}\label{L:ER-cvmp1}
Let $p= x_1 x_{2} ~\cdots~ x_{n-1}x_n$ be a CVMP  of length $n-1$  for a bipartite graph $BG'$. Then $ER(p) = \emptyset \Longleftrightarrow E(\pi)$ is a perfect matching in $BG'$ given by  \eqref{e:CVMP-perm2}, where $\pi$ is given by  \eqref{e:CVMP-perm1}.
\end{lemma}
\begin{proof}
The expression for $ER(p)$ in \eqref{e:ER-path} can be re-written as:
\begin{align*}
ER(p) =& \big( \bigcup_{x_i \in p} ER(x_i)\big) \bigcap \big( \{ e~|~ e \in x_i \in p\} - \{ SE(x_j x_k)~| ~x_j, x_k \in p \} )~\big)\\
=& \big(\bigcup_{x_i \in p} ER(x_i)~ \big)\bigcap E(\pi)
\end{align*}
Therefore, $ER(p) = \emptyset $ iff either\\
\hspace {1.7in}
(1) $\forall x_i \in p, ER(x_i)= \emptyset $, or\\
\hspace {1.7in}
(2) $\forall e \in E(\pi), ~e \notin \cup ER(x_i)$, and hence $e \in BG'$.
\par
Thus both cases lead to $E(\pi)$ being realized by $ BG'$.
\par
\end{proof}
\newpage
\vspace{-11pt}
\subsubsection*{A Pattern of ER Satisfiability}
\vspace{-011pt}
Following the enumeration scheme in \ref{a:L:VMPprodAtX}, each step of constructing incrementally larger length C\vmpset ~can reduce the $ER$ of its
 member CVMPs by at the most one edge. For example, each perfect matching in
$C\vmpset(m_1, m_{n-1})$ can allow at the most one missing edge in the perfect
matchings in $BG_1$ as illustrated in Figure \ref{FG:ER-Satisfiability} below.
\begin{figure}[h]
\center
\includegraphics[scale=0.50]{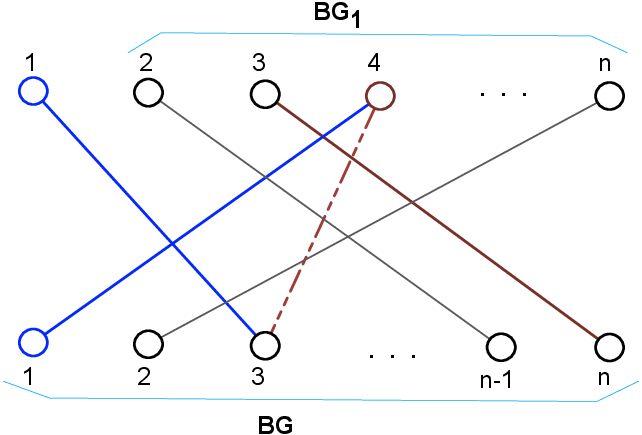}
   \caption{\textbf{ \emph{Incremental Reduction of ER}}}\label{FG:ER-Satisfiability}
\vspace{-10pt}
\end{figure}

 The following Property is a consequence of Lemma \ref{a:L:VMPprodAtX}.
\begin{property}\label{P:ER_MinSet}
 For every $(x_1, p_2) \in g(1) \times C\vmpset (m_2, m_{n-1})$, if $x_1 \centerdot p_2$ is a perfect matching in $BG$, then $|ER(p_2)| \leq 1$.
\end{property}
\vspace{-00pt}
\begin{proof}
The permutation group enumeration (Lemma \ref{a:L:VMPprodAtX}) requires that each
$p_2 \in C\vmpset (m_2, m_{n-1})$ be multiplied by exactly one
 generating element $x_1 \in g(1)$ in order to generate another member $x_1\centerdot p_2$ in $C\vmpset (m_1, m_{n-1})$.\\
Therefore, $ER(x_1\centerdot p_2) =\emptyset \Longrightarrow |ER(p_2)| \leq 1$.\\
\end{proof}
\subsubsection{MinSets: The VMPs of Common ER}
\vspace{-8pt}
Counting all the VMPs of a common ER for any given bipartite graph cannot be done in polynomial time because of the possibility of exponentially many ER sequences over a VMPSet. However, for certain small fixed patterns of the ERs, the VMPSet can be counted in polynomial time.
\par
 The above Property \ref{P:ER_MinSet} drives the definition of an ER-constrained set, called MinSet,
  which has a common ER for all the contained VMPs. A $C\vmpset(m_1, m_{n-1})$ can then be expressed as a polynomially bounded set of a sequence of P-time enumerable MinSets.
  \par
\begin{figure}[h]
 \center
\includegraphics[scale=0.45]{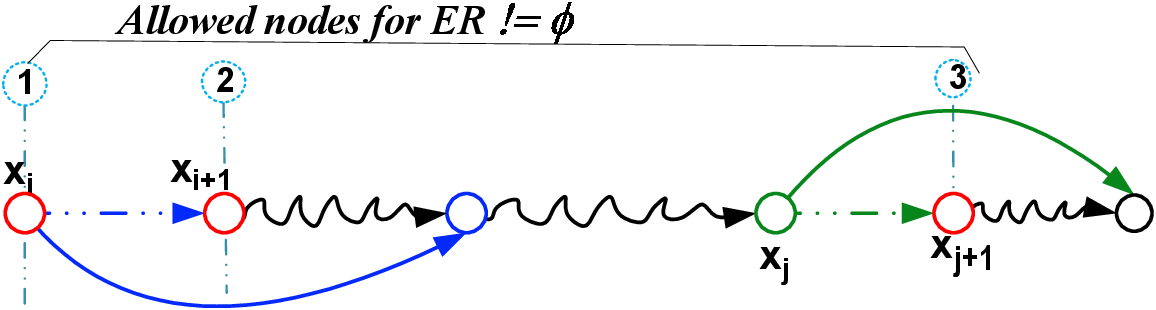}
\vskip 5pt
\caption{\textbf{An Abstract MinSet: $\mathbf {MinSet(mdag\less x_i \more, mdag\less x_j \more)}$}}\label{FG:DefnMinSet} \vspace{-18pt}
\end{figure}

Let $ER^p(x_j)$ denote the ER of a node $x_j$ covered by a VMP, $p$.
\begin{definition}\label{D:MinSet0}
A $MinSet(m_i, m_{j})$, $1 \le i < j \le n-1$, is the largest subset of
$\vmpset(m_i, \, m_{j})$, where each $p \in MinSet(m_i, \,m_{j})$
has a common ER, $ER(p)$, such that
\vspace{-15pt}
\begin{quote}
\vspace{-15pt}
 \item $\forall (p, x_k) \in MinSet(m_i, \,m_{j})$, the common ER, $ER^p (x_k) =\emptyset$
     except for the 3 common nodes, $x_i, ~~x_{i+1}$, and $x_{j+1}$, in 3 distinguished node partitions ($i, ~{i+1}$, and ${j+1}$) \\ (Fig \ref{FG:DefnMinSet}).
\end{quote}	

\end{definition}

\vspace{-10pt}
 \subsubsection*{Representation of a MinSet}
\vspace{-10pt}
 A MinSet has a representation similar to that of a VMPSet except for the additional attributes for the common ER and the incident $R$-edges. For notational convenience,  the common ER at the 3rd node $x_{j+1}$ is not being captured, although it can differ between any two ${MinSet(mdag\less x_i \more, mdag\less x_j \more)}$.
\par
$EdgesAtNode ~(\text{ \textbf{Node}, \{incident edges\}})$;\\
$NodePartition \text{ \textbf{Array}[ ] \emph{of }} EdgesAtNode$;\\
\vspace{-8pt}
\begin{equation}
\vspace{-8pt}
\begin{split}
\mathbf{MinSet(mdag\langle a_i \rangle, ~mdag\langle b_j\rangle) =} \textbf{ Struct } \{\\
&MdagPair \text{ (mdag\less $a_i$\more, ~mdag\less $b_j$\more);}\\
&PartitionList ~\text{\textbf{Array}[$i\,\cdot\cdot\,(j\!+\!1)$]\, \emph{{of}}}\,\text{NodePartition;}\\
&\text{ //ER at 3 distinguished node positions}\\
& CommonER ~ER(x_i);~ ~\\
& CommonER ~ER(x_{i+1});~ \\
& CommonER ~ER(x_{j+1});~ \\
&Count ~\textbf{integer}; \text{// the count of all the contained VMPs} \\
&\}
\end{split}
\vspace{-18pt}
\end{equation}
\vspace{8pt}
 \begin{definition}\label{D:MinSet1}
A $MinSet(m_i, m_t, m_{j})$, $1 \le i < j \le n-1$, is a distinguished $MinSet(m_i, m_{j})$ such that \\
\hskip 0.4in
(1) $MinSet(m_i, m_t, m_{j}) \subseteq prod\vmpset(m_i,\, m_t, \, m_{j})$, and\\
\hskip 0.4in (2) $ER( m_t) = ER(x_{t+1})$, where $m_t = mdag(x_t, x_{t+1}, x_r)$.
\end{definition}
\vspace{0pt}
 \begin{fact}

For $K_{n,n}$,\\ \vspace{-10pt}
\[
 C\vmpset(m_i, m_{n-1})= MinSet(m_i, m_{n-1}).
 \]

\end{fact}
\subsubsection*{Notation: The labeling of nodes and edges in $\Gamma(n)$}
\vspace{-6pt}
\emph{Assuming the nodes in $K_{n,n}$ are labeled from $N$ using decimal numbers,
a node $(iv,wi) \in \Gamma(n)$ is then labeled as $i.v,w.i$, while the $R$-edges $((iv,wi), (wv,tw))$ are labled by $+w.v$, where $``\cdot"$ is used as a delimiter to separate the node labels. When the node numbers are $0,1,2,\, \cdots \,9$, we will ignore this delimiter $``\cdot"$.}

\vspace{10pt}
\newpage
\subsubsection{Finding the MinSets of a C\vmpset}
\vspace{-05pt}
We begin with an example C\vmpset,\, showing how do the various MinSets compose a C\vmpset.
We show how a MinSet is determined for the various values of ERs of $x_4$ and $x_6$ in the subset of a $C\vmpset(m_1, m_5)$ in a bipartite graph.
We consider two cases: a) when $ER(x_6)= \{68\}$, and b) when $ER(x_4)= \{48\}, ER(x_6)= \{68\}$.

\begin{figure}[h]
\vskip -0.0in
\center
  \includegraphics[scale=0.450]{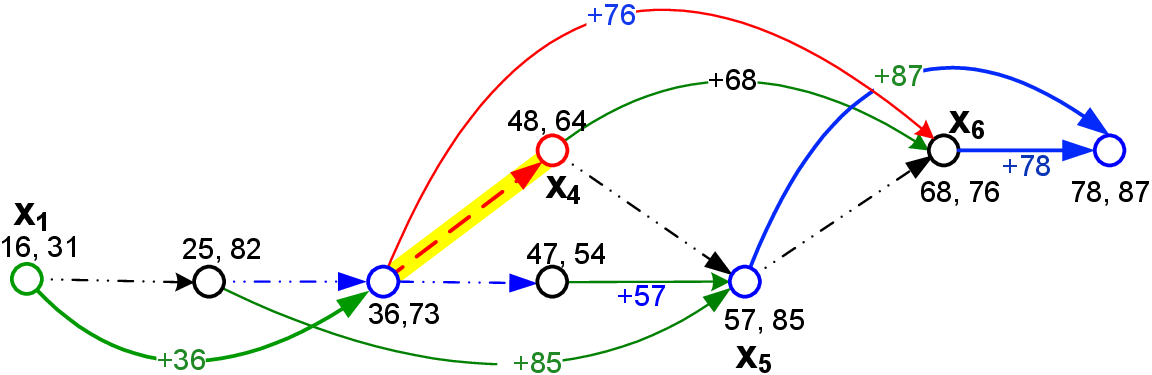}
  \center
\caption{\textbf{A C\vmpset\ Subset with 2 CVMPs}}\label{FG:MinsetExCases}
\end{figure}
\par
\underline{VMPSet Partition 1: $ER(x_6)= \{68\}$}\par
Here the two MinSets differ in the common $R$-edge incident at $x_6$.
 \par
$C\vmpset(m_1, m_5)= MinSet(m_1, m_5)+ MinSet(m_1, m'_5)$
\par
\vskip +0.3in
\center
  \includegraphics[scale=0.44]{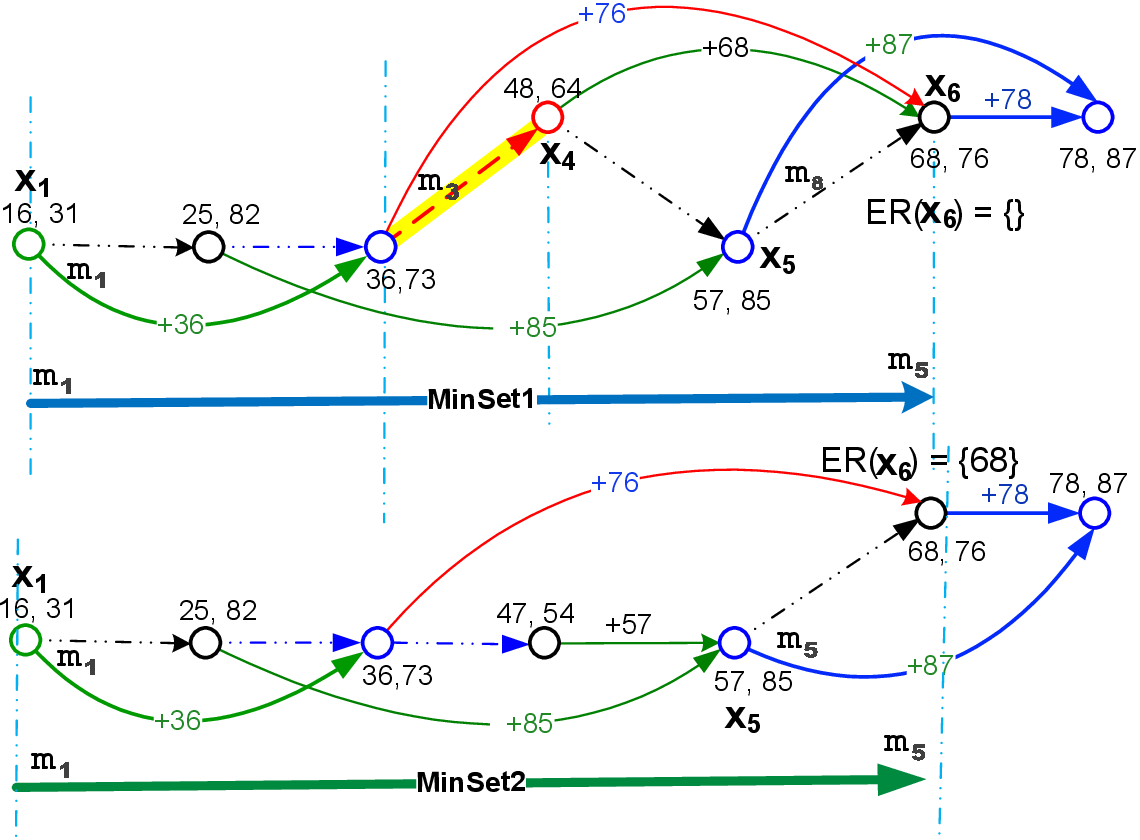}
\flushleft
\newpage
\underline{VMPSet Partition 2: $ER(x_4)= \{48\}, ER(x_6)= \{68\}$}\\
 \
 \par
$\vmpset(m_1, m_5)=\! MinSet(m_1,m_3) \centerdot MinSet(m_3, m_5) + MinSet(m_1, m_5)$\\

\center
\includegraphics[scale=0.4300]{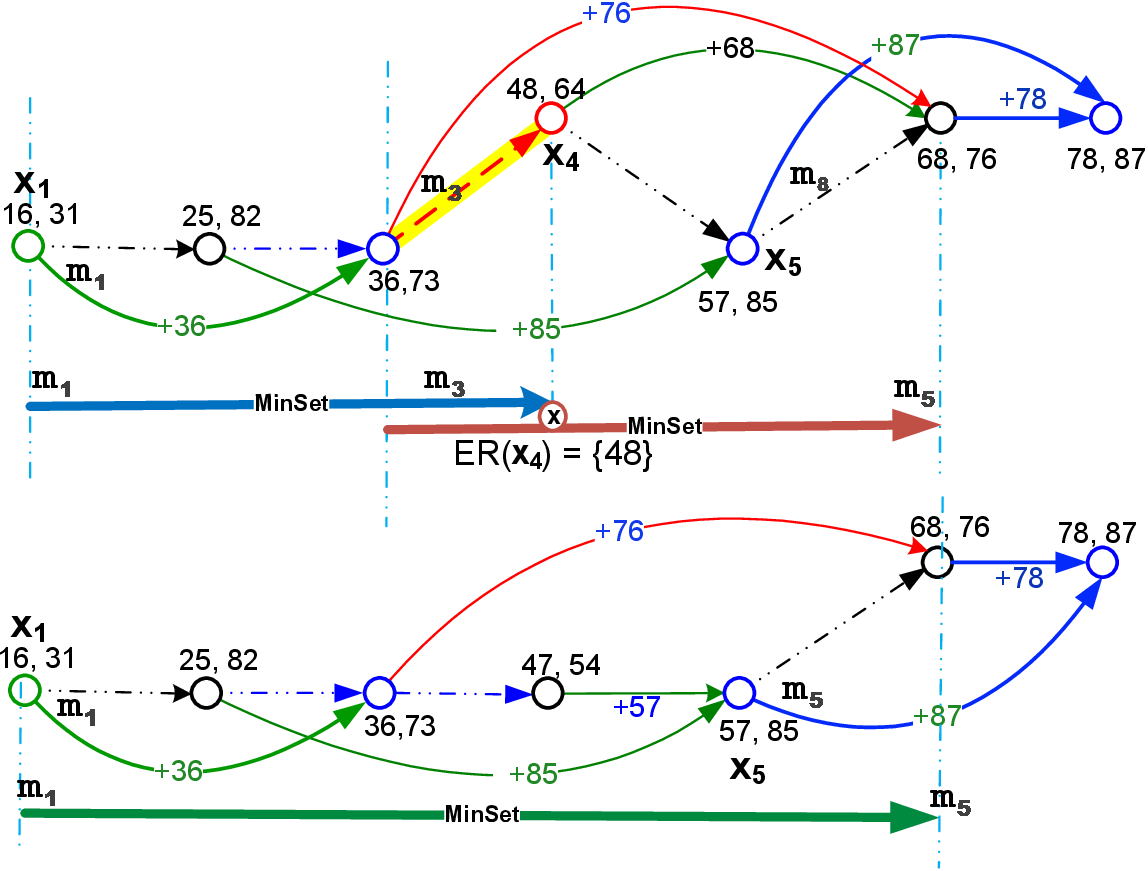}\\
\vskip -0.050in
 \flushleft

\flushleft
\begin{lemma}\label{L:boundMinSet1}
 For each $\vmpset(m_r, m_s)$ there are at the most 4 MinSets, $MinSet(m_r, m_s)$.
 \end{lemma}
  \begin{proof}
 By Definition \ref{D:MinSet1}, for a fixed $(m_r, m_s)$,
   two MinSets in $\{MinSet(m_r, m_s)\}$ can differ only due to 3rd distinguished node because of the incident $R$-edges can make its ER to be different. For any other node at positions $\{r+2, r+3, \cdots ~ s\}$, a non-null ER would imply MinSets of shorter length.\\
    Moreover, the first two nodes do not have any $R$-edges incident on them. Thus, the 3rd node is the only one left that can induce subsets of $\vmpset(m_r, m_s)$ due to its two possible ERs resulting from the incident $R$-edges.\\
\par
   Now the ER of this 3rd distinguished node, $ER(x_{s+1})$, can take at the most 4 values corresponding to the 4 possible subsets of the edge pair $x_{s+1} \in g(s+1)$ in $m_s$.
Therefore, for any fixed $(m_r,m_s)$,
\vspace{ -5pt}
$$\big |\{MinSet(m_r, m_s)\}\big|~ \le ~4 \vspace{ -15pt}$$.
\vspace{ -5pt}
\end{proof}
 \textbf{Remark}. Clearly, any $\vmpset(m_r, m_s)$ of length more than 3 can have an empty MinSet, $MinSet(m_r, m_s)$, because of the non null edge requirements.
  \par
  \begin{lemma}\label{L:boundMinSet2}
The maximum number of $MinSet(m_i, m_t, m_{n-1})$, $1 \le i \le t \le n-1$, for a given $C\vmpset(m_i, m_{n-1})$ is bounded by $O(n^6)$.
\end{lemma}
\begin{proof}
The bound follows from the bound on the maximum number of mdags covered by
any $C\vmpset(m_i, m_{n-1})$ in a given node partition, where there can be $O(n^3)$ mdags, $mdag\less x_i \more$ at any node $x_i$.
 \end{proof}
\newpage
\subsubsection{The Structure of a C\vmpset~Partition}
 \vspace{-10pt}
 We will now show the exact composition of a C\vmpset~ in terms of its components, the MinSets.
\vspace{-10pt}
 \subsubsection*{The Covering MinSet- a subset of CVMPSet}
\vspace{-10pt}
Now we show how each $C\vmpset(m_i, m_{n-1})$ can be partitioned
 into disjoint subsets which are equivalence classes represented by a sequence of MinSets. We first define this MinSet sequence and then show how does it cover a $C\vmpset(m_i, m_{n-1})$.\\
\begin{figure}[h]
\center
 \includegraphics[scale=0.452]{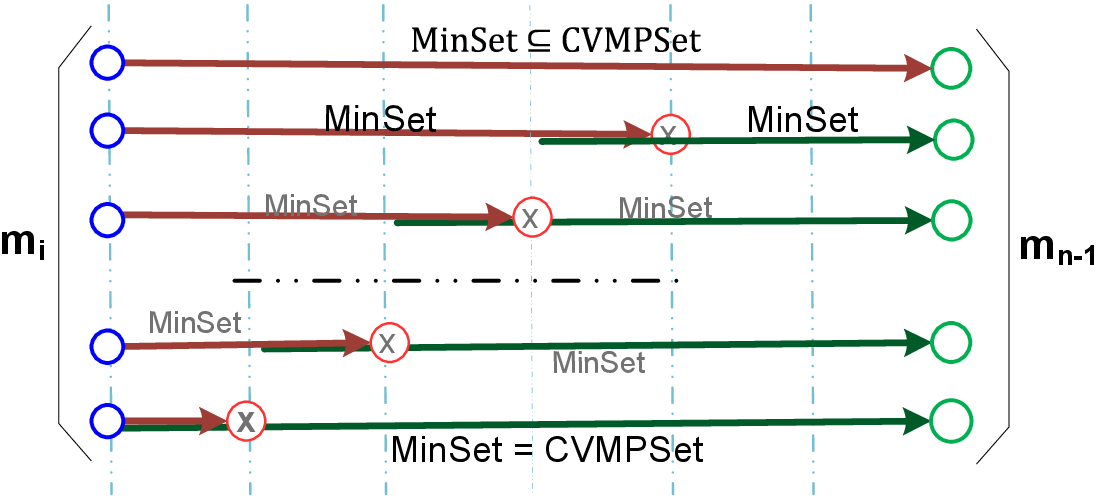}
\vspace{0pt}
 \caption{Sequences of 1-2 MinSets:} \label{FG:The MinSets}
 \vspace{-5pt}
 $C\vmpset(m_i, m_{n-1})= \biguplus MinSet(m_{i}, m_t) \centerdot MinSet ( m_{t}, m_{n-1})$
\end{figure}

\vspace{5pt}
Let $\prod$ denote the product of two or more adjacent MinSets, similar to the product of VMPSets.
\begin{definition}
A \emph{covering minset}, $CMS_{it} (r)$, represents a subset of $\vmpset(m_i, m_{t})$ by
a sequence of $r$ MinSets for the given $\vmpset(m_i, m_{t})$.
\vspace{6pt}
 That is,
\vskip 0.0in
$CMS_{it} (r) \defn \{MinSet ( m_{i}, m_{j_1}), MinSet ( m_{j_1}, m_{j_2}), ~\cdots, ~MinSet ( m_{j_{r-1}}, m_{t})\},$\\

\hskip -0.0in
such that
$$
\prod_{i_j \in I} MinSet ( m_{i_{j}}, m_{i_{j+1}}) ~\subseteq \vmpset(m_i, m_{t}),
$$
where $ I = \{i, j_1, j_2, \cdots,~j_{r-1}\}$ is an index set representing the various node partitions induced by the $ER \ne \emptyset$ nodes in $\vmpset(m_i, m_{t})$ such that $|I| =r$, $1 \le r \le t-i$.
\end{definition}
\par
The following Lemma \ref{L:CountofCMS} precisely states the composition of a $C\vmpset (m_i, m_{n-1})$ in terms of its MinSet sequences.
\begin{lemma}\label{L:CountofCMS}
 Let $CMS_{in}(r)$ be a MinSet sequence of length $r$ representing a subset of $C\vmpset (m_i, m_{n-1})$, where $1 \le r \le n-2$, $1 \le i \le n-2$.\\
 Further, let
$ I = \{i, r_1, r_2, \cdots,~ n-1\}$ be an index set representing the indices to various node partitions induced by the $ER \ne \emptyset$ nodes in $CMS_{in} (r)$ such that $|I| =1+r$.
 Then, for all $i, 1\le i\le n-2$,
 \begin{equation}\label{EQ:CMS-partition}
  C\vmpset (m_i, m_{n-1}) = \biguplus_{\hspace{-1pt}r=1}^{n-2} \hspace{0pt}\prod_{\hspace{03pt}\substack{\cms\hspace{-02pt}_{in}(r),\\ i_j \in I}} \hspace{-15pt}MinSet(m_{i_j}, m_{i_{j+1}})
\vspace{ -5pt}
 \end{equation}
\end{lemma}
(The proof is deferred until after the counting algorithm)

\vspace{-10pt}
\newpage
\subsubsection*{The Equivalence Class- MinSet Sequences}
\vskip -8pt
We define a new equivalence class induced by the following equivalence relation $\mathbb{\mathbf{\Re}}$ over the \\set C\vmpset.
\vspace {-14pt}
\subsubsection*{The Equivalence Relation $\mathbb{\mathbf{\Re}}$}
\vskip -8pt
\begin{definition}
\vspace {-10pt}
 \begin{align*}
 \vspace {-20pt} \forall (p_i, p_j) \in C\vmpset (m_i, m_{n-1}), ~p_i \mathbb{\mathbf{\Re}} p_j \Longleftrightarrow & ~\exists \text{ a prefix $ MinSet(m_i, m_t)$ common to the }\\
 &\text{ sequences in \eqref{EQ:CMS-partition} containing $p_i$ and $p_j$}.
\end{align*}
\end{definition}
\vskip 0pt
\begin{claim}\label{CL:equi-mss}
 The relation $\mathbb{\mathbf{\Re}}$ is an equivalence relation over the set
C\vmpset$((m_i, m_{n-1})$.
\end{claim}
\begin{proof}
 Follows from Property \ref{P:MS-Unique}.
\end{proof}

\vspace{-10pt}
 \begin{property}\label{P:MS-Unique}
Every MinSet sequence $CMS_{in}(r)$ in $C\vmpset(m_i, m_{n-1})$ is unique.
   \end{property}
   \vspace{-10pt}
 \begin{proof}
Note that each mdag in the mdag sequence $\{m_{i_j}, m_{i_{j+1}}, \cdots, m_r\}$ induced by the nodes with $ER \ne \emptyset$ is unique because of the uniqueness of the missing edges in the bipartite graph.\\
 Now one can view each MinSet sequence, $CMS_{in}(r)$, as words composed out of unique mdags with $ER \ne \emptyset$, and without repetitions. \\
Thus, two MinSet sequences are identical iff the two sequences of the ``delimiting" mdags are identical. Hence each sequence of unique mdags gives rise to a unique sequence in $CMS_{in}(r)$.
\end{proof}

\vskip 15pt
\par

\begin{property}\label{P:prefixMinSet}
  There are $O(n^{6})$ prefix MinSets, $MinSet(m_i, m_{t})$, for any $C\vmpset(m_i, m_{n-1})$,
covering all the exponentially many MinSet sequences in \eqref{EQ:CMS-partition}, where $1 \le i <t \le n-1$, .
\end{property}
\begin{proof}
 Note that for each fixed mdag $m_i, ~i < t \le n-1$, $|\{ (m_i, m_{t})\mid i < t \le n-1\}| = O(n^6)$,\\ and $| \{ MinSet(m_i, m_{t})\}| = c| \{ \vmpset(m_i, m_{t})\}|,~ c \le 4$ (Lemma \ref{L:boundMinSet1}).
\end{proof}
\vskip 10pt
\subsubsection*{Partitions Induced by the MinSet Prefixes}
\vskip -8pt
The following Lemma is another version of \eqref{EQ:CMS-partition} representing a C\vmpset~ partition induced by the prefix MinSets. \\
Let $PFIX(l)$ denote a class of MinSets, $MinSet(m_1, m_l)$, containing \emph{VMPs} of length $l$. Also, we define all MinSet sequences of zero length , $CMS(0)$, to be an identity,
$I$, such that\\
 $$PFIX(l) \centerdot \hspace{-10pt}\prod_{\hspace{03pt}CMS(0)} \hspace{-10pt}MinSet(m_{i_{j}}, m_{i_{j+1}}) = PFIX(l).$$
\vskip -10pt
\begin{lemma}\label{L:ptnByPrefixMinSet}

Each $C\vmpset (m_1, m_{n-1})$ for a bipartite graph can be partitioned into $O(n^{6})$
 equivalence classes induced by the $O(n^{6})$ prefix MinSets,
$PFIX(l)$, $2 \le l \le n-1$ in \eqref{EQ:CMS-partition}, such that:
 \begin{equation}\label{EQ:prefix-partition}
  C\vmpset (m_1, m_{n-1}) =
 \biguplus_{l=2}^{n-1} \Bigl( \biguplus_{\hspace{-1pt}r=1}^{n-l-1} \hspace{0pt} PFIX(l) \centerdot \hspace{-7pt}\prod_{\hspace{03pt}\substack{\cms\hspace{-02pt}_{in}(r),\\ {l=i_j\in I}}}^{s_{r-1}} \hspace{-15pt}MinSet(m_{i_{j}}, m_{i_{j+1}}) \Bigr),
\vspace{ -5pt}
 \end{equation}
where $l$, $2 \le l \le n-1$, is the first index in any index set,
 $I =\{1, s_1\!=\!l, s_2, \cdots, s_{r-1}, n-1\}$, representing all the node partitions induced by the $ER \ne \emptyset$ nodes in $CMS_{in} (r)$.
 \vskip -0.1in
\end{lemma}
\vskip -5pt
\begin{proof}
The proof follows from Lemma \ref{L:CountofCMS}, Property \ref{P:MS-Unique} and Property \ref{P:prefixMinSet}.
\end{proof}
\vskip 0.0in
The following figures ([Fig \ref{FIG:Ptn-CVMP} and Fig \ref{FIG:minSetCMS}]) depict the partition implied by \eqref{EQ:prefix-partition}, i.e., how $O(n^{6})$ prefix MinSets induce polynomially many ($O(n^{6})$) equivalence classes containing exponentially many $CMS_{in}(r)$ sequences in $C\vmpset (m_i, m_{n-1})$.
\par
\vskip 10pt
\begin{figure}[h]
 \center
  \includegraphics[scale=0.50]{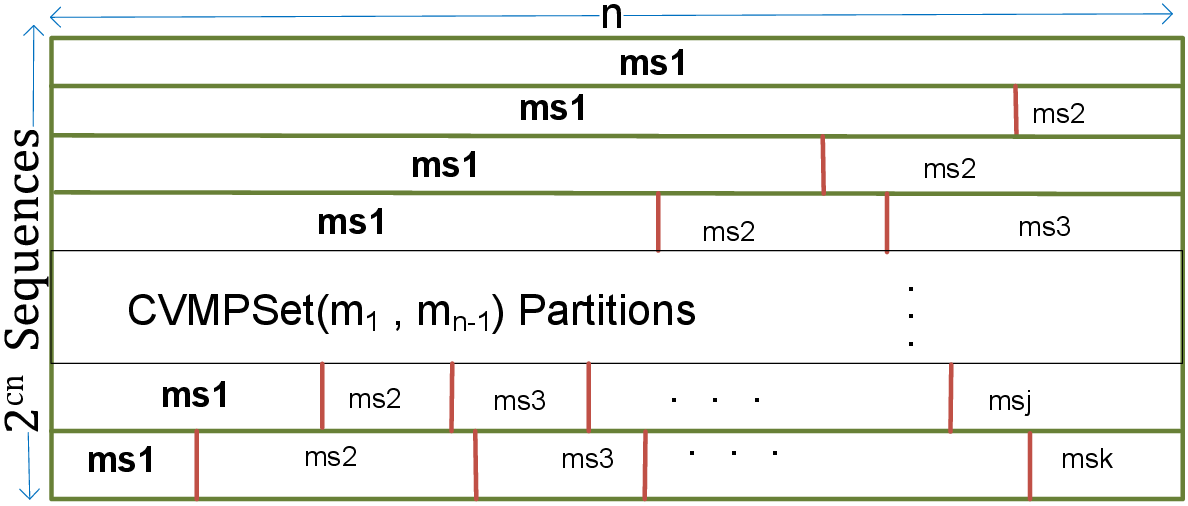}
\vskip 5pt
  \caption{\textbf{CVMP Set Partitions by $CMS(r) $ Sequences}}\label{FIG:Ptn-CVMP}
  \end{figure}
 \vskip 00pt
 \begin{figure}[h]
 \center
 \includegraphics[scale=0.60]{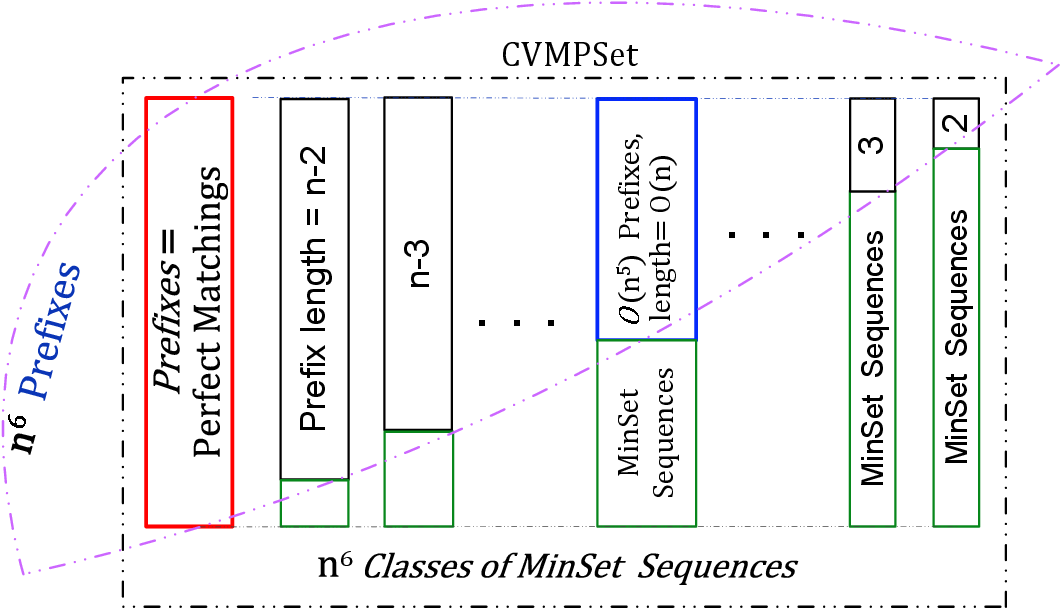}
\vskip 5pt
\caption{\textbf{Partition of a CVMPSet Induced by the MinSet Prefixes}}\label{FIG:minSetCMS}
\vskip -0pt
\end{figure}

\subsubsection*{A Generating Set for the MinSet Sequences}
\vskip -08pt
Before we can describe the counting algorithm, we need one more generating set to consolidate the generation of all the MinSets shared by the various CVMPSets through their CMS partitions. To accomplish this we define a \emph{generating set}, called $\gms$, for generating the Covering MinSets which constitute a partition of C\vmpset.

  \vspace{-0pt}
\begin{definition}
 A generating set for MinSet sequences, $\gms(i,n), ~ 1 \le i \le n-2$, for a bipartite graph on $2n$ nodes is a set of MinSets defined as \\
 \[
 \gms(i,n) \defn \big\{MinSet( m_{r}, m_{s})\, \big |\, (r,s) \in [i\! \,\cdot\cdot ~n\!-\!2] \times [i+1\,\cdot\cdot ~n\!-\!1], ~r <s \big \},
 \]
 where $\{(m_r, m_s)\}$ covers $g(r) \times g(s)$.

  \end{definition}
Note that $\gms(i,n)$ contains all the $O(n^{11})$ MinSet prefixes for exponentially many sequences,
$CMS_{in}(r)$, for all $C\vmpset(m_i, m_{n-1})$ in $\Gamma(n)$.
\par
The following two properties follow from the definitions of \gms and CMS.
 \begin{property}
    \begin{flalign}
 \gms(1,n)~ &\supseteq \hspace {-10pt} \bigcup_{r,(m_1, m_{n-1})}\hspace {-15pt} CMS_{1n}(r)
\end{flalign}

\end{property}

\begin{property}
An upper bound on the size of $\gms(1,n)$ is $O(n^{12})$.
\end{property}
\vspace{-5pt}
\begin{proof}
 Note that $\gms(1,n)$ is precisely the set of all the $MinSet(m_r, m_s)$ which connect the $O(n^{12})$ pair of mdags, $\{(m_{r}, m_{s})\}$, by all the VMPs in $\{\vmpset(m_{r}, m_{s})\}$ in $\Gamma(n)$, where $(r,s) \in [1~\cdot\cdot~n-2] \times [r+1~\cdot\cdot~n-1]$.\\ Thus, this bound is same as the bound on the number of edges in a graph with $O(n^6)$ nodes.\\
  Therefore, by Lemma \ref{L:boundMinSet1},
   \[
   \big | \gms(1,n)\big| \le 4\big |\{\vmpset(m_{r}, m_{s})\} \big| \le O(n^{12}).
   \vspace{-10pt}
   \]
\vspace{-15pt}
\end{proof}
\vspace{-00pt}
\newpage
\section{A Polynomial Time Enumeration Algorithm}\label{S:polybound}
 \vspace{-9pt}
 Based on the concepts of MinSet sequences developed in the previous Section, we have the following enumeration Algorithm \ref{ALG:countMatchings} which counts all the perfect matchings in a given bipartite graph $BG$ on $2n$ nodes, $n \ge 3$.
\begin{algorithm}[h]
  \caption{ countPerfectMatchings$( BG)$}
 \begin{algorithmic}\label{ALG:countMatchings}
  \STATE \textbf{Input:} a bipartite graph $BG$ on $2n$ nodes, $n \ge 3$;
  \STATE \textbf{Output:} count of the \emph{perfect matchings} in $BG$;
  \end{algorithmic}
  \hrule
   \begin{description}
  \item[Step 0: \emph{Initialize- Compute the Initial Generating Set of all the MinSet Sequences}]\ \\
   \end{description}
   \vspace{-25pt}
   \begin{algorithmic}[1]
 \STATE $i = n-3$; \COMMENT{ $i$ is the current node partition};
  \STATE \emph{Compute the generating set} $E_M = \{g(r)\,\mid \,1 \le r\le n\}$;
  \STATE \emph{Compute the generating set} $\gms(i+1,n)=\{MinSet(m_{n-2}, m_{n-1})\}$;\COMMENT{the set of all
the MinSet Sequences; each $C\vmpset(m_{n-2}, m_{n-1})$ is a $MinSet\in \gms(n\!-\!2,n)$, with a total count of 6 CVMPs.}
 \end{algorithmic}
 \vspace{5pt}
\hrule
  \begin{description}
  \item[Step 1: \emph{Count}] \ \\
  \hskip +0.35in
  \textbf{if} $(i=0)$ \textbf{then} /\!/ $\gms(1,n)$ may contain the set $\{MinSet(m_1,\, m_{n-1})\} $
 \vspace{-5pt}
 \hspace{0.100in}
 \[ \vspace{-5pt}
\hspace{-0.300in} \text{perfect matching \emph{count} }=\hspace{-10 pt} \sum_{\substack{ER=\emptyset,\\ ( m_1,\, m_{n-1})}} \hspace{-15 pt}MinSet( m_1,\, m_{n-1})\centerdot Count;\ \\
\vspace{-1pt}
\vspace{-15pt}
\]
\vspace{010pt}
\hskip +0.88in
\textbf{return};\\
\hskip +0.68in
\vspace{-005pt}

\vspace{-10pt}

\item[Step 2: \emph{Increment \& Join the MinSet Sequences}] \ \\
\hskip +01.68in  \hspace{0.350in}$incrementMSS (\gms(i\!+\!1,n))$; /\!/ assuming $n \ge 3$\\
\hskip +0.68in (Follows the structures in Figure \ref{FIG:minSetCMS})\\
\text{\emph{decrement} $i$};\\


\vspace{-10pt}
 \vspace{-00pt}\textbf{\hspace{-00pt}repeat} Steps 1-2;\\
 \end{description}\vspace{-15pt}
\hspace{-0.0000in} \textbf{End.}
 \vspace{-00pt}

\vspace{5pt}
  \end{algorithm}
\vspace{-7pt}
\subsection{The Polynomial Time Bound}
 \vspace{-5pt}
 \begin{claim} \label{CL: PTimebound}
The time complexity of Algorithm \ref{ALG:countMatchings} is $O(n^{45}\log n)$.
\end{claim}
\vspace{-7pt}
\begin{proof}
 Although a tight upper bound would require more details of the algorithm, a fairly loose upper bound should be easy to establish as follows. Let $T(ops)$ denote the time complexity of the operation $ops$.\par
  We note that
   Step 2 calls Algorithm  \ref{ALG:incrementMSS}, $incrementMSS()$, having a time complexity $O(n^{44}\log n)$, and dominates:
   \par
    $T (Step~ 2: \emph{Increment}) = O(n^{44}\log n)$\\
    $T (Step ~1: \emph{Count}) = O(n^8)$.
\par
      Steps 1-2 are iterated $O(n)$ times, and thus the time complexity of the counting algorithm\\ is $O(n^{45}\log n)$.
\vspace{-10pt}
\end{proof}
\newpage
 \subsection{Correctness of the Count}
 \vskip -09pt
 \begin{lemma}\label{L: correctCount}
 All the perfect matchings in a bipartite graph $BG$ on $2n$ nodes can be correctly enumerated by \ref{ALG:countMatchings} in polynomial sequential time $O(n^{45}\log n)$.
\end{lemma}
 \vspace{-5pt}
\begin{proof} The correctness of the count follows from the Lemmas \ref{L:CountofCMS} and \ref{L:corrctenss-incrementMS} which prove the following two assertions:
\vskip -011pt
\begin{enumerate}
\vskip -09pt
\item All $MinSet( m_1,\, m_{n-1})$ with $ER=\emptyset$ are contained in $\gms(1,n)$.
	\vskip -010pt
 \item The perfect matching \emph{count} is:
$$
\sum_{\substack{ER=\emptyset,\\ ( m_1, m_{n-1})}} \hspace{-10 pt}MinSet( m_1, m_{n-1})\centerdot Count$$
\vskip -12pt

\vskip -010pt
\end{enumerate}
\end{proof}

\begin{proof} \textbf{{(of Lemma \ref{L:CountofCMS}})}\\
 The proof is by induction on the length $r$ of the MinSet sequence, $CMS_{in}(r)$. We will consider VMPSets as a general representation of CVMPSets. \\
Let $m_j = mdag(x_j, x_{j+1}, x_d)$, where $x_j \in g(j),x_{j+1} \in g(j+1)$, and $x_d \in g(d)$.
 \par
 \underline{Case: $r =1$}\\
 The length of a $MinSet$ sequence for a $\vmpset(m_r, m_s)$ is one when all the VMPs in the MinSet are of the same length as those in $\vmpset(m_r, m_s)$. This can happen
 when the $ER\ne \emptyset$ only at the 3 allowed nodes, viz., at $x_i$ or at $x_{i+1}$ in $m_i$ and $x_{j+1}$ in $m_j$. And then, we have 2 sub cases, that is,
\\ either
 \vspace{-0pt}
\begin{subequations}
   \begin{equation}\label{e:CountofCMS1}
   \vmpset(m_i, m_{j}) = MinSet(m_i, m_{j}), \text{ when $ER \ne \emptyset$ at $x_i$ or at $x_{i+1}$,}
   \end{equation}
or\\
\vspace{-10pt}
  \begin{equation}\label{e:CountofCMS2}
   \vmpset(m_i, m_{j}) = \biguplus_{m_j} MinSet(m_i, m_{j}) = \biguplus_{m_j} ~prod\vmpset( m_i, \,m_j, \,m_{j}),
   \vspace{-5pt}
  \end{equation}
\hskip 0.6in when $ER(x_{j+1}) \ne \emptyset$ for some $p \in \vmpset(m_i, m_j)$ at the last node $x_{j+1}$ in $m_j$.

\end{subequations}
 \vspace{10pt}
 \par
\underline{Case: $r \le 2$}\par Let $\cms_{ij} (r) =\{MinSet ( m_{i}, m_{t}), MinSet ( m_t, m_{j}) \} $,
where the two MinSets have a common $m_t$ with $ER(m_t) \ne \emptyset $, $i <t<j$.
Then, the first MinSet could be a subset of $\vmpset ( m_i, \,m_t)$,
governed by \eqref{e:CountofCMS2}, whereas the second MinSet is the corresponding $\vmpset( m_t, m_{j})$ by \eqref{e:CountofCMS1}.
\\
 Therefore, for each such sequence we can apply the result of $r=1$ to each MinSet and Property \ref{P:propVMPxCVMP}, \eqref{e:CVMP.VMP} to obtain:
 \vspace{-5pt}
 \begin{align*}
 MinSet(m_i, m_t) \centerdot MinSet(m_t, m_j) &\subseteq \vmpset ( m_i, \,m_t)\centerdot \vmpset (m_t, \,m_{j})\\
  &=prod\vmpset ( m_i, \,m_t, \,m_{j})
 \vspace{-5pt}
  \end{align*}
Further we note that each $m_t$ in each sequence of 2 MinSets must be disjoint with each other, irrespective of the node partition $t$ to which $m_t$ belongs. Or else, we will have $r >2$ because of the additional $ER \ne \emptyset$ nodes common to the same MinSet.\\
 Therefore,\\
 \vspace{-20pt}
 \begin{align*}
 \biguplus_{m_t} MinSet(m_i, m_t) \centerdot MinSet(m_t, m_{j}) &= \biguplus_{t}\biguplus_{m_t}prod\vmpset ( m_i, \,m_t, \,m_{j})\\
 &=\vmpset(m_i, m_{j}) .
\vspace{-10pt}
  \end{align*}
\vskip -00pt
 \underline{Induction: Sequence size $l =r+1$}
  \par
  If the induction hypothesis is true for all sequence sizes, $l \le r$, then each sequence of length $r+1$ can be partitioned into 2 subsequences, $MinSet(m_i, m_t)$ and $MinSet(m_t, m_{n-1})$, each of length less than or equal to $r$.
\\
 Therefore, again applying the above cases for $r \le 2$ we have\\
 \vskip -5pt
 \[
 \vspace{ -10pt}
 C\vmpset(m_i, m_{n-1}) = \biguplus_{m_t} MinSet(m_i, m_t) \centerdot MinSet(m_t, m_{n-1}).
 \]
 \vskip -00pt
\end{proof}


\vspace{-5pt}
\subsection{Examples- Incrementing and Joining the Adjacent MinSets }
\begin{figure}[h]
\vspace{10pt}
\flushright
\includegraphics[scale=0.75]{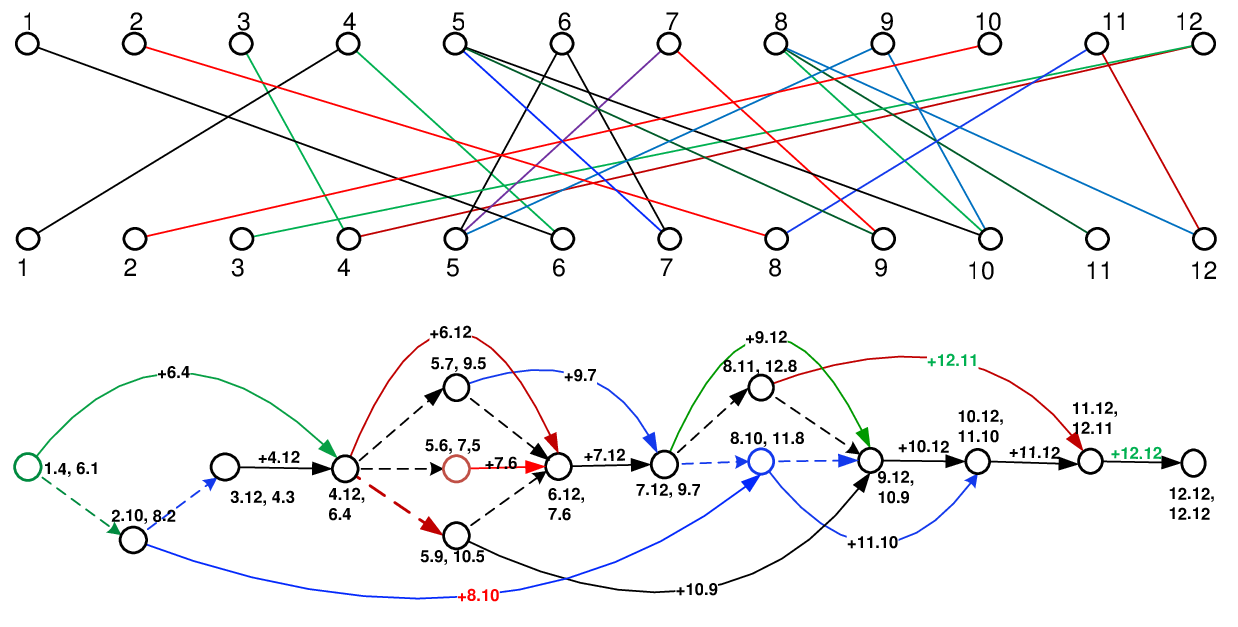}
\caption{\textbf{A Bipartite Graph and its CVMPs}\label{FG:FigCVMPSets}}
\end{figure}

\begin{figure}[h]
\vspace{5pt}
\flushright
\includegraphics[scale=0.7505]{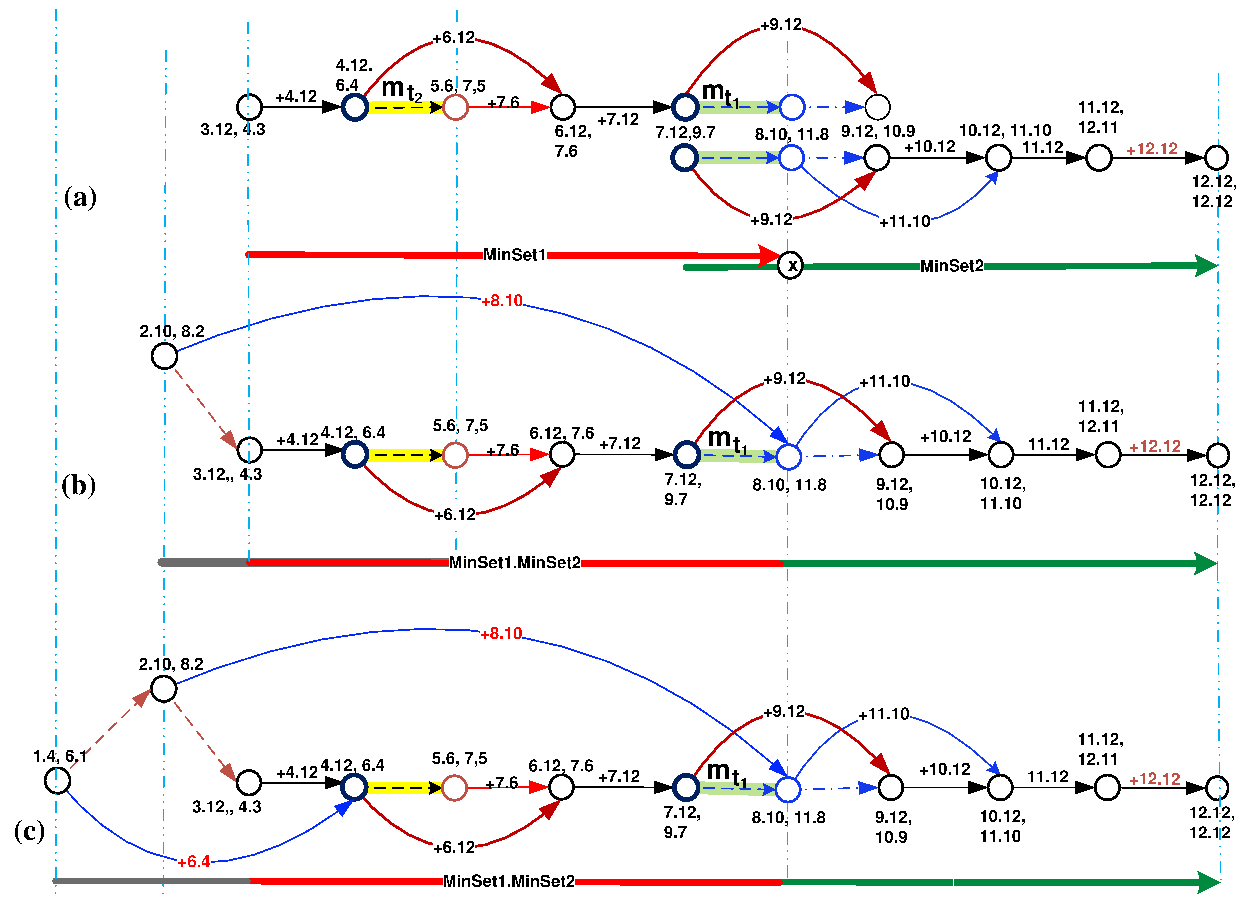}
\caption{\textbf{A Simple Increment \& Join of 2 MinSets}\label{FG:FigSimpleIncr}}
\vspace{-0pt}
\end{figure}

\vspace{-10pt}
 The following series of figures illustrate  the Step 2: ``Increment all the MinSet Sequences" of Algorithm \ref{ALG:countMatchings}.
 Figure \ref{FG:FigCVMPSets} shows a bipartite graph with 3 perfect matchings and the associated CVMPs that will generate the perfect matchings.
 Figure \ref{FG:FigSimpleIncr} shows a very simple Increment \& Join operation on the MinSets.
\\

\vspace{-5pt}
\newpage
Figure \ref{FG:FigSimpleIncr}(a) shows two MinSets each having a common $ER=\{8.10\}$ because of the missing edge $(8,10)$ in the bipartite graph.\\
Figure \ref{FG:FigSimpleIncr}(b) shows an increment of the MinSet1 by $x_2 = (2.10,8.2)$. This increment creates a jump $R$-edge between $x_2$ and $x_8$, and thus satisfying the $ER=\{8.10\}$ of $x_8$, and joining the two MinSets to produce $ p = MinSet1\cdot MinSet2$, which contains 3 CVMPs.  Note that $ER(p) = \emptyset$.\\
Figure \ref{FG:FigSimpleIncr}(c) shows a further increment of the CVMPs in $p$ by $x_1 =(1.4,6.1)$. This gives a set of 3 CVMPs with null ER to produce 3 perfect matchings.
\par
Note: The edge $(4,6)$  in the bipartite graph $BG$ is not needed for the resulting 3 perfect matchings but its presence simplifies the Step 2 of the algorithm.
\flushleft

\newpage
\subsection{The Basic Behavior of Incrementing a MinSet Sequence}
\vspace{-5pt}

 The incrementing process can become more intricate if there are other MinSets, $MinSet(m_t, m_s)$, adjacent to $MinSet(m_{i+1}, m_t)$, which may also have to be joined with the prefix MinSet.
A simplified view of this process is illustrated in the following Figures \ref{FG:FigMinSetParts}(a-c).
 These Figures show how a sequence of three adjacent MinSets is incremented and joined by
 the multiplying mdags at node partitions, $i=2$ and $i=1$.

\par
Figure \ref{FG:FigMinSetParts}(a) shows a sequence of three MinSets, $MS_1$, $MS_2$ and $MS_3$.\\
Figure \ref{FG:FigMinSetParts}(b) shows that $x_2 = (2.9, 8.2)$ increments the prefix MinSet,
$MS_1$ but with no join operation. \\
Figure \ref{FG:FigMinSetParts}(c) shows that $x_1=(1.6, 5.1)$ increments the prefix MinSet, $MS_1$ and joins all three MinSets in the sequence.
\par


\begin{figure}[H] \vspace{-5pt}
\includegraphics[scale=0.690]{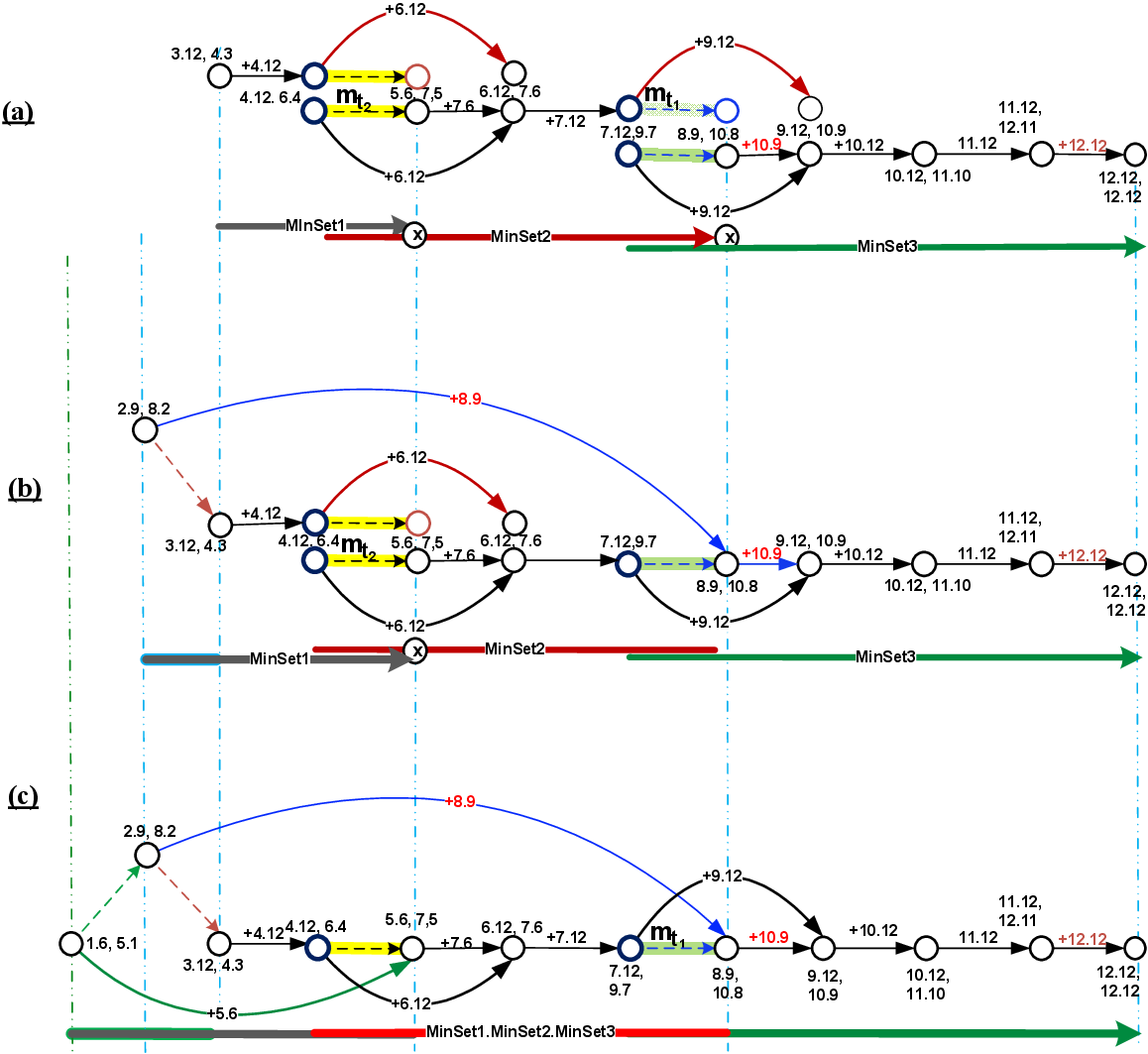}
\vspace{5pt}
 \caption{\textbf{Incrementing \& Joining the Adjacent MinSets}}\label{FG:FigMinSetParts}
\vspace{-15pt}
 \end{figure}

\newpage
\vskip -15pt
 \subsection{Incrementing the MinSet Sequences in $\mathbf{\gms}$}
 \vskip -09pt
 The above counting algorithm \ref{ALG:countMatchings} calls the following algorithm \ref{ALG:incrementMSS}, $incrementM\!SS()$, to increment all the MinSet Sequences in $\gms(i,n)$.
This algorithm uses two basic operations on MinSet, viz., increment a prefix MinSet, and join two or more adjacent MinSets into one, described by the associated algorithms in the next subsection.
\begin{algorithm} [h]
 \caption{ $incrementM\!S\!S( \gms(i+1, n))$}
 \begin{algorithmic}\label{ALG:incrementMSS}
  \STATE \textbf{Input:} $\gms(i+1, n)$;// contains all sequences $CMS_{i+1,n}(r), 1\le r \le n-2$
  \STATE \textbf{Output:} $\gms(i, n)$;
  \end{algorithmic}
  \hrule
 \vspace{-5pt}
 \begin{description}
  \item[Step (a): \emph{Increment all Prefix MinSets, $MinSet(m_{i+1}, m_s) \in \bigcup CMS_{i+1,n}(r)$}]
\end{description}
\vspace{-10pt}
\begin{algorithmic}[1]
\FORALL{$ x_i\in g(i)$}
  	\FORALL{$s \in [i\!+\!2~ \cdot\cdot~n\!-\!1], ~x_i R x_{t+1}, s>t+1$}
  		\FORALL[$O(n^6)$ MinSets covers all $CMS_{in}(r)$]{$MinSet ( m_{i+1}, m_s)\in \bigcup CMS_{i+1,n}(r)$}
		\STATE \emph{add} $incrMinSet(mdag\less x_i \more, MinSet(m_{i+1}, m_s))$ to $\gms(i, n)$;
  		\ENDFOR
 	\ENDFOR
  \ENDFOR
 \end{algorithmic}
 \vspace{-8pt}
\begin{description}
 \item[Step (b): \emph{Join all the Sub Sequences Selected by} $MinSet(m_{i}, m_t)$;] \ \\
     $ m_i = mdag\less x_i \more,$ $m_t = mdag\less x_t \more,~x_i R x_{t+1}$.
 \end{description}
 \vspace{-10pt}
 \begin{algorithmic}[1]
 \FORALL{$ x_i\in g(i)$}
 	\FORALL {$MinSet(m_i, m_{t}) \in \gms(i,n)$}
	\STATE {$updateSequence(MinSet(m_{i}, m_{t}), \gms(i,n))$};
 	 \ENDFOR
 \ENDFOR
\STATE \textbf{return} $\gms(i,n)$;
\end{algorithmic}
\end{algorithm}
   \vspace{-0pt}
\vspace{-5pt}
 \subsubsection{The Time Complexity }
 \vspace{-5pt}
\begin{claim}
The time complexity of Algorithm \ref{ALG:incrementMSS} is $O(n^{44}\log n)$.
\end{claim}
\vspace{-10pt}
\begin{proof}
It should be easy to see that the time in Step(b) dominates.\par
The For loop at line b(1) is iterated $O(n^2)$ times, and the For loop at line b(2) is iterated $O(n^6)$ times determined by the cardinality of
$\{(m_i, m_t)\}$ for a given $x_i$. \\
The time complexity at line b(3) of $updateSequence()$ is $O(n^{36}\log n)$ (Claim \ref{C:updateSequence}, Algorithm \ref{ALG:updateSequence}). \par

Therefore, the time complexity of the algorithm as determined by the Step(b) is\\
\vspace{-5pt}
$$T(Step(b)) = O(n^{8}* n^{36}\log n)= O(n^{44}\log n)$$.
\vspace{-15pt}
\end{proof}
  \subsubsection{Correctness of Algorithm \ref{ALG:incrementMSS}: incrementMSS()}
 \vspace{-0.1in}
 \begin{lemma}\label{L:corrctenss-incrementMS}
 For each $x_i \in g(i), i \ge 1$ the Algorithm \ref{ALG:incrementMSS} increments
 $\gms(i,n)$ to $\gms(i-1,n)$ to satisfy \eqref{EQ:CMS-partition} of Lemma \ref{L:CountofCMS}.
  \end{lemma}
 \vspace{-10pt}
 \begin{proof}

 The correctness follows from the fact that each increment operation by $ x_i \in g(i)$ applies either to a prefix or a subset of that prefix to each sequence of MinSets in $\gms(i,n)$. Moreover, all the $O(n^6)$ prefixes are incremented by the elements $x_i \in g(i)$.\par
Let $ps= MinSet(m_{i+1}, m_t)$ be a prefix to a sequence in $\gms(i, n), i+1 <t \le n-2$, and it is incremented by a unique $x_i \in g(i)$ in Step(a:4).
Then, either $x_i\cdot ps$ is in a MinSet in $\gms(i-1, n)$, or it exists as a new sequence of MinSets in $\gms(i-1, n)$, with a new prefix MinSet for the original MinSet sequence.
\par
Further, when increments by $x_i$ leads to selecting a subset of a MinSet sequence, such a subset of that sequence will also be a valid subset in $\vmpset(m_i, m_{n-1})$. Hence \eqref{EQ:CMS-partition} of Lemma \ref{L:CountofCMS} is always satisfied.
 \end{proof}
\newpage

\vspace{-5pt}
\subsection{Algorithm: Increment a MinSet}
\vspace{-5pt}
Increment of a MinSet, $MinSet ( m_{i+1}, m_{s})$ by an adjacent mdag, $mdag\less x_i \more$ would involve same kind of operations as in the case of $\vmpset ( m_{i+1}, m_{s})$. Unless the $R$-edge from $x_i$ is incident at the two distinguished nodes $x_{i+1}$ or at $x_{i+2}$, the multiplication would cover only a subset of VMPs in $MinSet ( m_{i+1}, m_{s})$. This subset is determined by the node partition $t$ in which the node $x_t$ in $MinSet ( m_{i+1}, m_{s})$ lies while $x_i R x_t$ holds true.
\begin{figure}[h]
\center
\includegraphics[scale=0.480]{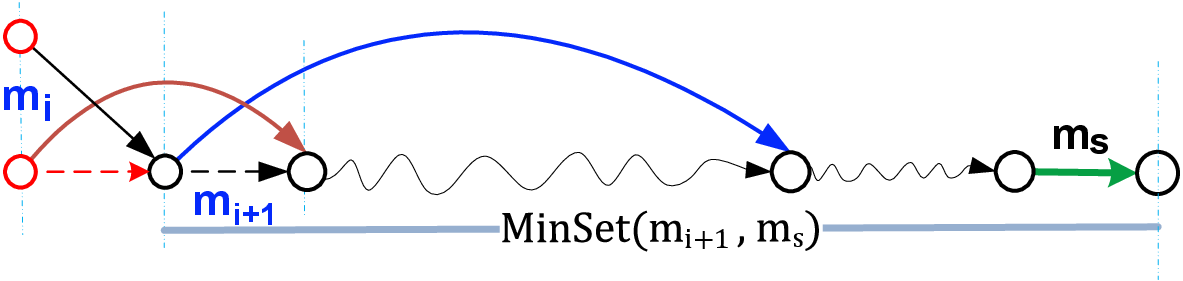}
\caption{Special Case: Incrementing a MinSet at a Common Node $x_{i+1}$}\label{FigIncrSp}
\vspace{35pt}
 \includegraphics[scale=0.55]{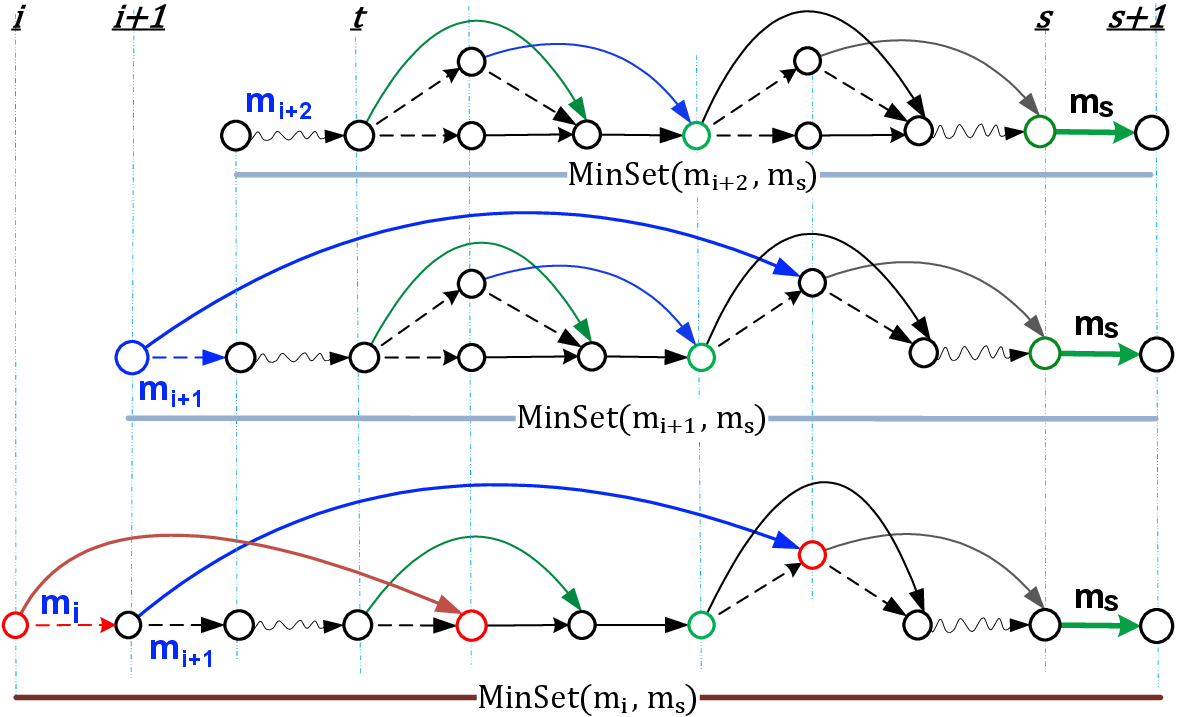}
 \caption{Two Successive Increments at $x_{i+2}$ and $x_{i+1}$}\label{FigIncrMinSet}
  \vspace{-10pt}
 \end{figure}
 \flushleft
  \par
The Step 0 in the algorithm \ref{ALG:incrMinSet} covers special cases, where the multiplying mdag multiplies all the VMPs in the MinSet using a common node (Figure \ref{FigIncrSp}). The Steps 1-2 show a general case (Figure \ref{FigIncrMinSet}) where the mdag can multiply only a subset of the VMPs from the original MinSet. This involves effectively re-constructing the whole MinSet by the revised list of the multiplying mdags in each node partition.
\vspace{-5pt}

 \ 
\begin{algorithm}
 \caption{ $incrMinSet( mdag\langle x_i\rangle, MinSet ( m_{i+1}, m_{s}))$} \begin{algorithmic}\label{ALG:incrMinSet}
  \STATE \textbf{Input:} $ m_i = mdag(x_{i}, x_{i+1}, x_{t+1}),~ m_{i+1} = mdag(x_{i+1}, x_{i+2}, x_{t+r}), r \ge 1;~MinSet ( m_{i+1}, m_{s})$
  \STATE \textbf{Output:} $MinSet ( m_{i}, m_{s})$ or $\{MinSet ( m_{i}, m_{i+1}),\, MinSet ( m_{i+1}, m_{s}) \}$
\vskip +5pt
 \hrule
  \end{algorithmic}
  \vskip -10pt
   \begin{description}
   \vskip -15pt
  \item[\ \ Step 0: \emph{Initialization and Special Cases}]
  \end{description}
  \vspace{-10pt}
 \begin{algorithmic}[1]
 \STATE let $x_i R x_{t+1}$; given $m_{i+1} = mdag(x_{i+1}, x_{i+2}, x_{t+1})$;
 \IF {$ER(x_{i+2}) \ne \emptyset$}
 	\STATE return $\{ MinSet( m_i, m_{i+1}),~ MinSet ( m_{i+1}, m_{s}) \}$;
  \ENDIF
 \IF { ($x_i R x_{i+1}$ OR $x_i R x_{i+2}$)}
  	\STATE return $ MinSet ( m_{i}, m_{s}) = m_i\cdot MinSet ( m_{i+1}, m_{s}) $;
  \ENDIF
 \end{algorithmic}
   \hrule
  \begin{description}
  \item[Step 1: \emph{Determine the candidate mdags in each node partition of $MinSet ( m_{i+1}, m_{s})$};]
      assumption: $i < t \le s$;
  \end{description}
  \vspace{-10pt}
 \begin{algorithmic}[1]

  \FORALL{$nodePartition~ j \in [i+1\, \cdot\cdot~ s-1]$}
    \STATE $mdagList[j] = \{mdag\langle x_j \rangle\}$;
   	\ENDFOR
    \STATE \emph{remove} each $mdag( x_t, x_{t+1}, x_r)$ from $mdagList[t]$ where $x_i {R\hspace {-7pt}\mathbf{/}} \hspace {+2pt} x_{t+1}$;
 \end{algorithmic}
  \hrule
 \vspace{-5pt}
  \begin{description}
  \item[Step 2: \emph{Rebuild the whole MinSet with the updated mdags}]\ \\\emph{(sequentially Increment the vmpSets by the mdag List)}
 \end{description}
  \vspace{-10pt}
 \begin{algorithmic}[1]
  \STATE $vmpSetList = \{mdag\langle x_{s-1} \rangle \cdot m_s\}$;
  \FORALL{$nodePartition~ j= s-2~downto~ i$}
     \FORALL{ $mdag \in mdagList[j]$}
		\STATE $newList = \emptyset$;
   	 	\FORALL{$vmpSet \in vmpSetList$ and adjacent to $mdag$}
 				\STATE \textbf{update:} \emph{add} $mdag \centerdot vmpSet$ to $newList$;
         \ENDFOR
 	 	\STATE $vmpSetList \Longleftarrow newList$;
    \ENDFOR
 	 \IF {($|mdagList[j]| =1$)}
      	\STATE $vmpSetList \Longleftarrow mergeMinSet(vmpSetList)$;
   	\ENDIF
  \ENDFOR 
 \STATE \textbf{output} $vmpSetList = MinSet ( m_{i}, m_{s})$; //
 \end{algorithmic}
 \vspace{-15pt}
  \begin{description}
  \item[End.]
 \end{description}
\vspace{-5pt}
  \end{algorithm}

 \flushleft
\newpage
\vspace{-15pt}
\begin{lemma}\label{L:incrMinsets}
 The time complexity of Algorithm \ref{ALG:incrMinSet}, $incrMinset()$, is $O(n^{12}\log n)$.
 \end{lemma}
 \begin{proof}
 The dominant time comes from Step 2. Each of the FOR loops at lines 3 and 5 are iterated $O(n^5)$ times and $O(n^3)$ times, determined by the bound $O(n^5)$ on the cardinality of the mdags in any node partition in $MinSet ( m_{i+1}, m_{s})$. \\
 The merge operation at line Step (2:11) takes at the most $O(n^{11}\log n)$ steps, and it dominates in Step 2. \\
 The FOR loop at line Step(2:2) can be iterated $O(n)$ times.\\
 Therefore, the time complexity of the above algorithm is $O(n*n^{11}\log n) = O(n^{12}\log n)$.
 \vskip -0.2in
 \end{proof}

\vspace{-10pt}
\subsubsection{Joining two Adjacent MinSets}
\vspace{-10pt}
The following algorithm \ref{ALG:joinMinSet} is essentially an iterative increment of the second MinSet, $MinSet ( m_{t}, m_{s})$, by the available adjacent mdags in the successive partitions of the first MinSet, $MinSet ( m_{i}, m_{t})$.\\
 At each iteration, a new set of VMP sets is created and which could also be merged into one VMP set.
 \begin{figure}[h]
   \center
 \includegraphics[scale=0.550]{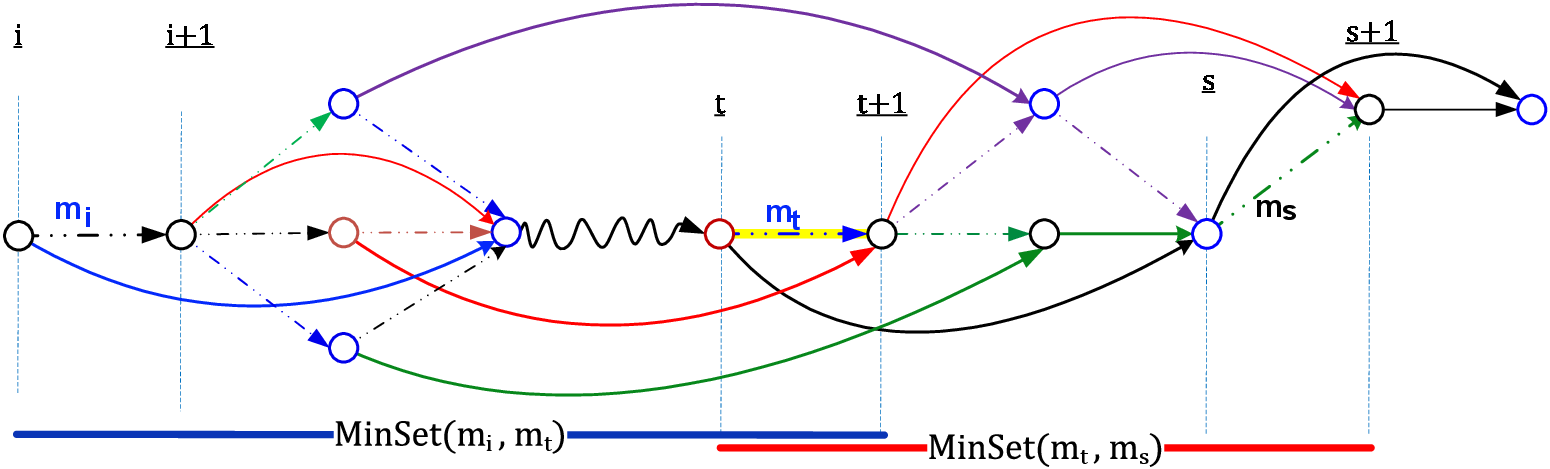}
 \caption{\textbf{Joining two Adjacent MinSets}\label{FG:FigJoinMinSet}}
 \end{figure}
 \vspace{-5pt}
\hskip -0.3in
 \begin{algorithm}[h]
 \caption{ $joinMinSet( MinSet ( m_{i}, m_{t}), MinSet ( m_{t}, m_{s}))$} \begin{algorithmic}\label{ALG:joinMinSet}
  \STATE \textbf{Input:} $MinSet ( m_{i}, m_{t}), MinSet ( m_{t}, m_{s})$;
  \STATE \textbf{Output:} $MinSet ( m_{i}, m_{t}, m_{s})$;
  \vskip +5pt
  \end{algorithmic}
  \vskip -10pt
  \begin{description}
  \vskip -10pt
  \item[ \emph{Compute the Product $MinSet(m_{i}, m_t)\centerdot MinSet ( m_{t}, m_{s})$}]
  \hrule
   \end{description}
\vspace{-10pt}
   \begin{algorithmic}[1]
	\FORALL{$nodePartition~ j \in [i\, \cdot\cdot~ t]$}
    \STATE $partition[j] = \{ x_j | x_j \in MinSet(m_{i}, m_t)\}$;
   	\ENDFOR

 \STATE $vmpSetList = \{MinSet ( m_{t}, m_{s}) \}$;
 \FORALL{$nodePartition~ j= t-1 ~downto~ i$}
  \FORALL{ $mdag\langle x_j \rangle$ in $partition[j]$}
 	\STATE $newList = \emptyset$;
    \FORALL{$vmpSet \in vmpSetList$ and adjacent to $mdag\langle x_j \rangle $}
         \STATE add $incrMinSet( mdag\langle x_j \rangle , vmpSet)$ to $newList$;
 	\ENDFOR
 \ENDFOR
 	\STATE $vmpSetList \Longleftarrow newList$;
  \IF {($|partition[j]| =1$)}
      \STATE {$mergeMinSet(vmpSetList)$};
  \ENDIF

  \ENDFOR 
 \STATE \textbf{output} $vmpSetList$;
 \end{algorithmic}
 \end{algorithm}

 \newpage
 \begin{lemma}\label{L:joinMinsets}
The time complexity of $joinMinSet()$ by Algorithm \ref{ALG:joinMinSet} is $O(n^{21}\log n)$.
 \end{lemma}

 \begin{proof} \ \\
  In the above Algorithm \ref{ALG:joinMinSet},\\
  the FOR loop on line 5 is iterated $O(n)$ times,\\
  the FOR loop on line 6 is iterated $O(n^5)$ times, determined by the cardinality of $\{mdag\less x_j \more\}$, \\
  the FOR loop on line 8 is iterated $O(n^3)$ times, determined by the subset of $vmpSetList$ adjacent to each mdag, and\\
  the time complexity of $incrMinSet()$ at line 9 is $O(n^{12}\log n)$.
 \par
  The merge operation at line 11 can be done in time $O(n^{11}\log n)$.\\
  Clearly the time between the lines 6-11 dominates and which is: $O(n^5*n^3*n^{12}\log n)= O(n^{20}\log n)$. \\

  Therefore, the total time over $O(n)$ iterations between the lines 2-13 is $O(n^{21}\log n)$.
 \end{proof}
 \subsubsection{Updating the MinSet Prefixes}
\vspace{-7pt}
 Joining two MinSets by an $R$-edge can become overly intricate when the $R$-edge from the incrementing node $x_i \in g(i)$ at step (a) of the algorithm $incrementMS()$ is incident at MinSets that are not adjacent to $x_i$. The algorithm $incrementMS()$ defers that "join" until an incrementing node $x_i$ which will join the MinSets adjacent to it is found.
 \par
 The following Figure \ref{FG:FigUpdateSequence} extends the previous Figure \ref{FG:FigMinSetParts} (Incrementing \& Joining Adjacent MinSets) to capture the basic behavior of this algorithm. It shows a chain of joining operations in a MinSet sequence induced by the $R$-edges stemming from the prefix MinSet, $MinSet(m_{i}, m_{t})$.
 \begin{figure}[h]
\vspace{-0pt}
 \center
\includegraphics[scale=0.75]{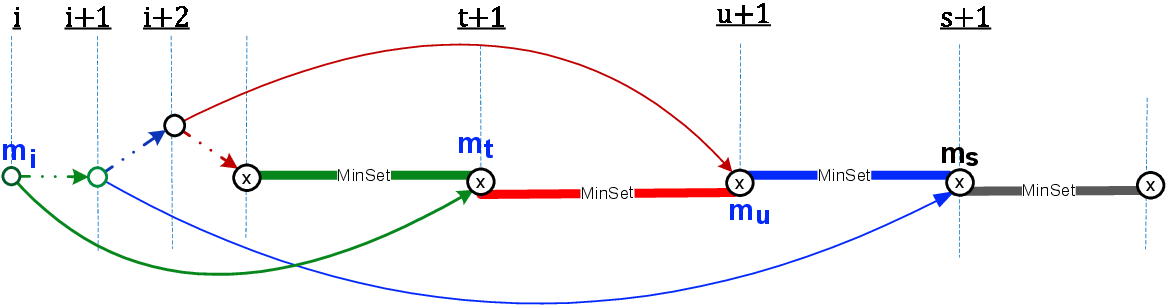}
 \caption{\textbf{Joining a Chain of MinSet Sequences by the Incrementing mdags}} \label{FG:FigUpdateSequence}
\vspace{-5pt}
 \end{figure}

 \begin{algorithm}[H]
\caption{ $updateSequence(MinSet(m_{i}, m_{t}), \gms(i, n))$}
 \begin{algorithmic}\label{ALG:updateSequence}
  \STATE \textbf{Input:} the prefix $MinSet(m_{i}, m_{t}), \gms(i, n)$;
  \STATE \textbf{Output:} \emph{updated} $\gms(i, n)$ with incremented MinSet prefixes in each sequence;
  \end{algorithmic}
  \hrule
 \vspace{-5pt}
 \begin{description}
  \item[Step (a): \emph{Find all the free jump edges in each partition in the Prefix $MinSet(m_{i}, m_t)$}]
 \end{description}
\vspace{-10pt}
\begin{algorithmic}[1]
   \STATE $freeRList := \{(x_j, x_k) \in MinSet(m_{i}, m_{t})\,|\, k \ge t;~ x_j R x_k \}$;
\COMMENT { $|freeRList| = O(n^4)$};
 	\STATE sort $freeRList$ in ascending $k$;
	\end{algorithmic}
 \vspace{-8pt}
\begin{description}
 \item[Step (b): \emph{Join the Sub-Sequences induced by the $R$-edges in $freeRList$}]
\end{description}
 \vspace{-10pt}
 \begin{algorithmic}[1]
 \STATE $mtSet := \{m_t\}; ~ prefixSet := \{MinSet(m_i, m_t)\}$;
\WHILE {$freeRList \ne \emptyset$}
	\STATE get next free jump edge $(x_j, x_k) \in freeRList$;
 	\WHILE {$mtSet \ne \emptyset $}
		\STATE remove next $mt$ from $mtSet$;
     	\STATE find next $prefix = MinSet(m_i, m_t) \in prefixSet$;
 		\IF {($mdag\less x_k \more = mt$)}
			\FORALL {$minset = MinSet(mt, m_u) \in \vmpset(mt, m_{n-1})$, where $m_u = mdag\less x_u\more$ }
				\STATE $prefix := joinMinSet( prefix, MinSet(mt, m_u))$;							 \STATE add $mt := m_u$ to $mtSet$;
				\STATE add $prefix = MinSet(m_i, m_t)$ to $prefixSet$;
				\STATE recompute $freeRList$ using Step(a);
            \ENDFOR	
		 \ENDIF
	 \ENDWHILE
\ENDWHILE
  \STATE replace all prefix MinSets in $\gms(i,n)$ with $prefixSet$;
 \end{algorithmic}
\end{algorithm}
   \vspace{-0pt}
\newpage
 \begin{claim}\label{C:updateSequence}
The time complexity of $updateSequence()$ in Algorithm \ref{ALG:updateSequence} is $O(n^{36}\log n)$.
\end{claim}
\vspace{-5pt}
\begin{proof}
It should be easy to see that the time in Step(b) dominates.\\
The loop at line b(2) is iterated $O(n^4)$ times.\\
The While loop at lines b(4) is iterated $O(n^5)$ times.\\
The For loop at line b(8) is iterated $O(n^6)$ times, determined by the number of mdags to be searched for $ER \neq \emptyset $.\\
 The time complexity at line b(9) of $joinMinSet()$ is $O(n^{21}\log n)$ and dominates all other operations even outside the For loops.\\
 Therefore, the time complexity of $updateSequence()$ is,\\
 \center
$T(Step(b)) = O(n^{4}*n^{5}* n^{6}*n^{21}\log n)= O(n^{36}\log n)$.

 \end{proof}

 \subsubsection{The Merge Operation}
 \vspace{-10pt}
 The following are high level algorithms for the merge operation used in the above algorithms, \ref{ALG:incrementMSS} and \ref{ALG:joinMinSet}.
\subsubsection*{Merging two MinSets }
This is called by $mergeMinSets()$ to merge all the MinSets in a given list.
 \vspace{-5pt}
\begin{algorithm} [H]
\caption{ $merge2MinSets (minSet1, minSet2)$}
 \begin{algorithmic}\label{ALG:mergeMinSet}
  \STATE \textbf{Input:} $minSet1= MinSet (m_{r}, m_{s})$, $minSet2=MinSet(m_{r}, m_{s})$;
  \STATE \textbf{Output:} $MinSet ( m_{r}, m_{s})$;
  \hrule
  \vskip +5pt
  \end{algorithmic}
  \begin{algorithmic}[1]
  \STATE \emph{union \& merge} each node partition pair for the MinSets,
  $minSet1$\\ and $minSet2$;
  \STATE {\emph{add} the counts if the MinSets can be merged: \vspace {-10pt}
  \[\vspace {-10pt} MinSet ( m_{r}, m_{s}).Count := minSet1.Count + minSet2.Count; \vspace {-10pt}
  \]}
  \STATE { \textbf{return} $MinSet(m_{r}, m_{s})$;}
\end{algorithmic}
\end{algorithm}

  \begin{claim}
 The time complexity of $merge2MinSets()$ in Algorithm \ref{ALG:mergeMinSet} is $O(n^6 \log n)$.
  \end{claim}
  \vspace{-5pt}
 \begin{proof}
 This merge operation is essentially a union operation of the $O(n^5)$ distinct mdags in each node partition. Assuming that all the node partitions have been pre-sorted, the search for any mdag in a partition with $O(n^5)$ mdags can be done in $O(\log n)$ time. Thus the union operation in each node partition requires $O(n^5 \log n)$ time.\\
Clearly, the add operation at line 2 is not dominating, taking only $O(n^3)$ time.
 \end{proof}

\newpage
  \subsubsection{Merge all MinSets }
 \vspace{-10pt}
 The following algorithm mergers all the MinSets, $MinSet(m_i, m_j)$, given in a list of MinSets.
 \vspace{-5pt}
   \begin{algorithm} [h]
 \caption{ $mergeMinSets (MinSetList)$}
 \begin{algorithmic}\label{ALG:mergeMinSetList}
  \STATE \textbf{Input:} A list of $MinSet(m_r, m_s)$ in $MinSetList$;
  \STATE \textbf{Output:} $\{MinSet ( m_{r}, m_{s})\}$;
  \hrule
  \vskip +5pt
  \end{algorithmic}
  \begin{algorithmic}[1]
    \STATE mergeSet := MinSetList[1]; l := length(MinSetList);
  	\FOR {j := 2 to l}
		\STATE $mergeSet := merge2MinSet(mergeSet, MinSetList[j]);$
	 \ENDFOR
     \STATE return $mergeSet$:
   \end{algorithmic}
  \end{algorithm}
  \begin{claim}
 The time complexity of Algorithm \ref{ALG:mergeMinSetList} is $O(m n^6 \log n)$, where $m$ is the size of $MinSetList$.
  \end{claim}
 \section{Conclusions}\label{S:conclusion}
\vspace{-10pt}
\subsection{Collapse of the Polynomial Hierarchy}
 We can re-state Lemma \ref{L: correctCount} as the following Theorem in terms of the class $\mathbf{FP}$ which is defined as the class of functions
  $f: \{0, 1\}^* \rightarrow \mathbb{N} $ computable in polynomial time on a deterministic model of computation such as a deterministic Turing machine or a RAM.
\par
\begin{theorem}
The counting problem for perfect matching is in $\mathbf{FP}$, and therefore,
$\mathbf{\#P}=\mathbf{FP}$ and $\mathbf{NP} = \mathbf{P}$.

\end{theorem}
Based on the fact that every $\#P$-complete problem is also $NP$-hard, it follows that
 $\mathbf{NP \subseteq P^{\#P}}$. And therefore, the above Theorem implies that polynomial hierarchy $\mathbf{PH}$ collapses into $\mathbf{P}$. Needless to say that the main Theorem of Toda ~\cite{Toda89}, which states that the class $\mathbf{\#P}$ contains $PH$, is a re-confirmation of $\mathbf{PH}$ collapsing into $\mathbf{P}$.
 \vspace{-10pt}
\subsection{A Characterization of P-time Enumeration}
\vspace{-10pt}
The  forgoing enumeration technique gives rise to the following conjecture on a characterization of the polynomial time enumeration:
\vspace{-10pt}
\begin{quote}
A sufficient condition for the existence of a P-time algorithm for any enumeration problem is
 the existence of a partition hierarchy of the exponentially decreasing solution spaces, where each partition is polynomially bounded and the disjoint subsets in each partition are P-time enumerable for each $n \ge 1$, n being a problem size parameter.\vspace{-10pt}
\end{quote}
\vspace{-4pt}
Although we may have an existential proof for a sufficient condition for the P-time enumeration, a more fundamental question is if this condition is also necessary.
\par
An attempt to prove that necessary condition was made in an unpublished paper \cite{jaslam92}. The basic logic behind the proof was that any deterministic search must cover the entire solution space of the search problem, and hence must also be able to count all the solutions in essentially the same time bound. This logic lead to the enumeration model conjectured in this paper.
\par
A simultaneous polynomial bound on the depth as well as on the width (partition size) of the partition hierarchy creates a non-trivial relationship between the partitions at any two consecutive levels. This is simply because an exponentially large set can never be reduced to a constant size in polynomially many steps by the subset operations.
\par
This is an area of algorithm design which has not received much attention so far.
Some thoughts along this line have been covered in Appendix \ref{a:charactPTimeEnum}.
\vspace{-10pt}
\bibliographystyle{amsalpha}
\bibliography{permalgebraArxiv94}
\vspace{-10pt}
\newpage
\begin{appendices}
\section{}
\subsection{Permutation Multiplication- Proof of Theorem \ref{TH:multBasic1}}\label{ap:multBasic1}
\begin{proof}

Let $\psi = (j, k)$ be a transposition in $S_n$. Note that $\psi$ need not be realized by $BG'$, however, we will show that there are two unique edges in $BG'$ that represent $\psi$, and depend on $E(\pi)$ whenever $\pi\psi$ is realized by $BG'$.
\par
Let $i, t \in \Omega$ be the two points mapped by $\pi$ such that
 $i^{\pi} = j$, and $t^{\pi} = k$. Thus $E(\pi)$ covers the edges $v_i w_j$ and $v_t w_k$ in $BG'$.

 \par
 \underline{$\pi\psi \in M(BG') ~\implies~ $ a cycle of length 4}
\par
If the product $\pi\psi$ is realized by $BG'$, then we must have:
\begin{align*}
&i^{\pi\psi} = j^\psi = k, ~~\text{ and }\\
&t^{\pi\psi} = k^\psi = j.
\end{align*}
That is, the existence of the edges in $E(\pi\psi)$ dictates
that $BG'$ contain the edges $v_i w_j$ and $v_i w_k$ at the
vertex $v_i \in V$, and $v_t w_j$ and $v_t w_k$ at the vertex $v_t
\in V$. And hence, $BG'$ has a cycle $v_i w_j v_t w_k$ of length 4.
\par

\vspace{0.1in}
\underline{A cycle of length 4 $\implies \pi\psi \in M(BG')$}
\par
Let $C=v_i w_j v_t w_k$ be a cycle of length 4 in $BG'$ where $\pi$ is such that $i ^\pi = j$ and $t ^\pi = k$, and thus $\pi$ covers $v_i w_j$ and $v_t w_k$.\\
The new permutation $\pi_1 = \pi \psi$ can be realized by swapping the alternate edges of $C$ such that $\pi_1$ differs from $\pi$ only in two positions, viz., $i^ {\pi_1} = k$ and $t ^{\pi_1} = j$, corresponding to the edges $v_iw_k$ and $v_tw_j$. \\
Now we show how $\psi$ is encoded by the two alternate edges of $C$.
\par
Since $\psi = \pi^{-1} \pi_1$, we have
\begin{align*}
&j ^\psi = j^{\pi^{-1}\pi_1} = i ^ {\pi_1} = k,~~\text{ and}~~ \\
&k ^ \psi =k^{\pi^{-1}\pi_1} = t ^{\pi_1} = j.
\end{align*}
Therefore, $\psi = (j, k)$ is represented by the alternate edges, $v_i w_k$ and $v_t w_j$ in $C$
which effectively realizes $\pi\psi$. Clearly, the edges in $C$ representing $\psi$
 depend on $\pi$ by the mapping $t ^{\pi} = k$.\\
\end{proof}
\vskip 5pt
\par
The following Corollary of Theorem \ref {TH:multBasic1} generalizes the
multiplier $\psi$ to any permutation cycle not necessarily a
transposition.
\begin{corollary}\label{CR:multBasic1}
Let $\pi \in S_n$ is realized by a bipartite graph $BG'$ on $2n$ nodes. If $\psi \in S_n$ is a permutation cycle of length $r \le n$, then $\pi\psi $ is realized by a bipartite graph $BG'$ iff there exists a graph cycle of length $2r$ in $BG'$ such that the alternate edges in the cycle are covered by $\pi$ and $\pi\psi$.
\end{corollary}
\begin{proof}
The result can be easily proved by an inductive application of the above Theorem.
\end{proof}
\newpage
\subsection{Proof of Corollary \ref{CR2:cosetEdge}} \label{a:cosetEdge}

\begin{proof} Recall that $BG_i$ is a subgraph of the complete bipartite graph $BG=K_{n,n}$ induced by the subgroup $G^{(i)}$. That is, $\forall j \in \{1,2, \cdots, i\}$, and for each $E(\pi)$ in $BG_i$, $j^{\pi} = j $. Following Theorem \ref {TH:multBasic1} we can identify the cycle responsible for realizing the multiplication $\pi\psi$, and see how $\psi$ depends on $\pi \in G^{(i)}$.
\begin{align*}
\text{$\pi$ and $\pi\psi$ is realized by $BG_{i-1}$ } &\Longleftrightarrow i^{\pi\psi} = i^{\psi} = k \text{ and } t^{\pi\psi} = k^\psi = i\\
&\Longleftrightarrow \text{ edges $v_i w_k$, $v_t w_i \in BG_{i-1}$}, \text{ where $BG_0 = BG$}.
\end{align*}

\begin{figure}[h]
\center
  \includegraphics[scale=0.45]{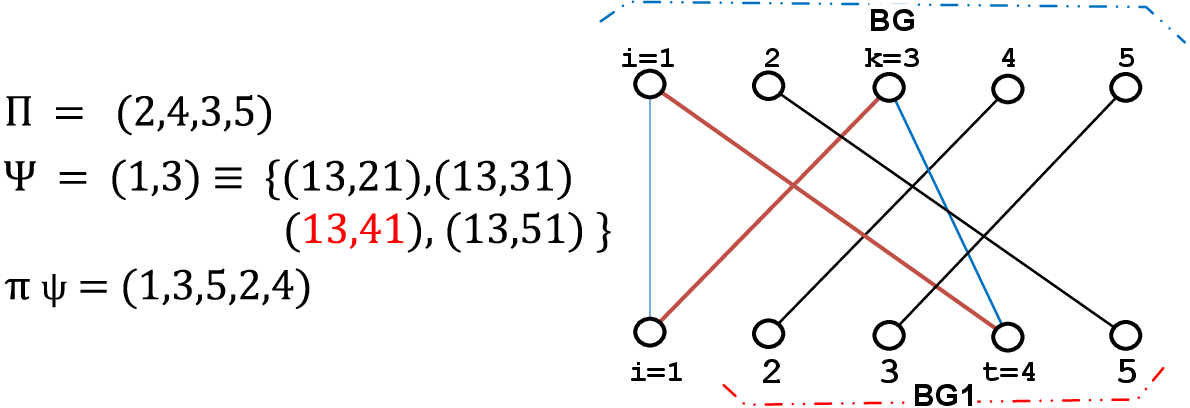}
   \caption{\textbf{Multiplication by a Coset Representative $\mathbf{\psi=(1,3)}$}}\label{FG:a:multCoset}
\end{figure}
\renewcommand{\abovecaptionskip}{5pt}

Clearly, the point $k$ is fixed by $\psi$ for a given $i$, and $t$ is then fixed by $\pi$. Therefore, each $(\psi, \pi)$ pair uniquely defines the edge pair $a_i( \psi, \pi) =(v_i w_k, v_t w_i)$.
Also, it is easy to see that the only edge pair that can form a cycle of length four with the edge pair $(v_i w_i, v_t w_k)$ is $(v_i w_k, v_t w_i)$, giving the cycle $(v_i w_k v_t w_i)$.
\end{proof}
\begin{remark}
One should note the analogy of forming the product $\pi\psi$ with the augmenting path concept in constructing a perfect matching ~\cite{kuhn55, egervary}. The cycle $(v_i,w_k,v_t,w_i)$ [Figure \ref{FG:a:multCoset}], which is used to multiply $\pi$ and $\psi$, always contains the augmenting path $(v_i,w_k,v_t,w_i)$ corresponding to the matched edge $v_tw_k$ in $E(\pi)$.
\end{remark}

\subsection{Permutation Multiplication Defined by an $R$-Cycle}
The following Lemma shows how does an $R$-cycle compose a sequence of coset representatives. It is an extension of Corollary \ref{CR2:cosetEdge}.
 \begin{lemma} \label{L:multR-cycle}
 Let $C_{ab}$ be an $R$-cycle, defining $aRb$, in a bipartite graph $K_{n,n}$, where
   $a\in g(i)$ and $ b \in g(j), ~ 1 \le i < j \le n$, and $x_{i_r} \in g(i_r)$, $1 \le r \le j-i$,
    are all the edge pairs covered by $C_{ab}$ such that $i = i_1 < i_2 \cdots <i_{r-1} < i_r <i_{r+1} =j$. Also let $\pi(b) \in G^{(j-1)}$ be a permutation realized by the bipartite graph $BG_{j-1}$.
 Then $C_{ab}$ represents a composition of the coset representatives leading to the permutation $\pi_a$ given by
 \begin{equation}\label{e:R-cycleMult}
 \pi_a = \psi(x_{i_{r}}) \psi(x_{i_{r-1}}) ~\cdots ~\psi(x_{i_2})\psi(x_{i_1}), \text{ where }\psi(x_{i_r}) = \psi_{i_r} \in U_{i_r},
 \end{equation}
 such that $\pi(b)\pi_a \in G^{(i-1)}$ covers $a$ and other alternate edges in $C_{ab}$.
 \end{lemma}
 \begin{proof}
 The proof is by induction on $r$, using the arguments in the proof of Corollary \ref{CR2:cosetEdge}.
 \ \end{proof}
The following Lemma provides a group theoretic semantics for the relation $R$. It correlates the permutation multiplication in U$K_{n,n}$ and the relation $R$ in $\Gamma(n)$ .
\begin{lemma}\label{TH:R-CycleMult-TR}
Let $a \in g(i),~ b \in g(j)$ be the edge pairs at the nodes $i$
and $ j$ respectively in $BG=K_{n,n}$, such that $G^{(j)} < G^{(i)}, ~1 \le i < j \le n$.
Let $aRb$ be realized by the transitivity over the intermediate nodes such that
$\forall k, ~ j > k \ge i ,~ \exists x_k \in g(k) ,~ x_{k+1} \in g(k+1)$ and $x_k R x_{k+1}$.
Then aRb represents a permutation
\begin{equation}\label{e:R-CycleMult-TR}
\pi_a = \psi(x_{j-1}) \psi(x_{j-2})~\cdots ~\psi(x_{i-1})\psi(x_i)
\end{equation}
 where $ \psi(x_r) = \psi_r \in U_r,~ i \le r \le j-1$, such that the product $\pi(b)\pi_a $ is realized by $BG_{i-1}$ and that it covers $a$, $b$ and other alternate edges of the $R$-cycle(s)
defined by $aRb$.\\
\end{lemma}
\begin{proof}
 The proof is essentially by induction on the number of $R$-cycles in the transitive chain $aRb$.
 When there is exactly one $R$-cycle defined by $aRb$, the result follows directly from the above Lemma \ref{L:multR-cycle}.
\par
  Whenever there are one or more ID nodes between $i$ and $j$, we have two or more disjoint $R$-cycles such that each cycle represents a permutation given by Lemma \ref{L:multR-cycle}.
\end{proof}

\subsection{More Properties of the Generating Graph}\label{a:genGraphProps}
We now present few basic properties and attributes of the generating graph.
\par
The $R$-\emph{in (out)degree} of a node $x \in \Gamma$ is defined as the number of $R$-edges incident (going out) on (from) $x$.
The $S$-\emph{ in (out) degree} of a node $x \in \Gamma$ is defined analogously.

\begin {property}\label{F:R-forall}
In every generating graph $\Gamma(n)$, $\forall i < n$ and $\forall x_i \in g(i),~\exists ~j \le n$ and
$ x_j \in g(j)$ such that $x_i R x_j$. Similarly, the reverse result is also true-- for all $x_j \in g(j)$ and $ \forall i < j$ there exists $x_i \in g(i),$, such that $x_i R x_j$.
\end{property}
\begin{proof}
The result is due to the completeness of the bipartite graph.\\
For all $x_i = (v_i w_k, v_j w_i) \in g(i)$, $ 1 \le i < j, k \le n$, there exist edges, $v_j w_k$ and $v_i w _i$ in $BG$, such that they form an $R$-cycle of length 4 with $x_i$ covering the edge $v_j w_k$. Therefore, we will always have either $x_i R x_j$ or $x_i R x_k$.
\end{proof}
\begin {property}\label{F:R-forall2}
In every generating graph $\Gamma(n)$, $\forall (i, j),~ 1 \le i < j \le n,$ $\exists ~x_i \in g(i)$ and $x_j \in g(j)$, such that $x_i R x_j$
\end{property}
\begin{proof}
Simply note that the edges needed for forming a cycle of length four with $x_i$ and one of the edges in $x_j$ are always available in $K_{n, n}$.
\end{proof}
\begin {property}\label{PR:R_or_S1}
Let $i$ and $j > i$ be any two node partitions in $\Gamma(n)$. Then $\forall x_i \in g(i)$, $x_i R x_j \implies \nexists y_j \in g(j)$ such that $x_i$ and $y_j$ are disjoint, and $x_i R x_j$ is false. Similarly $x_i$ and $y_j$ being disjoint, and $x_i R x_j$ being false implies $\nexists y_j \in g(j)$ such that $x_i R y_j$.
\end{property}
\begin{proof}
One should note that the condition for two edge pairs in $K_{n,n}$ being related by $R$ is mutually exclusive to the condition for the corresponding nodes in $\Gamma(n)$ being disjoint. In one case, when $x_i R y_j$ is true, the node pairs at $j$ overlap with the vertex of one of the edges in the edge pair $x_i$ in $BG$, and in the other case, $x_i R x_j$ being false, $j$ must be disjoint with the vertices at the node pairs covered by $x_i$.
\end{proof}
The following Property is essentially a complement of Property \ref{PR:R_or_S1}.
\begin {property}\label{PR:R_or_S2}
For all $ (i, j),~ 1 \le i < j < n,$ and $\forall x_i \in g(i)$, if $\exists x_k \in g(k), n\ge k > j$, such that $x_i R x_k$, then $\exists x_j\in g(j)$ such that $x_i$ and $x_j$ are disjoint.
\end{property}
\begin{proof}
An instance of this property can best be understood by looking at the layout of the edge pairs, $x_i, x_j$ and $x_k$ in $K_{n,n}$. The relation $x_i R x_k$ directly implies that the edge pairs in all the partitions in $\{t ~| ~ i < t<k\}$ have at least one edge pair $x_t$ available such that a perfect matching can be formed. This must be true since we have a complete bipartite graph. And hence $x_t$ must be disjoint to $x_i$ (although not necessarily to $x_k$).
\\
\end{proof}
\begin {property}\label{F:allRsamePartn}
All the $R$-edges coming from a given node in $\Gamma(n)$ go to the same node partition. Thus either all $R$-edges coming from a node are \emph{direct} edges, or all are \emph{jump} edges.
\end{property}
\subsection{Permutation Represented by an $R$-Path}
The following is a direct Corollary of Theorem \ref {TH:R-CycleMult-TR}, noting that the product $\pi(b)\pi_a$ is realized by $BG_{i-1}$. It provides a group theoretic semantics to an $R$-path in $\Gamma(n)$.
\begin{corollary}\label{CR:R-Mult1}
Let $p = x_i x_{i+1} ~ \cdots ~ x_{j-1} x_{j}$, $1 \le i < j \le n$, be an $R$-path in $\Gamma(n)$ defined by $ x_i R x_{j}$, where $x_i \in g(i)$, and let $\psi(x_k)$ be the transposition defined by the edge-pair $x_k$. Then $p$ defines a permutation cycle $\pi_p$ given by the product of the transpositions
 \begin{equation}\label{e:R-pathPermutation}
\pi_p =
 \psi(x_{j}) \psi(x_{j-1}) ~\cdots ~ \psi(x_{i+1})\psi(x_{i}),
\end{equation}
such that
$\pi_p$ covers $x_{i}$, $x_j$ and other alternate edges of the $R$-cycle(s) defined by $ x_i R x_{j}$.
\end{corollary}
\flushleft

\par
The above Corollary \ref{CR:R-Mult1} effectively describes how larger permutation cycles are composed by the $R$-paths which eventually lead to a perfect matching whenever that $R$-path covers all the $n$ node partitions in $\Gamma(n)$.

\par
\newpage

 \subsection{More VMP Properties}\label{a:VMPProps}
\begin{property}\label{PR:CVMP-2}
A VMP, $p = x_i x_{i+1} ~\cdots~ x_{t-1}x_j$ in $\Gamma(n)$, is a \underline{complete VMP} if it satisfies any one of the following conditions:
\begin{enumerate}
\item $p$ is an $R$-path with no jump edges.
\item The path $p= x_ip'$ obtained by incrementing a CVMP, $p'= x_{i+1} x_{i+2} ~\cdots~ x_j$, using a valid mdag, $mdag(x_i,~x_{i+1},~x_t)$, $x_t \in p'$, or by an $R$-edge $x_i x_{i+1}$.

\item $p = p_1 p_2$, where $p_1$ and $p_2$ are CVMPs.
\end{enumerate}
\end{property}
\begin{proof}
The proof of the above three properties is as follows.
\begin{enumerate}
\item $p$ is an $R$-path:
Obvious.
\item $p = x_i p'$ is a CVMP:\label{PR:CVMP-2-INCR}\\
Clearly, the new path $p$ is a VMP by virtue of the valid mdag, $mdag(x_i,~x_{i+1},~x_t)$, and this mdag is covered by $p$.
\item $p = p_1 p_2$:\\
Simply note that the concatenation behavior of two or more CVMPs is exactly same as that of the $R$-edges-- except that in CVMPs there may be two $R$-edges meeting at the starting node of $p_2$.

\end{enumerate}
\end{proof}

\subsection{The Permutation Represented by a CVMP}
The following Lemma provides a group theoretic semantics of a CVMP, showing how a CVMP represents a product of coset representatives that would multiply any element of the associated subgroup. Further, it shows how that product is represented by a set of matched edges.
\par
Let $E'(\pi)$ represent a subset of the matched edges in $E(\pi)$.
\begin{lemma}\label{a:L:CVMP-perm}
Every $CVMP$, $p= x_i x_{i+1} ~\cdots~ x_{j-1}x_j$ in $\Gamma(n)$, represents a permutation $\pi \in G^{(i-1)}$, and a matching $E'(\pi)\subseteq E(\pi)$ \emph{(}on the nodes $i, i+1, ~\cdots, ~ j$ in $K_{n,n}$\emph{)} given by
\begin{equation} \label{e:CVMP-perm3}
\pi = \psi(x_j) \psi(x_{j-1}) ~\cdots ~ \psi(x_{i-1}) \psi(x_{i})
\end{equation}
where $1 \le i < j \le n$, and $x_i \in g(i)$.
\par
Note. It is implicit that whenever $j <n$, $\exists x_k$ such that $x_j R x_k$, where $~ j < k \le n$. Therefore, by Theorem \ref{TH:R-CycleMult-TR}, $\pi$ would multiply all the
permutations $\pi'(x_k) \in M(BG_{k-1})$, to give rise to $\pi'(x_k)\pi \in M(BG_{i-1})$.
\end{lemma}
\begin{proof}
The proof is by induction on the length, $l=|p|$ of the CVMP, $p$. For notational convenience we can assume each edge pair $x_i$ to be a set of two edges.
\par
\underline{Basis}
\newline
For $l=1$ the CVMP is an $R$-edge, $x_i x_{i+1}$, which represents the permutation,
$\pi = \psi(x_{i+1}) \psi(x_{i})$ (Corollary \ref{CR:R-Mult1}).\\ For $l=2$ the CVMP is either an $R$-path of length 2, or an mdag, $mdag(x_i, x_{i+1}, x_{i+2})$, which represents $\pi =\psi(x_{i+2})\psi(x_{i+1}) \psi(x_{i})$.
\par
\underline{Induction}\newline
Let \eqref{e:CVMP-perm3} be true for all $ p$, $2 \le |p| \le l< n-1$, that is, we have a CVMP, $p$, of length $j-i$ that realizes the permutation $\pi$ and a matching $E'(\pi)$. Let the new CVMP of length $j-i+1$ be $x_{i-1}p$, $x_{i-1}\in g(i-1)$, and let $x_t \in p$ be such that $x_{i-1}R x_t$. It will suffice to show that the new CVMP realizes the permutation $\pi \psi(x_{i-1})\in G^{(i-2)}$.
\par
\emph{\textbf{Note}}: We assume that $x_{i-1} $ is not an ID node, i.e., $x_{i-1} \ne id_{i-1}$, otherwise the result would be trivially true.
\par
Since the new CVMP $p'$ of length $l+1$ is derived from Property \ref{PR:CVMP-2}(2), there is an mdag, $mdag(x_{i-1}, x_i, x_t)$, or an $R$-edge $x_{i-1}x_i$, such that
 $\psi(x_{i-1})= (i-1, k)$, and $k^\pi = t$ . Therefore, by Corollary \ref{CR2:cosetEdge}, the cycle defined by $x_{i-1}R x_t$ realizes the product $\pi \psi(x_{i-1})\in G^{(i-2)}$.
\end{proof}

\subsection*{The Matching Represented by a CVMP}
\begin{lemma}\label{a:L:CVMP-ER}
Every $CVMP(m_i, m_j)$, $p= x_i x_{i+1} ~\cdots~ x_{j-1}x_j$ in $\Gamma(n)$, represents a matching $E'(\pi)\subseteq E(\pi)$ \emph{(}on the nodes $i, i+1, ~\cdots, ~ j$ in $K_{n,n}$\emph{)} given by
\begin{equation} \label{e:CVMP-er1}
E'(\pi) = \{ e~|~ e \in x_i \in p\} - \{SE(x_s x_t)| x_s, x_t \in p\},
\end{equation}
where $\pi \in G^{(i-1)} < S_n$, $1 \le i < j \le n$, and $x_i \in g(i)$.
\par
Note. It is implicit that whenever $j <n$, $\exists x_k$ such that $x_j R x_k$, where $~ j < k \le n$. Therefore, by Theorem \ref{TH:R-CycleMult-TR}, $\pi$ would multiply all the
permutations $\pi'(x_k) \in M(BG_{k-1})$, to give rise to $\pi'(x_k)\pi \in M(BG_{i-1})$.
\end{lemma}
\begin{proof}
The proof is by induction on the length, $l=|p|$ of the CVMP, $p$. For notational convenience we can assume each edge pair $x_i$ to be a set of two edges.
\par
\underline{Basis}
\newline
For $l=1$ the CVMP is an $R$-edge, $x_i x_{i+1}$, which represents the permutation,
$\pi = \psi(x_{i+1}) \psi(x_{i})$, and a matching
$E'(\pi) = x_i\cup x_{i+1} -\{SE(x_ix_{i+1})\}$. For $l=2$ the CVMP is either an $R$-path of length 2, or an mdag, $mdag(x_i, x_{i+1}, x_{i+2})$, which represents $\pi =\psi(x_{i+2})\psi(x_{i+1}) \psi(x_{i})$. The matched edges can be deduced from the $SE(e)$ of associated $R$-edge $e$. That is, we have\\
 either
\[
 E'(\pi) =x_1\cup x_2 \cup x_3 -\{SE(x_1x_2), SE(x_2x_3)\},
 \]
or
\[
 E'(\pi) =x_1\cup x_2 \cup x_3 -\{SE(x_1x_3), SE(x_2x_3)\}.
\]
\par
\underline{Induction}\newline
Let \eqref{e:CVMP-er1} be true for all $ p$, $2 \le |p| \le l< n-1$, that is, we have a CVMP, $p$, of length $j-i$ that realizes the matching $E'(\pi)$. Let the new CVMP of length $j-i+1$ be $x_{i-1}p$, $x_{i-1}\in g(i-1)$, and let $x_t \in p$ be such that $x_{i-1}R x_t$. It will suffice to show that the new CVMP realizes the permutation $\pi \psi(x_{i-1})\in G^{(i-2)}$, and the new matching $E'(\pi\psi(x_{i-1})) = E'(\pi) \cup x_{i-1} -\{SE(x_{i-1}x_t)\}$.
\par
\emph{\textbf{Note}}: We assume that $x_{i-1} $ is not an ID node, i.e., $x_{i-1} \ne id_{i-1}$, otherwise the result would be trivially true.
\par
Since the new CVMP $p'$ of length $l+1$ is derived from Property \ref{PR:CVMP-2}(2), there is an mdag, $mdag(x_{i-1}, x_i, x_t)$, or an $R$-edge $x_{i-1}x_i$, such that
 $\psi(x_{i-1})= (i-1, k)$, and $k^\pi = t$ . Therefore, by Corollary \ref{CR2:cosetEdge}, the cycle defined by $x_{i-1}R x_t$ realizes the product $\pi \psi(x_{i-1})\in G^{(i-2)}$.
\par
The addition of the new node $x_{i-1}$ to $p$ adds the corresponding edge pair $x_{i-1}$ in the bipartite graph to the matched edges. Moreover, the new $R$-edge $x_{i-1}x_t$ in $p'$ will remove the edge $SE(x_{i-1}x_t)$ from the set $E'(\pi) \cup x_{i-1}$. Therefore,
 \[
 E'(\pi\psi(x_{i-1})) = \{ e~|~ e \in x_i \in p'\} - \{SE(x_j x_k)| x_j, x_k \in p'\}.
\]
\end{proof}
\subsection{A Partitioning Scheme for CVMP Sets }

\par
\begin{claim}\label{a:CL:cosetCover}
For each $(\pi, \psi) \in G^{(i)} \times U_i$, there exists a unique pair
$(p, x_i) \in prod\vmpset ( m_{i+1}, m_t, m_{n-1}) \times g(i) $, where $x_i R x_{t+1}$,
and $x_{t+1} \in m_t$, such that\\
 the product $\pi\psi$ is uniquely realized by $ x_i \cdot p \in x_i\cdot prod\vmpset ( m_{i+1}, m_t, m_{n-1})$.
\end{claim}
\vspace{ -6pt}
\begin{proof}
The result follows from the existence of a unique $R$-edge incident from $x_i$ to $p$ whenever the associated permutation product $\pi \psi$ is realized by $x_i.p$. [Corollary \ref{CR2:cosetEdge}].
\par
Note that there exists a mapping\par
\hskip 0.4in $f:G^{(i)} \times U_i \rightarrow prod\vmpset ( m_{i+1}, m_t, m_{n-1}) \times g(i) $, such that\\
\hskip 0.4in $\forall (\pi, \psi) \in G^{(i)} \times U_i$, the product
$ x_i \cdot p \in x_i\cdot prod\vmpset ( m_{i+1}, m_t, m_{n-1})$ is realized by a unique $R$-edge such that $x_i R x_{t+1}$, where $x_{t+1} \in m_t$, $\psi(x_i) = \psi$ and
$\pi(p) = \pi$.
 \end{proof}
Let $x_{i}\centerdot prod\vmpset ( m_{i+1}, m_t, m_{n-1}) $ denote all the \emph{allowed} multiplication of the CVMPs in
$prod\vmpset ( m_{i+1}, m_t, m_{n-1}) $ by $x_{i} \in g(i) $, with the resulting CVMPs of length $n\!-i$ between the nodes $x_{i}$ and $x_n$. Also let $m_i =mdag\less x_i\more$.
\begin{lemma} \label{a:L:VMPprodAtX} 
Given all the partitions, $\{prod\vmpset ( m_{i+1}, m_t, m_{n-1})\}$, where $i\! \le \! t\! \le\! n\!-\!1$, of
$\{C\vmpset ( m_{i+1}, m_{n-1})\,\}$, the next larger CVMP sets, can be computed as follows:
\vspace{-5pt}
\begin{multline}\label{e:VMPprodAtX}
C\vmpset ( m_{i},\, m_{n-1}) = \biguplus_{ m_t} \Big\{\hspace{-3pt} (x_{i}\centerdot prod\vmpset ( m_{i+1}, m_t, m_{n-1}) )\, \big |\\
\vspace{-5pt}
 \, (x_{i} R x_{t+1} \text{ \textbf{and} } x_{i} Sx_{i+1})\text{\emph{ \textbf{or} }} x_{i} R x_{i+1}, \, x_i \in g(i) \,, i\! \le \! t\! \le\! n\!-\!1 \Big\}
\vspace{-19pt}
 \end{multline}
where $(m_{i}, m_{n-1})$ \emph{covers} $g(i)\!\times\! g(n-1)$, and $(m_{i+1}, m_t, m_{n-1})$ \emph{covers} $g(i+1)\!\times\! g(t)\times\!g(n-1)$.
\end{lemma}
\begin{proof}\ \\
Follows from Lemma \ref{L:mapGenset} and Claim \ref {a:CL:cosetCover}.

\end{proof}

 \begin{lemma}
All the $C\vmpset(m_i, m_{n-1})$, $1 \le i \le n-1$, in \eqref{e:VMPprodAtX}, can be constructed from the subsets, $prodVMPSet(m_{i+1}, m_t, m_{n-1})$, in polynomial time where each $C\vmpset(m_i, m_{n-1})$ uses only $O(n^6)$ $prodVMPSet(m_{i+1}, m_t, m_{n-1})$.
\end{lemma}
\begin{proof}
The proof on the size of these disjoint sets follows from \eqref{e:VMPxCVMPb}, noting that
$\mid \{ prodVMPSet(m_{i+1}, m_t, m_{n-1})\}\mid = O(n^6)$ for each pair $(m_{i+1}, m_{n-1})$.\\
The bound on the time follows from the polynomial bound on the "increment CVMPSet()".
\end{proof}

\vskip -8pt
\subsection{A Characterization of Polynomial Time Enumeration}\label{a:charactPTimeEnum}

\begin{figure}[h]
\center
\includegraphics[scale=0.3]{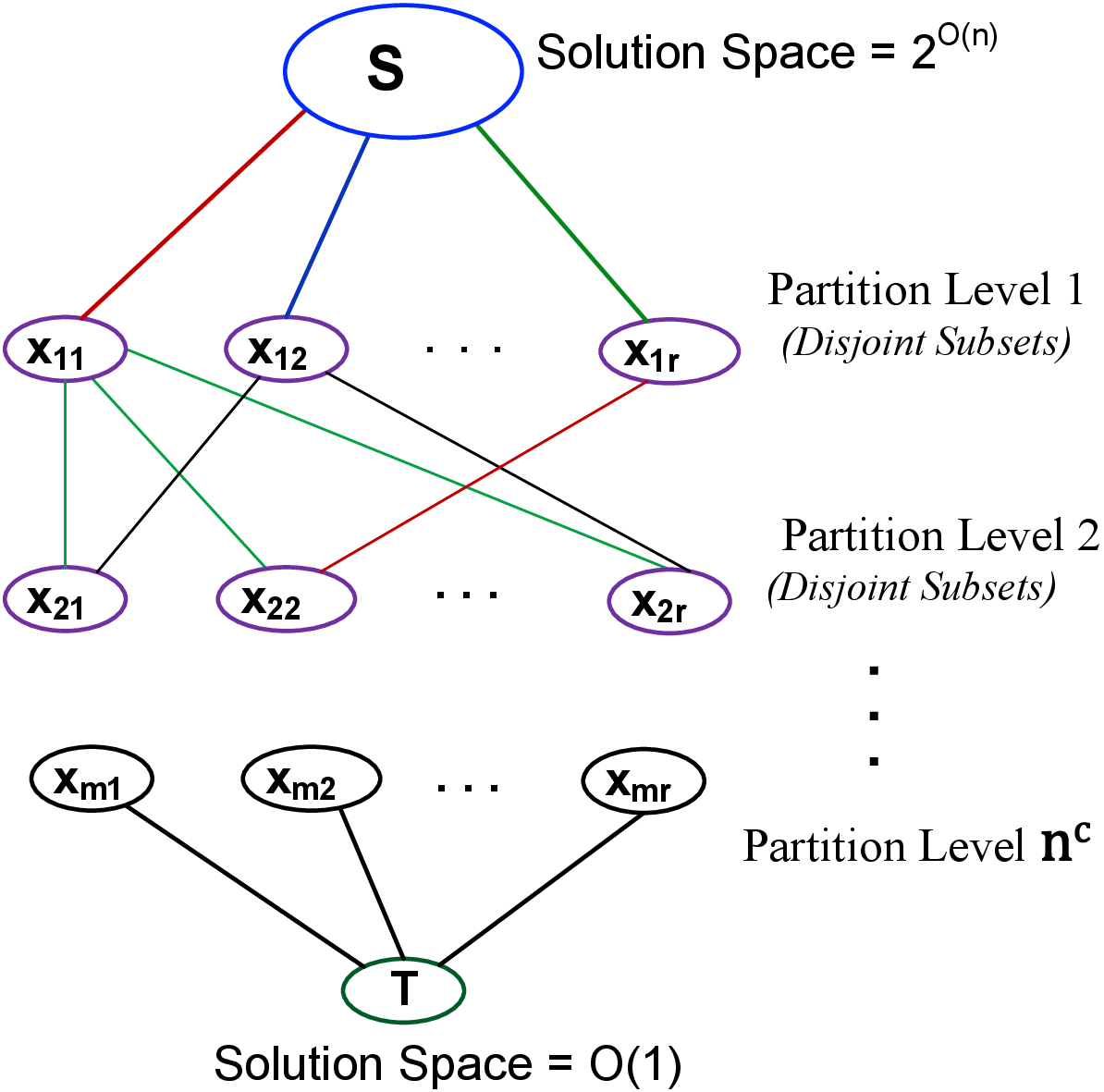}
\vskip 8pt
\caption{Partitions of the Solution Space of a Search Problem}
\end{figure}
\vskip -9pt
\subsubsection*{A Sufficient Condition for P-time Enumeration.}

\textbf{Conjecture 1.}
An enumeration problem $\mathbf{\chi}$ is in $\mathbf{FP}$ if there exists a hierarchy of the partitions of the solution space of $\mathbf{\chi}$ such that\\
 \vskip -0.1in
\begin{enumerate}
 \vskip -015pt
\item
 Each partition at level $i$ in the hierarchy is a polynomially bounded partition of the solution space of a subproblem represented by with mutually disjoint subsets.
 \vskip -015pt
 \item
 All the disjoint subsets at each level can be represented by a unique set of attributes-- that is, the partitioning is not recursive but represented by the generators of polynomial size.
\vspace{-5pt}
 \item
 The solution space at each level in the hierarchy decreases by a factor $c,~ c>1$.

\end{enumerate}
\vskip 0.35in
\subsubsection*{A Necessary Condition for P-time Enumeration.}
\vskip -0.05in
While the conditions listed under Conjecture 1 could be somewhat over restrictive, they must still hold true, in a perhaps more abstract form, in order to allow a P-time enumeration.
\vspace{-7pt}
\subsubsection*{Example: Directed $\mathbf{s-t}$ Paths in an $n$-partite graph }
\vskip -0.05in
Following is an example of an enumeration problem which meets the above characterization.\par

\begin{figure}[h]
\vskip -0.0in
\center
\includegraphics[scale=0.25]{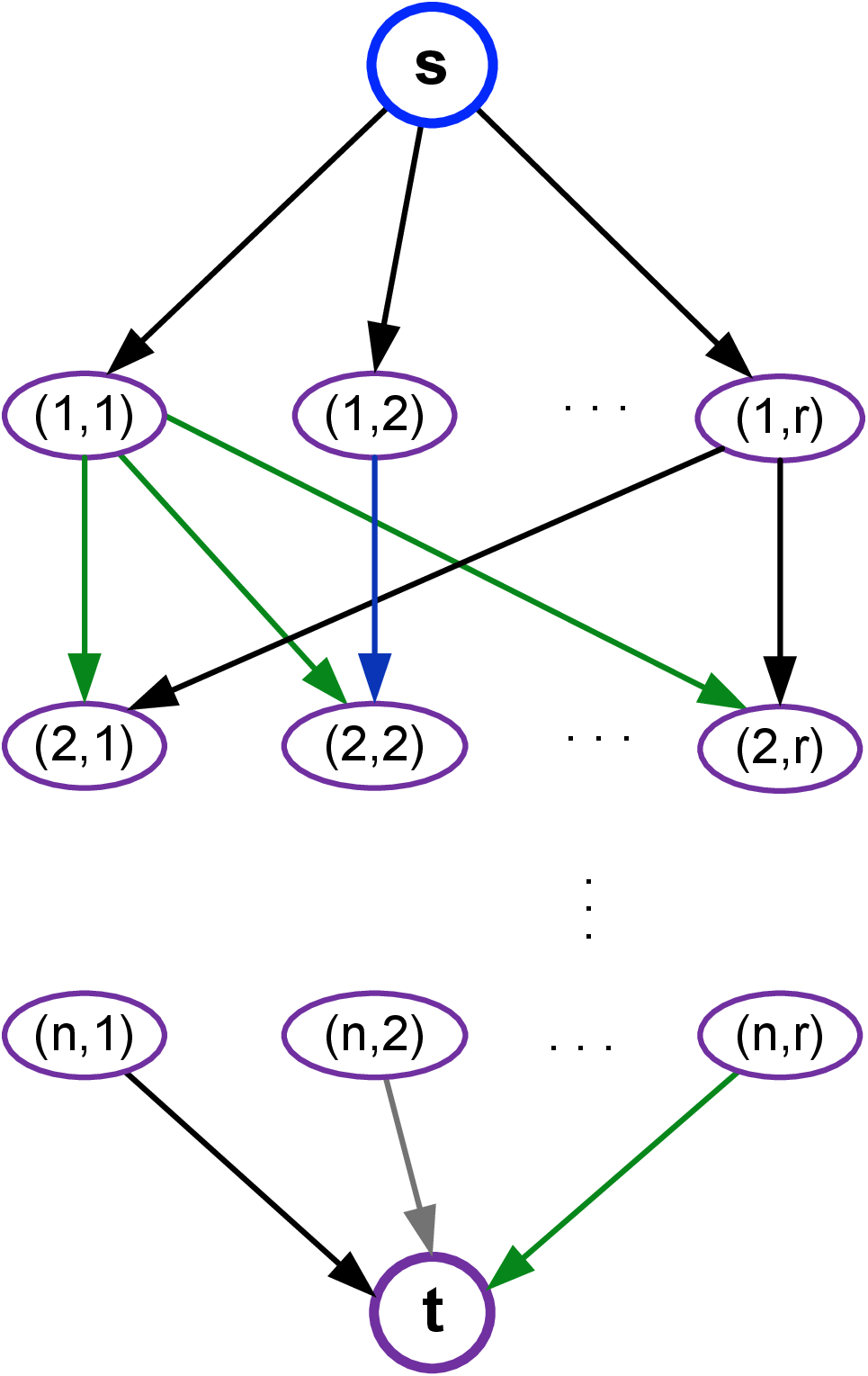}
\hfill
\caption{An $n$-partite Graph}\label{FIG:a:partiteG}
\end{figure}

 Consider an $n$-partite directed acyclic graph [Fig \ref{FIG:a:partiteG} ], $G= (V, E)$,\\ where
 $V = Vs \cup V_1 \cup V_2 \cdots\, \cup V_n \cup Vt$, and $E = \biguplus E_i$, where $E_i \subseteq V_i \times V_{i+1}$.\par

Further, let $x_i \in V_i = \{(i,1), (i,2), (i,3), \cdots\, (i,r) \mid\, r \le n^{O(1)}\}$.\\

Let $s \in Vs$, and $t \in Vt $. \\

Let $P(x_i)$ define the set of all paths between the node pair $(x_i,t)$ in $G$, i.e.,\\

 $$P(x_i) \defn \{ x_i x_{i+1}\, \cdots\, x_n \mid x_r \in V_r\} .$$
\par
Since each path $p \in P(x_i)$ covers exactly one distinct node at each level $i$, $P(x_i)$ can be written as:
$$ \mid P(x_i)\mid ~= ~\sum_{(x_i, x_{i+1}) \in E_i} \hspace{ -15pt}\mid P(x_{i+1})\mid \vspace{ -.1in}.$$

\par
Note that all $P(x_i)$ are disjoint at any level $i$, and hence, $P(x_i)$ is an equivalence class.
\par
The polynomial bound of $O(|V|^3)$, for enumerating $P(s)$ can certainly be achieved by a transitive closure of the $n$-partite graph, assuming the each edge $(x_{i}, x_{i+1}) \in E_i$ can be found in $O(1)$ time. In fact, an optimal bound of $O(|E|)$ can be determined by\\

 $$ T(P(x_i)) = O(|E_i|) + T(P(x_{i+1})).$$
\vskip 0.3in
\par
\textbf{Conjecture 2.}
\emph{The solution space of every enumeration problem of size $n$ is a subset of a universe which is a group isomorphic to a symmetric group of degree $n^{O(1)}$. This solution space is an equivalence class determined by the problem instance.}
	
\subsection{The Equivalence Classes in the Partition Hierarchy for Perfect Matchings}
Extending the original partitioning hierarchy means we have additional equivalence classes implied by the disjoint partitions at each partition level. These additional classes are described below.
\par
\subsubsection*{The Class C\vmpset}
\vskip -7pt
Consider the following relation $\equiv$ over the set, $\{C\vmpset(m_1, m_{n-1})\}$:\\
\begin{quote}
 For each $p, q\in \{C\vmpset(m_1, m_{n-1})\}$,\\
\hskip 0.3in $p\equiv q \,\Longleftrightarrow \,\exists\, m_1 =mdag\less x_1\more$ and $(p', q') \in \{C\vmpset(m_2, m_{n-1})\}$,\\
such that\\
\hskip 0.3in
$m_1\cdot p'= p$ and $m_1 \cdot q'=q$.
\end{quote}
Then the relation $\equiv$ is an equivalence relation giving $C\vmpset(m_1, m_{n-1})$ as an equivalence class.
\par
Other equivalence classes are:
\begin{itemize}
	 \item $prodVMPSet(m_i, m_t, m_{n-1})$ induced by the mdag pair $(m_i, m_t)$.
	\item the subset of MinSet sequences, $CMS_{in}(r)$ in $C\vmpset(m_i, m_{n-1})$, induced
by the MinSet, $MinSet(m_i, m_{t})$ such that $ER(x_{t+1}) \ne \emptyset$, where $x_{t+1} \in m_t$. \end{itemize}

\end{appendices}

\end{document}